\documentclass[11pt]{article}
\usepackage[T1]{fontenc}
\usepackage[utf8]{inputenc}
\usepackage{float}
\usepackage{mathrsfs}
\usepackage{mathtools}
\usepackage{enumitem}
\usepackage{amsmath}
\usepackage{amsthm}
\usepackage{amssymb}
\usepackage{geometry}
\usepackage{microtype}
\usepackage[bookmarks=false,
 breaklinks=false,pdfborder={0 0 1},backref=false,colorlinks=false]
 {hyperref}

\makeatletter

\floatstyle{ruled}
\newfloat{algorithm}{tbp}{loa}
\providecommand{\algorithmname}{Algorithm}
\floatname{algorithm}{\protect\algorithmname}

\usepackage{tikz}
\usetikzlibrary{positioning,arrows.meta,calc}
\usepackage{fullpage}
\usepackage{algorithmic}

\makeatother

\theoremstyle{plain}
\newtheorem{thm}{\protect\theoremname}
\newtheorem{lem}[thm]{\protect\lemmaname}
\newtheorem{fact}[thm]{\protect\factname}
\newtheorem{prop}[thm]{\protect\propositionname}
\theoremstyle{remark}
\newtheorem{rem}[thm]{\protect\remarkname}
\providecommand{\factname}{Fact}
\providecommand{\lemmaname}{Lemma}
\providecommand{\propositionname}{Proposition}
\providecommand{\remarkname}{Remark}
\providecommand{\theoremname}{Theorem}

\begin{document}
\global\long\def\EO{\mathsf{EO}}%

\global\long\def\R{\mathbb{R}}%

\global\long\def\E{\mathbb{E}}%

\global\long\def\B{\mathbb{B}}%

\global\long\def\calh{\mathscr{\mathcal{H}}}%

\global\long\def\cl{\operatorname{cl}}%

\global\long\def\Anc{\operatorname{Anc}}%

\global\long\def\SCC{\operatorname{SCC}}%

\global\long\def\lift{\operatorname{lift}}%

\global\long\def\polylog{\log^{O(1)}}%

\global\long\def\tr{\operatorname{tr}}%

\global\long\def\Diag{\operatorname{Diag}}%

\global\long\def\Hs{\mathcal{H}}%

\global\long\def\dom{\mathrm{dom}}%

\allowdisplaybreaks
\title{A Faster Strongly Polynomial Time Algorithm for Submodular Function Minimization}
\author{
  Alina Ene\thanks{Department of Computer Science, Boston University, \texttt{aene@bu.edu}.} \and
  Huy L. Nguyen\thanks{Khoury College of Computer and Information Science, Northeastern University, \texttt{hu.nguyen@northeastern.edu}.}
}
\date{}
\maketitle
\begin{abstract}
We give a strongly polynomial time algorithm for submodular function
minimization with overall running time $\widetilde{O}(n^{2}\mathsf{EO}+n^{3})$
time, where $\mathsf{EO}$ is the running time of a single evaluation
of the function. Our algorithm improves upon the previously fastest
strongly polynomial algorithms of Lee et al. (FOCS 2015), Dadush et
al. (SODA 2018), and Jiang et al. (SODA 2024) that run in time $\widetilde{O}(n^{3}\mathsf{EO}+n^{4})$.
Similarly to prior works, our work combines a cutting plane algorithm
with the combinatorial arc-finding framework of Fleischer et al. (JACM
2001). We achieve our faster running time by developing an algorithm
that runs the cutting plane method end to end, without restarting
after discovering arcs and contracting strongly connected components.
\end{abstract}

\section{Introduction}

Submodular functions are a general class of functions that capture
objective functions in a wide range of applications from graph and
hypergraph cuts to coverage and entropy functions, among many others.
Both submodular minimization and maximization are extensively studied
and they have led to the development of many classical techniques
in combinatorial optimization. Developments in algorithms for submodular
optimization continue to inspire progress in optimization algorithms
more broadly. In this work, we focus on the classical problem of minimizing
a submodular function given as input via an evaluation oracle. A fundamental
result is that submodular functions can be minimized in polynomial
time: the Lovász extension of a submodular function is a convex function
and minimizing it over the unit hypercube gives an optimal solution
to the discrete problem via a simple rounding strategy \cite{Lov83}.
This connection naturally leads to algorithms with weakly polynomial
running times via convex optimization techniques. Remarkably, submodular
functions can also be minimized in strongly polynomial time. The first
strongly polynomial algorithm for submodular minimization was developed
by Grötschel, Lovász and Schrijver \cite{GLS88} in a pioneering work
with many classical techniques powering a vast number of modern algorithms.
Following this work, there has been a significant line of work with
many developments, including the first combinatorial algorithms and
progressive improvements in the running time \cite{Sch00,IFF01,FI03,Iwa03,Orl09,LSW15,DVZ21,Jia21,JLSZ24}.
The currently fastest strongly polynomial time algorithm was developed
by Jiang et al. \cite{JLSZ24}. They give an algorithm with $O(n^{3}\log n)$
function evaluations and $O(n^{4}\log n)$ additional time, which
improved over the previously fastest algorithms due to Lee et al.
\cite{LSW15} in the logarithmic factors. Jiang \cite{Jia21} gives
an algorithm with an improved number of $O(n^{2}\log n)$ function
evaluations but with an exponential additional time of $2^{O(n)}$.
On the hardness side, the currently best lower bound is an $\Omega(n\log n)$
function evaluations \cite{CGJS22}. In this work, we make progress
toward bridging the remaining gap and provide a faster strongly polynomial
time algorithm with $O(n^{2}\log n)$ function evaluations and $O(n^{3}\polylog n)$
additional time. To achieve a more efficient implementation, our algorithm
uses the approach based on inverse maintenance \cite{LS15} and leverage
score maintenance \cite{JLSW20}, leading to a randomized algorithm.
We also give a simpler implementation of the algorithm that is deterministic
and it has the same $O(n^{2}\log n)$ function evaluations but with
a higher additional time of $O(n^{\omega+1}\polylog n)$, where $\omega$
is the matrix multiplication constant. The algorithm assumes exact
real arithmetic.
\begin{thm}
\label{thm:main} Let $h$ be a submodular function on a ground set
of size $n$ represented via a function value oracle. Let $\EO$ be
an upper bound on the running time of a single call to the value oracle.
There is a randomized algorithm that, with high probability, it returns
an exact minimizer of $h$ using $O(n^{2}\log n)$ value oracle calls
and total running time $O(n^{2}\log n\EO+n^{3}\polylog n).$ 
\end{thm}

Our work is most related to the work of Lee et al. \cite{LSW15} and
Dadush et al. \cite{DVZ21} that developed strongly polynomial time
algorithms with $O(n^{3}\log^{2}n\EO+n^{4}\polylog n)$ time. Similarly
to these works, our work combines a cutting plane algorithm with the
combinatorial arc-finding framework of Fleischer et al. \cite{IFF01,FI02,Fuj05}.
We achieve our faster running time by developing an algorithm that
runs the cutting plane method end to end, without restarting after
discovering arcs and contracting strongly connected components. We
give an overview of our algorithm in Section \ref{sec:map}.

\textbf{Acknowledgements.} The proofs were developed with GPT 6.1
Sol and GPT 6 Astra. ChatGPT was used to develop the ideas and write
the initial proofs. We verified and rewrote all the proofs. 

\section{Preliminaries}

Let $U$ be a finite ground set of size $\left|U\right|=$$n$. We
refer to $\varnothing$ and $U$ as endpoints. The function $h\colon2^{U}\to\R$
is submodular if it satisfies $h(X)+h(Y)\geq h(X\cap Y)+h(X\cup Y)$
for any two sets $X,Y\subseteq U$. We assume that $h$ is given as
input via a value oracle that, on input $X\subseteq U$, returns the
value $h(X)$. We let $\EO$ denote an upper bound on the running
time of a single call to the value oracle. We consider the problem
$\min_{X\subseteq U}h(X)$ of finding an exact minimizer of $h$.

For a given vector $a$, let $a^{-}$ be the vector such that $a^{-}_{i}=\min(a_{i},0)$.
Let $[x]_{+}=\max(x,0)$. For a positive definite matrix $M$, we
let $\left\Vert x\right\Vert _{M}=\sqrt{x^{\top}Mx}$. The $\widetilde{O}\left(\cdot\right)$
notation suppresses factors that are poly-logarithmic in $n$.

\subsection{Base polytopes and greedy vectors}

\label{sec:base-polytope}

Let $f\colon2^{V}\to\R$ be a submodular function on a ground set
$V$ of size $N$ that is normalized, i.e., it satisfies $f(\varnothing)=0$.
The base polytope of $f$ is 
\[
\B(f)=\{a\in\R^{V}:\ a(V)=f(V),\ a(X)\le f(X)\text{ for all }X\}.
\]
Every ordering (i.e., permutation) $\pi$ of $V$ defines a vector
in the base polytope as follows. for each $i$, set $g_{\pi(i)}=f(P_{i})-f(P_{i-1})$
where the sets $P_{j}=\left\{ \pi(1),\dots,\pi(j)\right\} $ are the
prefixes of the permutation. Submodularity can be used to show that
$g\in\mathbb{B}(f)$. The vector $g$ is a vertex of the base polytope.
We refer to $g$ as the \emph{greedy vector} for the ordering $\pi$.

Our algorithm will use the following types of greedy vectors. Let
$w\in\R^{V}$ be a given weight vector and let $\pi$ be the permutation
arising by sorting the weights in non-increasing weight:
\[
w_{\pi(1)}\ge w_{\pi(2)}\ge\cdots\ge w_{\pi(N)}.
\]
As before, for every $1\leq i\leq N$, set $g_{\pi(i)}=f(P_{i})-f(P_{i-1})$
where $P_{j}=\left\{ \pi(1),\dots,\pi(j)\right\} $ for every $0\leq j\leq N$.
The greedy vector $g$ maximizes the linear function $\left\langle w,a\right\rangle $
over the base polytope, i.e. it is an optimal solution to $\max_{a\in\mathbb{B}(f)}\left\langle w,a\right\rangle $~\cite{Edm70}.

The Lovász extension $F\colon\R^{V}\to\R$ is the support function
of the base polytope:
\[
F(x)=\max_{a\in\B(f)}\left\langle x,a\right\rangle .
\]
While the Lovász extension is more commonly defined on the domain
$[0,1]^{V}$, the above definition is valid for $\R^{V}$. A fundamental
result that enables the application of cutting-plane methods to submodular
function minimization \cite{GLS81,GLS88} is that Greedy vectors are
subgradients of the Lovász extension~\cite{Lov83}.

We will also use Edmonds' min--max theorem that states that $\min_{X}f(X)=\max\{a^{-}(V):\ a\in\B(f)\}$,
where we view a vector $a\in\R^{V}$ as the set function $a(X)=\sum_{i\in X}a_{i}$
for all $X\subseteq V$. In particular, for a function $f$ that is
equal to $0$ at both endpoints $\varnothing$ and $V$, $f\ge0$
exactly when $0\in\B(f)$.

\subsection{Self-concordant functions and Newton's method}

Here we present the definitions and properties of self-concordant
functions and Newton's method that we will use in our cutting plane
method. Let $X\subset\R^{n}$ be a closed convex domain. A function
$f\colon\operatorname{int}X\to\R$ is self-concordant (\cite[Definition 2.1]{NT08})
if (1)$f$ is three times continuously differentiable, convex, and
$f(x_{k})\to\infty$ if $x_{k}\to x\in\partial X$ and (2) for every
$x\in\dom(f)$ and $h\in\R^{n}$, we have $\left|D^{3}f(x)\left[h,h,h\right]\right|\leq2\left(D^{2}f(x)\left[h,h\right]\right)^{3/2}$.
We note that the constant $2$ is arbitrary, and it can be changed
to any constant without changing the class of self-concordant functions
up to a rescaling of the function. On the other hand, the constant
$3/2$ in the exponent is not arbitrary and it ensures that self-concordant
functions are affine-invariant. Next, we state the properties of self-concordant
functions that we will use in our cutting plane method.

\textbf{Restriction to a subspace. }Let $S\subseteq\R^{n}$ be a subspace
of dimension $N$, and let $J\in\R^{n\times N}$ be a matrix whose
columns are a basis for $S$. Then we can write the restriction $f_{\vert S}$
of $f$ to $S$ as $g(y)=f(Jy)$. The derivatives of $g$ are $\nabla g(y)=J^{\top}\nabla f(Jy)$,
$\nabla^{2}g(y)=J^{\top}\nabla^{2}f(Jy)J$, and $D^{3}g(y)\left[h,h,h\right]=D^{3}f(Jy)\left[Jh,Jh,Jh\right]$.
Thus, if $f$ is self-concordant, then so is $g$.

\textbf{Local Newton's method for self-concordant functions.} We use
Newton's method to approximately minimize a self-concordant function
$f$. The algorithm starts with an initial solution and iteratively
updates it via the following update. Let $x$ be the current iterate
satisfying $\nabla^{2}f(x)\succ0$. The next iterate is
\[
x^{+}=x-\left(\nabla^{2}f(x)\right)^{-1}\nabla f(x).
\]
We refer to the above update as a full Newton step. The Newton decrement
at $x$ is defined as follows:
\begin{align*}
\mathfrak{d}_{f}(x) & =\sqrt{\nabla f(x)^{\top}(\nabla^{2}f(x))^{-1}\nabla f(x)}=\left\Vert \nabla f(x)\right\Vert _{\left(\nabla^{2}f(x)\right)^{-1}}
\end{align*}
The following properties hold if $r=\mathfrak{d}_{f}(x)<1$:
\begin{enumerate}
\item Hessian comparison \cite[Section 2.1.1]{NT08}: $(1-r)^{2}\nabla^{2}f(x)\preceq\nabla^{2}f(x^{+})\preceq(1-r)^{-2}\nabla^{2}f(x)$
\item Decrement bounds distance to the minimum \cite[Equation 2.6c]{NT08}:
Let $x^{*}$ be the minimizer of $f$ and $\calh=\nabla^{2}f(x)$.
We have $\left\Vert x^{*}-x\right\Vert _{\calh}\le\frac{\mathfrak{d}_{f}(x)}{1-\mathfrak{d}_{f}(x)}$.
\item Decrement bounds objective value \cite[Equation 2.6b]{NT08}: $f(x)-f(x^{*})\le\rho(r)$
where $\rho(s)=-\ln(1-s)-s$.
\item Full Newton steps converge quadratically \cite[Section 9.6.4, Equation 9.55]{BoydVandenberghe2004}:
\[
\mathfrak{d}_{f}(x^{+})\le\left(\frac{\mathfrak{d}_{f}(x)}{1-\mathfrak{d}_{f}(x)}\right)^{2}.
\]
\end{enumerate}

\section{Overview of the algorithm}

\label{sec:map} 
\begin{figure}[t]
\centering \begin{tikzpicture}[
  node distance=5mm and 8mm,
  box/.style={draw, rounded corners=2pt, align=center, font=\small, inner sep=4pt, text width=#1},
  box/.default=8.4cm,
  arr/.style={-{Stealth[length=2.2mm]}, thick},
  lab/.style={font=\footnotesize}
]
\node[box] (cal) {\textbf{Calibrate} the implication DAG\\ using $O(n^2)$ queries over the whole run (Sec. 4.1)}; 
\node[box, below=of cal] (ctr) {Compute \textbf{center} $z$ of the barrier $\Psi_{\mathrm fl}$ on the current subspace $S$ (Sec. 7.1)};
\node[box, below=of ctr] (orc) {Query \textbf{oracle} at $z$: receive graph vector (no query)\\ or greedy vector(at most $N-1$ queries) (Sec. 4.2)};
\node[box, below=of orc] (br) {\textbf{Branch} on $\upsilon=\sqrt{b^{\mathsf T}\calh_{fl}(z)^{-1}b/\gamma}$};
\node[box=4.3cm, right=8mm of orc] (wide) {\textbf{Wide step } $\upsilon\ge\delta$): insert vector as new constraint; $\Phi$ rises by at least $\gamma/8$ (Sec. 5.4)}; 
\node[box, below=of br] (floor) {\textbf{Thin event } ($\upsilon<\delta$): the convex body is thin;\\ create a certificate from gradient of $\Psi_{fl}$ at $z$ (Sec 5.6)};
\node[box, below=of floor] (ep) {\textbf{Arc-addition episode}: find head of arc, find tail using binary search;\\ one arc per $O(\log n)$ queries; repeat while acyclic (Sec. 4.3)};
\node[box, below=of ep] (con) {\textbf{Contract} an SCC: same barrier, smaller subspace,\\ so $\Phi$ can only rise; warm start by a penalty path\\ paid for by that rise (Sec. 7)};
\node[box=4.3cm, right=8mm of ep, fill=gray!12] (stop) {\textbf{Stop} and return the witness: zero greedy vector, $\Delta\ge\beta$, or $N=1$};
\draw[arr] (cal) -- (ctr);
\draw[arr] (ctr) -- (orc);
\draw[arr] (orc) -- (br);
\draw[arr] (br.east) -| node[lab, pos=0.25, below] {$\upsilon\ge\delta$} (wide.south);
\draw[arr] (wide.north) |- (ctr.east);
\draw[arr] (br) -- node[lab, right] {$\upsilon<\delta$} (floor);
\draw[arr] (floor) -- (ep);
\draw[arr] (ep) -- node[lab, right] {SCC formed} (con);
\draw[arr] (ep.east) -- (stop.west);
\draw[arr] (con.west) -- ++(-6mm,0) |- (cal.west);
\end{tikzpicture}\caption{Overview of the algorithm. Only greedy vector generation, arc searches
and calibration query the oracle; everything else works on stored
data. $\Phi=\min_{P\cap S}\Psi_{fl}$ is the potential of Theorem~\ref{thm:fresh}.}
\label{fig:flow}
\end{figure}
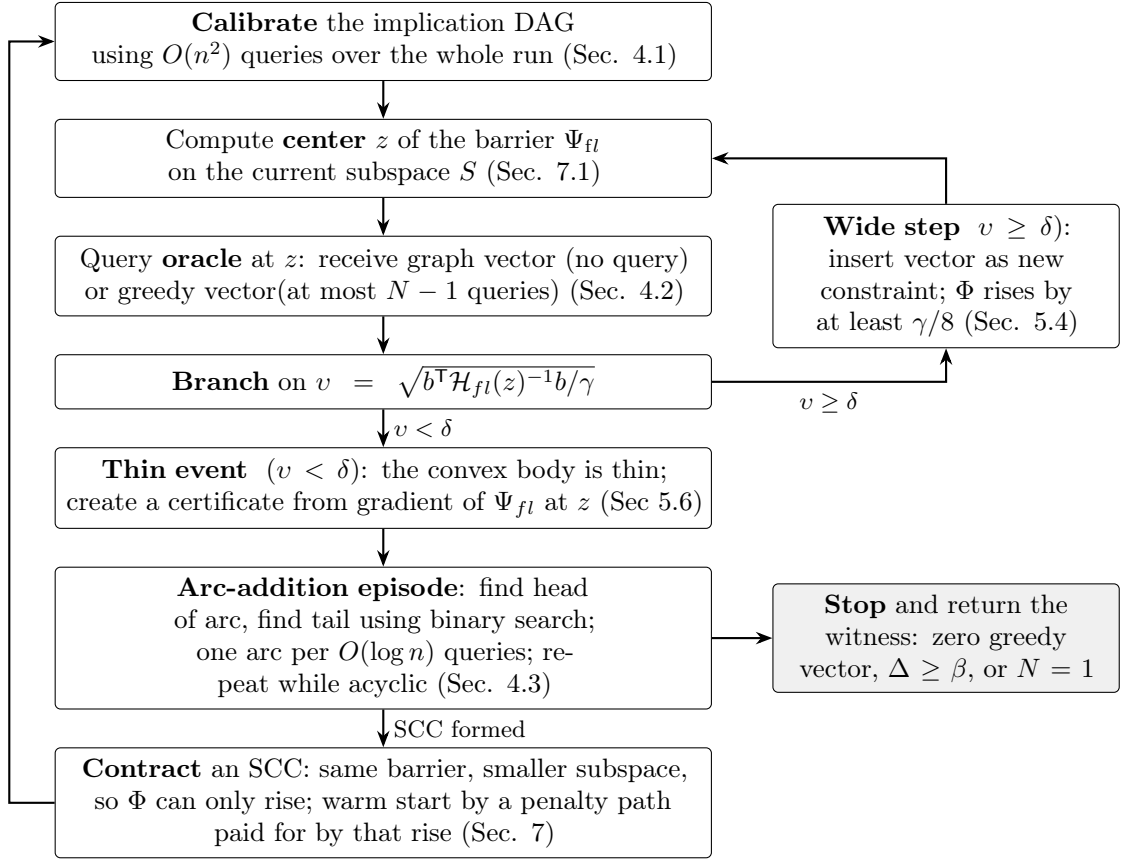

Figure~\ref{fig:flow} shows an overview of the
algorithm. At a high level, the algorithm follows the cutting plane
method for strongly polynomial submodular function minimization (SFM)
initiated by Grötschel, Lovász and Schrijver~\cite{GLS88}. Our method
most closely follows the more recent works by Lee, Sidford and Wong~\cite{LSW15}
and by Dadush, Végh and Zambelli~\cite{DVZ21}. It uses the cutting
plane method to find the minimum of the Lovász extension over the
unit hypercube. The cutting plane method maintains a barrier function
and queries the center of this barrier. The objective function is
modified appropriately to produce a cut/hyperplane through this point,
thus reducing the feasible set (the region where the minimizer must
lie). Eventually the feasible set becomes very thin and the algorithm
produces a certificate for this thinness. The certificate is turned
into a nearly cancelling combination of base vectors, which yields
new implications $u\to z$ satisfied by every minimizer (if a minimizer
contains $u$ then it must also contain $z$). Each implication shrinks
the family of sets satisfying all the implications, and implications
forming a cycle reduce the dimension of the family. Simultaneously
this also reduces the dimension of the subspace where we find the
minimizer of the Lovász extension. Intuitively when we show that the
feasible set is sufficiently thin, we can show that it must be a lower
dimensional subspace. This process is repeated until the dimension
is reduced to 1, or we obtain a certificate showing that the minimizer
is already found. Compared with prior works, the algorithm is more
efficient in several ways.
\begin{itemize}
\item The cutting-plane algorithms of~\cite{LSW15,DVZ21} restart the cutting-plane
method after each dimension reduction. In our method, there is only
one run of the cutting plane method from the beginning to the end
with one barrier function. The minimum of the barrier function does
not decrease when we reduce the dimension of the feasible set. Every
time that we obtain a new cut, either the barrier minimum increases
significantly or it leads to a dimension reduction. The work to identify
the new center after each dimension reduction is also relatively cheap
since the barrier function is the same, just with a different domain.
\item The same certificate obtained from the thin region can be reused to
obtain implications repeatedly until the next dimension reduction.
Each implication can be obtained cheaply using only $O(\log n)$ queries.
Since there are at most $O(n^{2})$ implications throughout, the total
number of queries is bounded.
\end{itemize}

The deterministic arithmetic bound has two components, the geometric
part and the combinatorial part. The combinatorial part uses data
structures to maintain incremental reachability as we add arcs, and
summation of weights on ancestors of each item. The total runtime
for the combinatorial part is $\widetilde{O}(n^{3})$. The geometric
work is done by exact Newton steps, each costing $\widetilde{O}(n^{\omega})$,
and there are only $\widetilde{O}(n)$ of them. Thus the total arithmetic
bound is $\widetilde{O}(n^{\omega+1})$. The randomized runtime improves
upon the deterministic bound by using inverse maintenance and leverage
score maintenance instead of exact Newton steps. 

The one quantity that is used to bound both the number of cutting
planes and the cost of recentering after contraction is the potential
\[
\Phi(S,P)=\min_{P\cap S}\Psi_{fl},
\]
the minimum of a single barrier $\Psi_{fl}$ over the current section
of the cutting-plane body $P$. Wide steps raise it by a constant,
contractions never lower it, and it never exceeds $O(n\log n)$. The
same potential is used to bound the number of separation oracle calls
and the cost of updating the center after contractions.

\section{Inferring arcs from the cutting plane certificates}

Similarly to previous work, we build on the combinatorial framework
first introduced by Iwata et al. \cite{IFF01} and subsequently used
in the works \cite{IO09,LSW15,DVZ21}. In this framework, we maintain
a directed graph in which every arc $u\to v$ represents the implication
that any solution containing $u$ must also contain $v$. We refer
to this graph as the \emph{implications graph}. Initially, the implications
graph has vertex set $U$ and an empty arc set, and new arcs are added
as the algorithm proceeds. Before describing how the algorithm identifies
new arcs, we first introduce some notation that we will use throughout.
We define the \emph{quotient DAG} to be the acyclic graph obtained
by contracting each strongly connected component of the implications
graph to a single vertex. We refer to the vertices of the quotient
DAG as \emph{blocks}. A set $S\subseteq U$ is \emph{closed} when
$u\in S$ and the implication $u\to v$ is in the graph imply $v\in S$.
Note that a closed set is a union of SCCs in the implications graph,
and thus a union of blocks. The family of closed sets is a ring family,
i.e., it is closed under union and intersection. For a set $X\subseteq V(D)$
of vertices of the quotient DAG $D$, the closure of $X$, denoted
by $\cl_{D}(X)$, is the smallest set containing $X$ that is closed.
The lift of $X\subseteq V(D)$, denoted by $\lift(X)$, is the union
of the blocks corresponding to the vertices in $X$. We refer to the
$\varnothing$ and $V(D)$ as the endpoints.

In the remainder of this section, we describe the combinatorial component
of our algorithm that identifies new arcs from a certificate returned
by our cutting plane method.

\subsection{Algorithm invariant and calibration}

\label{sec:calibration}

The algorithm maintains a value $m$ and a set $W$ satisfying $h(W)=m$.
The set $W$ is the set with smallest $h(\cdot)$ value that the algorithm
has evaluated so far and $m:=h(W)$ is its value. We refer to $W$
as the witness for the value $m$. The algorithm initializes $m$
and $W$ by querying $h(\varnothing)$ and $h(U)$ and setting
\[
W=\arg\min_{X\in\left\{ \varnothing,U\right\} }h(X)\qquad m=h(W).
\]
\textbf{The sliding parameter.} The value $m$ decreases as the algorithm
progresses. It is convenient to reparameterize $m$ so that we work
with a parameter $t$ that increases as the algorithm progresses.
The parameter $t$ is defined as follows:
\[
t:=h(\varnothing)+1-m.
\]
We refer to the parameter $t$ as the \emph{sliding parameter}. The
maximum possible value of $t$ is
\[
t^{*}:=h(\varnothing)+1-\min_{X}h(X).
\]
Every time the algorithm evaluates a new set $S$ with $h(S)<m$,
it immediately updates $m=h(S)$, $t=h(\varnothing)+1-m$, and $W=S$.
Thus, throughout the algorithm, $m$ is the minimum $h(\cdot)$ value
among all of the sets evaluated by the algorithm so far and $t=h(\varnothing)+1-m$.

 \textbf{Algorithm invariant.} Next, we define an invariant that
the algorithm will maintain. For a parameter $s$, let $h_{s}\colon2^{U}\to\R$
be the function such that $h_{s}(X)=h(X)-h(\varnothing)-1+s$ for
every $X\subseteq U$. Note that, at the current parameter, we have
$h_{t}(X)=h(X)-m$. The algorithm maintains the following invariant:
\begin{equation}
\begin{gathered}\text{For every }s\ge t,\text{ every set \ensuremath{X\subseteq U} with }h_{s}(X)\le0\\
\text{is a closed set in the current implications graph.}
\end{gathered}
\label{eq:invariant}
\end{equation}
Note that, if $X$ is a minimizer of $h$, for every $t\le s\le t^{*}$,
we have $h_{s}(X)\leq h_{t^{*}}(X)=0$ where the first inequality
follows from $s\leq t^{*}$ and the definition of $h_{s'}$ for every
$s'$, and the second equality follows from $h_{t^{*}}(X)=h(X)-m^{*}$
where $m^{*}$ is the minimum $h(\cdot)$ value. Thus the invariant
implies that every minimizer is a closed set in the current implications
graph.

\textbf{The slided submodular function.} Following previous work \cite{Fuj05,DVZ21},
for a parameter $t$, we define a function $H_{t}$ as follows. Recall
that we contract each SCC of the implications graph to obtain the
current quotient DAG $D$, with $N$ vertices corresponding to $N$
blocks of elements. For a parameter $t$, we define the following
function $H_{t}\colon2^{V(D)}\to\R$ on the vertex set of the quotient
DAG:
\begin{equation}
H_{t}(X)=\begin{cases}
0, & X=\varnothing\text{ or }X=V(D),\\
h(\lift(X))-h(\varnothing)-1+t, & \text{otherwise.}
\end{cases}\label{eq:normalized}
\end{equation}
The function $X\mapsto h(\lift(X))-m$ is submodular on all subsets
of $V(D)$. Its endpoint values are nonnegative, since $h(\varnothing),h(U)\geq m$,
and replacing those values by zero preserves submodularity. Hence
$H_{t}$ is submodular and vanishes at both endpoints. The same argument
applies to $H_{s}$ for every $s\ge t$.

For $s\ge t$, we have $H_{s}=H_{t}+(s-t)F$, where $F(X)=1$ for
every $\varnothing\subsetneq X\subsetneq V(D)$ and $F(X)=0$ for
$X\in\left\{ \varnothing,V(D)\right\} $. Thus $F\ge0$. Over time,
the value of $m$ only decreases and as a consequence, the parameter
$t$ and $H_{t}$ only increase. The algorithm terminates when we
can certify that $H_{t}(X)\ge0$ for all closed sets $X$. Together
with invariant \eqref{eq:invariant}, this shows that the witness
for the current parameter $t$ is a minimizer of $h$.

\textbf{Calibration.} For a block $z$, let $\Anc_{D}(z)$ be the
set of vertices that can reach $z$ in the quotient DAG, including
$z$ itself. Let
\[
A_{z}=V(D)\setminus\Anc_{D}(z),\qquad B_{z}=A_{z}\cup\{z\},\qquad c_{z}=H_{t}(A_{z}).
\]
Both $A_{z}$ and $B_{z}$ are closed sets. 

The function $H_{t}$ is called \emph{calibrated} if 
\begin{equation}
H_{t}(A_{z})\ge0,\qquad H_{t}(B_{z})\ge0\quad\text{for every }z.\label{eq:calibration}
\end{equation}
Before using the current DAG, the algorithm enforces these calibration
constraints by increasing the parameter $t$ as necessary. Evaluate
any uncached $h(\cdot)$ value of $\lift(A_{z}),\lift(B_{z})$. If
any value is smaller than $m$, we update $m$ and increase $t$ accordingly.
All other cached values can be updated to the current $H_{t}(\cdot)$
without additional function evaluations because $H_{t}(\cdot)$ is
an affine function of $t$. The sets $\varnothing$ and $V(D)$ get
value $0$ by definition and need no query. Both constraints in \eqref{eq:calibration}
are used below: $H_{t}(B_{z})\ge0$ gives the marginal lower bound
in Fact~\ref{fact:cal}(a), and $H_{t}(A_{z})\ge0$ makes every $c_{z}$
nonnegative, as Lemma~\ref{lem:no-cancel} requires.
\begin{lem}[Calibration running time]
\label{lem:cal-cache} All calibration steps can be implemented using
$O(n^{2})$ function evaluations and $O(n^{3})$ additional time.
\end{lem}

\begin{proof}
We maintain the reachability relation on original items as new arcs
are added to the graph. By the data structure of \cite{Ita86}, this
takes $O(n)$ amortized time per arc insertion and $O(n^{3})$ overall.
Here we describe a simple method to get within our time bound. On
inserting $x\to y$, scan the at most $n$ known predecessors $u$
of $x$ and enqueue each new pair $(u,y)$. When a once-only new pair
$(u,v)$ is processed, scan the original outgoing edges $v\to w$
and enqueue each new $(u,w)$. There are at most $n^{2}$ inserted
edges and $n^{2}$ successful pair insertions, each scanning $O(n)$
entries. This gives $O(n^{3})$ work and $O(n^{2})$ space. Rebuilding
SCC/quotient metadata in $O(n^{2})$ after each merge also costs $O(n^{3})$
in total.

For an original representative $z$, the two lifted sets $\lift(A_{z})$
and $\lift(B_{z})$ are exactly 
\[
U\setminus\Anc(z),\qquad U\setminus\bigl(\Anc(z)\setminus\SCC(z)\bigr).
\]
where $\Anc(z):=\lift(\Anc_{D}(z))$ is the lift of the ancestors
of $z$ (i.e. the set of original items that can reach $z$) and $\SCC(z)$
is the strongly connected component containing $z$. The first set
changes only when $z$ obtains a new ancestor $u$. The second set
can lose $u$ on that event or gain $u$ when $u,z$ first become
mutually reachable. Each of these two kinds of event occurs at most
once. Including initialization, the total number of changed calibration
sets is therefore $O(n^{2})$. Building a changed lift explicitly
costs $O(n)$. Parameter changes alone do not change these original
sets.
\end{proof}

The key properties of the calibrated function are the following. An
ordering $\pi$ of $V(D)$ is descendant-first if, for every $u\to v$
arc in $D$, $v$ appears before $u$ in $\pi$.
\begin{fact}[Calibration facts]
\label{fact:cal} Let $D$ be calibrated at parameter $t$ and let
$s\ge t$ (calibration persists, since $F\ge0$). 
\begin{enumerate}[label=(\alph*)]
\item If $T$ and $T\cup\{v\}$ are closed and $v\notin T$, then $H_{s}(T\cup\{v\})-H_{s}(T)\ge-c_{v}$,
where now $c_{v}=H_{s}(A_{v})$. 
\item If $\varnothing=T_{0}\subset T_{1}\subset\dots\subset T_{N}=V(D)$
are the prefixes of a descendant-first order $\pi$, then $H_{s}(T_{j})\le\sum_{k>j}c_{\pi(k)}\le\sum_{z}c_{z}$. 
\end{enumerate}
\end{fact}

\begin{proof}
(a) A closed $T$ with $v\notin T$ contains no ancestor of $v$,
so $T\subseteq A_{v}$. Diminishing returns gives $H_{s}(T+v)-H_{s}(T)\ge H_{s}(B_{v})-H_{s}(A_{v})\ge-c_{v}$.
(b) By construction, all $T_{i}$ are closed. Telescope (a) backwards
from $H_{s}(V(D))=0$ and use $c\ge0$. 
\end{proof}

\subsection{Implementation of the separation oracle for the cutting plane method
}

\label{subsec:separation-oracle}

Next, we describe the separation oracle that we will use to implement
our cutting plane method. Let $D_{i}$ be the current quotient DAG.
Let $t$ be the current sliding parameter and assume that $H_{t}$
is calibrated (as in Section \ref{sec:calibration}). The cutting
plane method queries the oracle using a point $w\in\R^{V(D_{i})}$
with coordinates corresponding to the vertices of $D_{i}$. Given
$w$, the oracle constructs a vector as we describe below. For intuition,
the returned vector $v$ specifies a homogeneous hyperplane $\left\{ x\in\R^{V(D_{i})}\colon\left\langle v,x\right\rangle =0\right\} $
that weakly separates $w$ from the indicator vectors of all $D_{i}$-closed
sets with negative $H_{t}$ values: we have $\left\langle v,w\right\rangle \ge0$
and $\left\langle v,\mathbf{1}_{X}\right\rangle \leq0$ for every
$D_{i}$-closed set $X\subseteq V(D_{i})$ with $H_{t}(\lift(X))\le0$.
However, this property is not necessary for the subsequent analysis.

We now describe how to construct the vector returned by the oracle
on query $w$. First test $w_{u}\le w_{v}$ for every arc $u\to v$
in $D_{i}$. If there is an arc $u\to v$ with $w_{u}>w_{v}$, we
return the vector $g_{uv}=e_{u}-e_{v}$, which satisfies $\left\langle g_{uv},w\right\rangle >0$.
This costs no function evaluations and $O(n|V(D_{i})|)$ arithmetic
operations. We refer to this vector $g_{uv}$ as a \emph{graph vector}.

If there are no violated arcs, we return a vector $a$ constructed
as follows. Let $\pi$ be the ordering of the vertices of $D_{i}$
by decreasing weight
\[
w_{\pi(1)}\ge w_{\pi(2)}\ge\cdots\ge w_{\pi(|V(D_{i})|)},
\]
breaking ties as follows: if $v$ is a descendant of $u$ in $D_{i}$,
we place $v$ before $u$ in the ordering.

Let $P_{0}=\varnothing,P_{1},\ldots,P_{N}=V(D_{i})$ be the prefixes
of the ordering. We query the values $h(P_{j})$ for all $P_{j}$
that have not been evaluated previously. With each evaluation, we
update the minimum value $m$ and its witness set if $h(P_{j})<m$,
and update $t$. This costs at most $|V(D_{i})|-1$ function evaluations
and $O(n|V(D_{i})|)$ arithmetic operations. Note that, after the
update, we have $H_{t}(P_{j})\geq0$ for all $j$. After performing
the evaluations and the updates, we define the vector $a$ as follows:
$a_{\pi(j)}=H_{t}(P_{j})-H_{t}(P_{j-1})$ for every $j$. The vector
$a$ is an optimal solution to $\max_{a'\in\mathbb{B}(H_{t})}\left\langle w,a'\right\rangle $,
where $\mathbb{B}(H_{t})$ is the base polytope of $H_{t}$ (Section
\ref{sec:base-polytope}). Note that we have
\[
\left\langle a,w\right\rangle =\sum^{|V(D_{i})|-1}_{j=1}(w_{\pi(j)}-w_{\pi(j+1)})H_{t}(P_{j}).
\]
Since $w_{\pi(j)}-w_{\pi(j+1)}\ge0$ and $H_{t}(P_{j})\ge0$ for all
$j$, we have $\left\langle a,w\right\rangle \geq0$. If $a=0$ then
$0\in\mathbb{B}(H_{t})$, which certifies that $H_{t}$ is nonnegative
and the algorithm terminates as the witness set is a minimizer of
$h$. Otherwise the algorithm returns $a$ together with the ordering
$\pi$, DAG $D_{i}$, partition of the ground set into blocks given
by $D_{i}$, and timestamp $t_{i}=t$ where $t$ is the value of the
sliding parameter after the update above. The ordering, the DAG, the
partition into blocks, and the timestamp are stored with $a$ and
are never changed. We refer to this vector $a$ as the \emph{greedy
vector} (for the ordering $\pi$), and we refer to $\pi$ as the \emph{historical
ordering} for $a$. The vertices of the DAG and the blocks will change
as we perform contractions, and the historical ordering refers to
the fact that it is the ordering with respect to the DAG vertices
at the time the vector was defined.

In both cases, the oracle returns a vector $v$ ($v=g_{uv}$ or $v=a$)
such that $\left\langle v,w\right\rangle \ge0$, and its homogeneous
halfspace is $\left\{ x:\left\langle v,x\right\rangle \le0\right\} $.
For a greedy vector, this halfspace contains the indicator vector
of every closed set $X$ with $H_{t}(X)<0$, since $a(X)\le H_{t}(X)<0$.

We conclude this section with the following lemma that shows that
cuts (separating hyperplanes) constructed by prior oracle calls remain
valid after parameter updates and SCC contractions.
\begin{lem}[Historical cuts remain valid]
\label{lem:historical-cuts-valid} Let $\widetilde{a}$ be a stored
greedy vector constructed by a prior oracle call, mapped to the vertices
of the current DAG by summing the values on the vertices that were
merged during contractions. Let $\widetilde{t}$ be $\widetilde{a}$'s
timestamp. For every proper nonempty $D_{i}$-closed set $X$, we
have
\[
\widetilde{a}(X)\leq H_{t_{i}}(X)-(t_{i}-\widetilde{t}).
\]
If $\widetilde{a}=0$ then $H_{t_{i}}$is non-negative on all $D_{i}$-closed
sets and the algorithm can terminate since a minimum has been found.

For a stored graph vector $\widetilde{g}_{uv}$ mapped to the current
DAG by summing the values on the vertices that were merged during
contractions, we have $\widetilde{g}_{uv}(X)\leq0$ for every $D_{i}$-closed
set $X$. If $\widetilde{g}_{uv}=0$, the vector can be discarded.
\end{lem}

\begin{proof}
Since we only add arcs over time, each SCC in the current DAG $D_{i}$
is a union of SCCs in the DAG $\widetilde{D}$ when $\widetilde{a}$
was constructed. Thus, if $X$ is $D_{i}$-closed now, its representation
as a union of vertices of $\widetilde{D}$ is a $\widetilde{D}$-closed
set and has the same lift as $X$. The stored greedy vector belongs
to $\B(H_{\widetilde{t}})$, and thus
\[
\widetilde{a}(X)\le H_{\widetilde{t}}(X)=H_{t_{i}}(X)-(t_{i}-\widetilde{t}).
\]
If $\widetilde{a}=0$, for every $D_{i}$-closed set $X$, we have
\[
0=\widetilde{a}(X)\leq H_{t_{i}}(X)-(t_{i}-\widetilde{t}).
\]
Since $t_{i}\geq\widetilde{t}$, it follows that $H_{t_{i}}(X)\geq0$
for every $D_{i}$-closed set $X$, and the algorithm can terminate
since a minimum has been found. 

Suppose $\widetilde{g}_{uv}$ is a graph vector. The are two possibilities:
either $u\to v$ is an arc of $D_{i}$ or its endpoints been merged
in a contraction. In the latter case, $\widetilde{g}_{uv}=0$ since
we obtained $\widetilde{g}_{uv}$ by summing the entries of $g_{uv}=e_{u}-e_{v}$.
In the former case, $\widetilde{g}_{uv}(X)\leq0$ for every $D_{i}$-closed
set $X$ since any closed set that contains $u$ also contains $v$.
\end{proof}

\subsection{Deriving arcs from the cutting plane certificate}

\label{subsec:arcs-from-certificate}

As in Lemma \ref{lem:historical-cuts-valid}, we map each stored vector
output by the separation oracle to a vector on the vertices of the
current DAG by summing the values on the vertices that were merged
during contractions. 

A certificate is a convex combination of nonzero stored vectors mapped
to the current DAG, both greedy vectors and graph vectors, with coefficients
$\alpha_{i}$ summing to one. Every stored vector returned by the
separation oracle defined in Section \ref{subsec:separation-oracle}
sums to zero, since greedy vectors lie in base polytopes and $g_{uv}=e_{u}-e_{v}$;
the mapping to the current DAG preserves these sums. Let
\[
y=\sum_{i}\alpha_{i}a_{i},\quad L=\sum_{i}\alpha_{i}\|a_{i}\|_{1}>0,\quad\beta_{X}=\sum_{v\in X}[-y_{v}]_{+},\quad\beta=\beta_{V(D)}=\|y\|_{1}/2,
\]
where the last equality uses $y(V(D))=0$. Define the following quantities
that only concern greedy vectors: 
\[
\theta=\sum_{i\text{ greedy}}\alpha_{i},\qquad\Delta=\sum_{i\text{ greedy}}\alpha_{i}(t-t_{i}),
\]
where $t$ is the current sliding parameter and $t_{i}$ is the timestamp
of $a_{i}$ (the value of the sliding parameter at the time $a_{i}$
was constructed, which was stored with $a_{i}$).

Let $r$ be the number of terms $\alpha_{i}>0$. All coordinates are
known. The certificate, its coefficients, and its stored vectors stay
fixed until the next SCC contraction. Only $t$ and $\Delta$ change.
By Lemma~\ref{lem:historical-cuts-valid}, 
\begin{equation}
y(X)\le\theta H_{t}(X)-\Delta\quad\text{for proper nonempty current closed }X.\label{eq:archive-ring}
\end{equation}
The following lemmata show that intuitively, if $\left\Vert y\right\Vert _{1}$
is small compared with the weighted average length of the terms then
some calibration value $\theta c_{z}$ must be large, and hence we
can identify a new arc.
\begin{lem}[Acyclic no-cancellation]
\label{lem:no-cancel} For a calibrated current DAG with $N$ blocks
and $\bar{c}=\sum_{z}c_{z}$, 
\begin{equation}
L\le4N\theta\bar{c}+4N\beta.\label{eq:no-cancel}
\end{equation}
\end{lem}

\begin{proof}
Choose a descendant-first order with closed prefixes $T_{j}$, and
set $d_{j}=H_{t}(T_{j})-H_{t}(T_{j-1})$. By Fact \ref{fact:cal},
calibration gives $d_{j}\ge-c_{\pi(j)}$ and $H_{t}(T_{j})\le\bar{c}$.
Consider a term $a_{i}$ and let $p_{j}=a_{i}(T_{j})$. If $a_{i}$
is a greedy vector then by Lemma \ref{lem:historical-cuts-valid},
$p_{j}\le H_{t}(T_{j})\le\bar{c}$. If $a_{i}$ is a graph vector
then by Lemma \ref{lem:historical-cuts-valid}, $p_{j}=a_{i}(T_{j})\le0$.
Use $\bar{c}_{i}=\bar{c}$ for greedy terms and $\bar{c}_{i}=0$ for
graph terms. We have 
\[
\|a_{i}\|_{1}=\sum^{N}_{j=1}\left|p_{j}-p_{j-1}\right|\le\underbrace{\left|p_{0}\right|+\left|p_{N}\right|}_{0}+2\sum_{j<N}|p_{j}|\le4(N-1)\bar{c}_{i}-2\sum_{j<N}p_{j}.
\]
The last inequality uses the identity $|x|=2[x]_{+}-x$. After averaging
with coefficient $\alpha_{i}$, we have 
\[
L=\sum_{i}\alpha_{i}\|a_{i}\|_{1}\le4(N-1)\theta\bar{c}-2\sum_{j<N}y(T_{j})
\]
The sum $\sum_{j<N}y(T_{j})$ is a linear function of coordinates
of $y$ with coefficients between zero and $N-1$, hence bounded by
$N\|y\|_{1}=2N\beta$ in absolute value. This proves \eqref{eq:no-cancel}. 
\end{proof}

If $\theta>0$ and $\Delta\ge\beta$, then \eqref{eq:archive-ring}
implies $H_{t}(X)\ge0$ on all proper closed sets, which certifies
that the current witness is an optimal solution, and the algorithm
terminates. Otherwise define the uniform historical extraction guard
\begin{equation}
G=[r(n-1)-1](\beta-\Delta),\qquad0\le\Delta<\beta.\label{eq:guard}
\end{equation}
The formula uses the original $n-1$ and not the current $N$ because
some historical order may have up to $n$ blocks.
\begin{lem}[A block containing the head of a new arc ]
\label{lem:target} If $\theta>0$, $\Delta<\beta$, and 
\begin{equation}
L>4N^{2}G+4N\beta,\label{eq:safety}
\end{equation}
there is a block $z$ with $\theta c_{z}>G$. Its $A_{z}$ is nonempty
and a proper subset of the ground set. 
\end{lem}

\begin{proof}
If every $\theta c_{z}\le G$, then \eqref{eq:no-cancel} gives $L\le4N^{2}G+4N\beta$,
a contradiction. An empty $A_{z}$ has $c_{z}=0$. $A_{z}$ always
excludes $z$ so it does not contain all blocks. 
\end{proof}

\begin{lem}[One new arc from the certificate]
\label{lem:extract} Given $\theta c_{z}>G$, one can find a block
$u\in A_{z}$ such that every closed set $X$ containing $u$ but
not containing $z$ has $H_{t}(X)>0$ (i.e. we can add the arc $u\to z$
to the graph) using $O(\log(n+2))$ EO and $O(rn+n^{2}+n\log(n+2))$
arithmetic operations. This leaves the certificate unchanged. 
\end{lem}

\begin{proof}
Let $A=A_{z}$ and $y^{+}(A)=\sum_{v\in A}[y_{v}]_{+}$, so that $y(A)=y^{+}(A)-\beta_{A}$.
All closed sets not containing $z$ are exactly the closed subsets
of $A$.

If $\Delta>\beta_{A}$, then every nonempty closed $X\subseteq A$
has $\theta H_{t}(X)\ge-\beta_{A}+\Delta>0$, so any $u\in A$ works. 

Otherwise $0\le\Delta\le\beta_{A}\le\beta$. Define the following
gap for each term in the convex combination of the certificate:
\[
\mathrm{gap}_{i}=\alpha_{i}[H_{t_{i}}(A)-a_{i}(A)]\quad\text{(greedy)},\qquad\mathrm{gap}_{i}=-\alpha_{i}g_{i}(A)\quad\text{(graph)}.
\]
For each greedy term $i$, every set $A$ that is closed now was also
closed at time $t_{i}$ when the greedy vector was created (after
mapping it to the vertices of the DAG at that time). The greedy gap
value can be computed without EO because $H_{t_{i}}(A)=c_{z}-(t-t_{i})$
is computable from the current $c_{z}$, $t,t_{i}$. The graph gap
value can also be computed without EO since it only depends on whether
the end points are in $A$. The sum of the gaps is $\theta c_{z}-\Delta-y(A)$.

Choose the term $j$ with the largest gap so that $\mathrm{gap}_{j}\ge\frac{\theta c_{z}-\Delta-y(A)}{r}$.
If the maximum gap is $0$, select a greedy term with gap $0$. First
consider the case term $j$ is a greedy term. Let $a_{j}$ be the
corresponding greedy vector, and let $D_{j}$, $\pi_{j}$, and $t_{j}$
be the DAG, ordering, and timestamp at the time $a_{j}$ was constructed
(recall that they are stored with $a_{j}$). Thus $a_{j}\in\R^{V(D_{j})}$
is a greedy vector in the base polytope of $H_{t_{j}}$ with ordering
$\pi_{j}$. Create a new vector $a_{j}'$ in the base polytope of
$H_{t_{j}}$ by pushing as follows. Map the current set $A$ to a
set $A'$ of $k\le n-1$ old blocks in $D_{j}$. Create a new ordering
$\pi_{j}'$ from $\pi_{j}$ where all blocks in $A'$ come first,
preserving their relative order in $\pi_{j}$, and then all remaining
blocks, preserving the relative order in $\pi_{j}$. Every prefix
of $\pi_{j}'$ is the intersection or union of $A'$ with a prefix
of $\pi_{j}$, so it is closed in $D_{j}$. Let $a_{j}'$ be the greedy
vector of $H_{t_{j}}$ for this new ordering $\pi_{j}'$. Both $a_{j}$
and $a_{j}'$ belong to the same base polytope $\B(H_{t_{j}})$. The
increment $\zeta=\alpha_{j}(a_{j}'-a_{j})$ is nonnegative on every
old block in $A'$ since these blocks are moved earlier in the ordering
and $H_{t_{j}}$is submodular and thus it has diminishing returns.
Since $A'$ is a prefix of $\pi_{j}'$, we also have $a_{j}'(A')=H_{t_{j}}(A')$
and $\zeta(A')=\mathrm{gap}_{j}$. For the proof, map $a_{j}'$ and
$\zeta$ to the blocks of the current DAG by summing the coordinates
that merged, and replace $a_{j}$ in the convex combination with the
mapped vector $a'_{j}$.

Second, consider the case term $j$ is a graph term. A positive gap
means its edge crosses from outside $A$ into $A$. Deleting term
$j$ from the convex combination gives the increment $\zeta=-\alpha_{j}g_{j}$,
which is $\alpha_{j}$ at its head and negative only at its tail outside
$A$. Thus, $\zeta(A)=\mathrm{gap}_{j}$.

In either case we have $\zeta(A)=\mathrm{gap}_{j}$. The vector $y+\zeta$
is still a combination of greedy and graph vectors with the same coefficients
on the greedy vectors so $y+\zeta$ still satisfies \eqref{eq:archive-ring}.
Furthermore, 
\begin{align*}
y^{+}(A)+\zeta(A) & \ge y^{+}(A)+\frac{\theta c_{z}-\Delta-y(A)}{r}\\
 & \ge\frac{\theta c_{z}-\Delta+\beta_{A}}{r}>(n-1)(\beta_{A}-\Delta).
\end{align*}
The middle step uses $y(A)=y^{+}(A)-\beta_{A}$ and $r\ge1$; the
last uses \eqref{eq:guard}, $\beta_{A}\le\beta$, and the nonnegativity
of $r(n-1)-1$.

For each block $v$ in the current DAG, we define the score $p_{v}=[y_{v}]_{+}+\zeta_{v}$.

First consider the case term $j$ is a greedy term, and let $D_{j}$
be the DAG at the time the vector was created. Let $\varnothing=T_{0}\subset T_{1}\subset\cdots\subset T_{k}$
be the first $k+1$ prefixes of $\pi_{j}'$, which are exactly the
intersections of $\lift(A)$ and the prefixes of $\pi_{j}$ ending
in a block in $V(D_{j})\cap\lift(A)$. For each block $v$ in the
current DAG, assign the mass $[y_{v}]_{+}$ to one old block of $D_{j}$
that is part of $v$. Let $P_{0}$ be the nonnegative vector whose
coordinates are the assignments from current blocks to the old blocks
(old blocks not assigned anything get zero). Each old block receives
a score from $P_{0}$ and a score from its coordinate of $\zeta=\alpha_{j}(a_{j}'-a_{j})$.
The cumulative score of each prefix is 
\begin{equation}
P(T_{\ell})=P_{0}(T_{\ell})+\alpha_{j}[a_{j}'(T_{\ell})-a_{j}(T_{\ell})]=P_{0}(T_{\ell})+\alpha_{j}[H_{t_{j}}(T_{\ell})-a_{j}(T_{\ell})].\label{eq:prefix-score}
\end{equation}
The sets $T_{\ell}$'s are closed in the old graph $D_{j}$: each
is the intersection of a closed prefix of $\pi_{j}$ with $A'$, which
is also closed. Evaluate $H_{t_{j}}(T_{\ell})$ using \eqref{eq:normalized}
on the saved partition. A prefix score costs one $\EO$, plus known
sums; $P(T_{0})=0$ and $P(T_{k})=y^{+}(A)+\mathrm{gap}_{j}$ need
none. The score for each old block is nonnegative because $P_{0}$
is nonnegative and $\zeta=\alpha_{j}(a_{j}'-a_{j})$ is nonnegative
on old blocks in $A'$. The total is $P(T_{k})$>$k(\beta_{A}-\Delta)$
so there exist a vertex of $D_{j}$ with score exceeding $\beta_{A}-\Delta$,
which we can find quickly using binary search. Binary interval search
keeps indices $\ell_{-}<\ell_{+}$ with $P(T_{\ell_{+}})-P(T_{\ell_{-}})>(\ell_{+}-\ell_{-})(\beta_{A}-\Delta)$,
starting from $\ell_{-}=0$ and $\ell_{+}=k$. For $\ell_{-}<\ell<\ell_{+}$,
we have $P(T_{\ell_{+}})-P(T_{\ell_{-}})=(P(T_{\ell_{+}})-P(T_{\ell}))+(P(T_{\ell})-P(T_{\ell_{-}}))$
so one of them still exceeds its length times $\beta_{A}-\Delta$;
one $\EO$ to evaluate$P(T_{\ell})$ decides which half to keep. When
$\ell_{+}=\ell_{-}+1$, the old block $\{v\}=T_{\ell_{+}}\setminus T_{\ell_{-}}$
has score exceeding $\beta_{A}-\Delta$. This takes $O(\log n)$ calls.
Because the score of every old block is nonnegative, if the old block
$v$ is part of the current block $u$ then the score $p_{u}=(y_{u})_{+}+\zeta_{u}$
is at least the score of $v$ and it exceeds $\beta_{A}-\Delta$.

Second, consider the case term $j$ is a graph term. In this case,
$\sum_{v\in A}p_{v}=y^{+}(A)+\zeta(A)>(n-1)(\beta_{A}-\Delta)$. By
averaging over at most $n-1$ blocks in $A$, there exists a block
$u\in A$ with $p_{u}=(y_{u})_{+}+\zeta_{u}>\beta_{A}-\Delta$. The
scores are all known and need no $\EO$.

In either case, we identify a block $u\in A$ such that $p_{u}>\beta_{A}-\Delta$.
For any closed $X$ containing $u$ and omitting $z$, $X\subseteq A$,
and 
\[
\theta H_{t}(X)\ge y(X)+\zeta(X)+\Delta\ge p_{u}-\beta_{A}+\Delta>0.
\]
Thus adding the arc $u\to z$ preserves \eqref{eq:invariant}. Since
$u\in A_{z}$, this arc is new. Every value query during this search
can only reduce $m$, which increases $\Delta$ further. The scores
in \eqref{eq:prefix-score} use the fixed historical timestamp $t_{j}$
and do not change. Keep the binary search threshold chosen at its
start; increasing $\Delta$ only makes the final test easier. If $\Delta\ge\beta$
is reached, $H_{t}$ is non-negative on every closed set, which certifies
that the current witness is optimal and the algorithm can terminate.
Report just one arc at a time and repeat with a new graph. The pushed
vector is only used in the proof and it is never stored. Selecting
the term with large gap involves processing $r$ known vectors, which
takes $O(rn)$ time. A faster implementation is discussed in Section
\ref{sec:cubic-bookkeeping}.
\end{proof}

\begin{lem}[One certificate suffices until an SCC is formed]
\label{lem:episode} If \eqref{eq:safety} remains true, keep the
fixed certificate, recalibrate after each new arc, and repeat Lemma~\ref{lem:extract}
until the first SCC contraction or the algorithm terminates since
a minimum was found. This always reaches one of those events. There
are at most $n-1$ nonterminal certificate episodes that end in an
SCC contraction.
\end{lem}

\begin{proof}
During acyclic updates $N,L,\beta,y,\theta$ are fixed and $\Delta$
only increases. Either $\Delta\ge\beta$ proves that the current witness
is optimal or $G$ decreases, so the condition \eqref{eq:safety}
persists. Lemma \ref{lem:no-cancel} holds for every calibrated DAG
with the same vertices and the certificate does not change, so we
can keep identifying new arcs. Each new arc adds a new ordered reachability
pair, of which there are fewer than $n^{2}$ globally. Each nonterminal
episode ends with a contraction, which reduces the number of vertices
by at least 1. 
\end{proof}

\section{Cutting plane method with center warm-starting}

\label{sec:geometry}

This section describe the cutting plane method driving the overall
algorithm. The method maintain a convex body in $\R^{n}$ approximated
by a barrier function, and a potential function that bounds the number
of homogeneous oracle calls. The point minimizing the barrier function
is called the (analytic) center. The potential function is the minimum
value of the barrier function. In each iteration, the method queries
the oracle with the (approximate) minimizer of the current barrier
function. Each oracle call can produce a cutting plane, consisting
of a normal vector in $\mathbb{R}^{n}$ plus an offset, which is added
to the barrier function. The method re-optimizes the modified barrier
function to find the new center and repeat. When the method notices
that the body is thin in the direction provided by the oracle, it
produces a certificate, which is used by the combinatorial component
to produce arcs, eventually leading to a contraction/dimension reduction.
A contraction reduces the dimension of the subspace on which the barrier
is minimized. A minimum over a smaller set can only be larger, so
contractions never decrease the potential. After the subspace changes,
some work is done to find the new center, warm-started from the old
center. The fact that contraction never decrease the potential allows
one to bound the total number of oracle calls (Theorem~\ref{thm:fresh}),
and the same bounded potential later pays for the warm starts of Section~\ref{sec:update-center}.

\textbf{Conversion between full space representation and subspace
representation.} The solutions and the cutting planes are represented
both in the original space $\R^{n}$ and in the subspace whose dimensions
$N$ correspond to the $N$ blocks. In $N$ dimensions, we also have
two different representations, one using coordinates in the geometric
subspace and one suitable for the combinatorial method. A key feature
of the cutting plane method is that the barrier function is in $\R^{n}$
and it works in $\R^{n}$ throughout whereas the combinatorial method
is best described in $\R^{N}$, where coordinates corresponds to the
current blocks.

The current blocks $C$ partition the original ground set $U$. For
each block $C$ let
\[
\xi_{C}=\frac{\mathbf{1}_{C}}{\sqrt{|C|}}\in\mathbb{R}^{n}.
\]
The blocks are disjoint, so these vectors are orthonormal. The matrix
$J=(\xi_{C})_{C}\in\mathbb{R}^{n\times N}$ therefore has orthonormal
columns, and its column space 
\[
S=\{x\in\mathbb{R}^{n}:\ x\text{ is constant on every block}\}
\]
is a subspace of dimension $N$, the number of blocks. A weight/query
point $w\in\R^{N}$ on the blocks is identified with the point $x=\sum_{C}w_{C}\mathbf{1}_{C}\in S$.
Its coordinates in the basis $J$ are $y_{C}=\sqrt{|C|}\,w_{C}$,
so that $x=Jy$ and $w_{C}=y_{C}/\sqrt{|C|}$. Note that $\left\Vert x\right\Vert ^{2}_{2}=\left\Vert y\right\Vert ^{2}_{2}$.

A compressed oracle vector $a\in\R^{N}$ is a greedy or graph vector
indexed by the vertices of the DAG returned by the oracle. To preserve
the dot product between compressed vector $a$ and weight $w$ in
the combinatorial representation when we switch to the geometric representation
with $y$ instead of $w$, we define the corresponding \emph{isometric
vector} $\widetilde{a}\in\R^{N}$ where
\[
\widetilde{a}_{C}=\frac{a_{C}}{\sqrt{|C|}},\qquad\text{so that}\qquad a\cdot w=\sum_{C}a_{C}w_{C}=\widetilde{a}\cdot y.
\]

Another interpretation of this definition is as follows. Let $\bar{a}\in\R^{n}$
be any vector such that the sum of coordinates in $\bar{a}$ over
each block $C$ is exactly $\sum_{u\in C}\bar{a}_{u}=a_{C}$. We have
$\widetilde{a}=J^{\top}\bar{a}$ i.e. $\widetilde{a}$ is the projection
of $\bar{a}$ onto $S$. Thus, $\widetilde{a}\cdot y=(J^{\top}\bar{a})\cdot y=\bar{a}\cdot x$.

When strongly connected components merge, every new block is a union
of old blocks, and the new subspace $S'$ is a subspace of $S$. Suppose
we have an old oracle vector $a$ with isometric vector $\widetilde{a}$
in old subspace $S$ with basis $J$ (with blocks $C$) and we want
to move to the new subspace $S'$ with basis $J'$ (with blocks $D$).
The new oracle vector $a'$ has coordinates $a'_{D}=\sum_{C\subseteq D}a_{C}$.
The corresponding isometric vector $\widetilde{a}'$ has coordinates
$\widetilde{a}'_{D}=\sum_{C\subseteq D}a_{C}/\sqrt{\left|D\right|}$.
Another way to compute this is by projecting the old (expanded) isometric
vector $J\widetilde{a}\in\R^{n}$ onto the span of $J'$ and obtain
the new coordinates $J'^{\top}J\widetilde{a}$. Because $\xi^{\top}_{C}\xi_{D}=\sqrt{|C|/|D|}\cdot\mathbb{I}[C\subseteq D]$,
we have $\left(J'^{\top}J\widetilde{a}\right)_{D}=\sum_{C\subseteq D}\frac{a_{C}}{\sqrt{\left|C\right|}}\cdot\sqrt{|C|/|D|}=\widetilde{a}'_{D}$.
Thus, when using isometric vectors and we need to process an SCC merge,
we can simply project old vector onto the new subspace.

\textbf{Inserted rows.} Suppose the oracle returns a nonzero vector
$a_{i}$ at the time when the basis is $J_{i}$ and the point has
isometric vector $\widetilde{a}^{{\rm cr}}_{i}$. Record its insertion
norm $\sigma^{{\rm cr}}_{i}=\|\widetilde{a}^{{\rm cr}}_{i}\|_{2}$
and its inserted row 
\[
v_{i}=J_{i}\widetilde{a}^{{\rm cr}}_{i}/\sigma^{{\rm cr}}_{i}\in\mathbb{R}^{n},
\]
a unit vector in the subspace of that time. The row $v_{i}$ is stored
in the original space and never changes. After later mergers its projection
onto the current subspace is $J^{\top}v_{i}=\widetilde{a}_{i}/\sigma^{{\rm cr}}_{i}$,
where $\widetilde{a}_{i}$ is the current isometric vector of the
compressed oracle vector mapped to the vertices of the current DAG
by summing up the values on the vertices that were merged during contractions.
If this projection is nonzero, put $\sigma_{i}=\|\widetilde{a}_{i}\|_{2}$;
then 
\[
\frac{J^{\top}v_{i}}{\|J^{\top}v_{i}\|_{2}}=\frac{\widetilde{a}_{i}}{\sigma_{i}}.
\]
Two scales must be kept apart. The insertion norm $\sigma^{{\rm cr}}_{i}$
is fixed forever and only determines the stored row. The current compressed
norm $\sigma_{i}$ is recomputed after every contraction, and it is
the one used in Section~\ref{sec:accuracy} when unit-direction coefficients
are converted into certificate coefficients. If the projection is
zero and the vector is a greedy vector, Lemma~\ref{lem:historical-cuts-valid}
proves optimality: the current parameter satisfies $t\ge t_{i}$,
so $0=a_{i}(X)\le H_{t}(X)-(t-t_{i})\le H_{t}(X)$ for every proper
current closed set $X$ and thus the current witness set is optimal.
A graph vector with zero projection carries no certificate mass and
is left out of certificates, but its row stays in the body.

\textbf{The retained body.} The convex body is specified by rows of
linear constraints of the form $v^{\top}_{i}x\le\rho_{i}$ where the
normal vector $\left\Vert v_{i}\right\Vert =1$ and the offset $\rho_{i}\ge\delta$.
The constraints start with the $2n$ rows of the original cube $\left[-1,1\right]^{n}$,
$(v,\rho)=(\pm e_{j},1)$ for $j=1,\ldots,n$. Every other row comes
from the answers of the homogeneous oracle from Section \ref{subsec:separation-oracle},
as described in Section \ref{subsec:Wide-steps}. Each inserted row
has $v_{i}$ coming from the inserted row above. Rows are kept permanently
even when its projection onto the current subspace becomes zero. Let
$M$ be the current number of rows and define 
\begin{equation}
P=\{x\in\mathbb{R}^{n}:v^{\top}_{i}x\le\rho_{i}\ (1\le i\le M)\},\qquad P_{S}=P\cap S.\label{eq:body}
\end{equation}
Every normal vector satisfies $\|v_{i}\|_{2}\le1$ and every offset
is at least $\delta$, so points in the ball $\delta B^{n}_{2}$ satisfy
every constraint. The cube rows confine $P$ to the cube. Hence 
\[
\delta B^{n}_{2}\subseteq P\subseteq[-1,1]^{n}\subseteq RB^{n}_{2},\qquad R=\sqrt{n}.
\]
In particular the origin is an interior point of $P$ and a relative
interior point of every section $P_{S}$, and every section is bounded.

The body and the certificates are kept separate. The certificate contains
only vectors returned by the homogeneous oracle, not the cube rows.
Conversely, an arc extracted from an certificate need not correspond
to any row of $P$: its graph vector enters the body only if the oracle
later returns it at a wide step. The body is used only to count oracle
calls and, at a thin event, to produce certificate coefficients (Lemma~\ref{lem:center-to-certificate}).

\subsection{Constants}

Fix the universal constants 
\[
\lambda=2^{-10},\qquad\gamma=2^{-28},\qquad\tau=\lambda\sqrt{\gamma}/16=2^{-28},\qquad K=2^{100}.
\]
Here $\lambda$ is the weight of the logarithmic terms in the barrier,
$\gamma$ fixes the leverage score of a newly inserted row (it is
$\gamma/(1+\gamma)$ right after insertion), and $\tau$ is the centering
tolerance. The original $n$ stays fixed throughout the run. Let 
\begin{equation}
\begin{gathered}p_{n}=n+1,\qquad\delta=(Kp_{n})^{-20},\\
\bar{k}_{n}=\frac{8n}{\gamma}\log\!\left(\frac{1+2/\gamma}{\delta^{2}}\right),\qquad Q_{n}=\bar{k}_{n}+n,\qquad\chi_{n}=Q_{n}\delta.
\end{gathered}
\label{eq:delta}
\end{equation}
The number $\delta$ is both the smallest offset ever given to a row
and the threshold that separates wide steps from thin events. The
number $\chi_{n}$ bounds the width of the body at a thin event. The
explicit number $\bar{k}_{n}$ bounds the number of insertions by
Lemma~\ref{lem:insertion-bound}. Consequently the retained row count
and the support size of the certificate are $O(n\log(n+2))$ before
thin events are counted. Theorem~\ref{thm:fresh} then proves that
$Q_{n}$ bounds all homogeneous oracle calls. The following numerical
facts are used repeatedly: 
\[
\frac{8\sqrt{\gamma}}{\lambda}=\frac{1}{2},\qquad40\sqrt{\gamma}<\frac{1}{8},\qquad\gamma\le\frac{\lambda\sqrt{\gamma}}{10},\qquad\lambda\le\frac{1}{32},\qquad\chi_{n}=Q_{n}\delta\le K^{-19}p^{-18}_{n}<1.
\]
The constants are generous on purpose; they affect only the constant
factors hidden in the final bounds.

\subsection{The barrier and certified centers}

Throughout, we use the subscript $fl$ to denote the full space $\R^{n}$
and we use the subscript $S$ to denote the subspace $S$. For $x\in\operatorname{int}P$
write 
\[
s_{i}(x)=\rho_{i}-v^{\top}_{i}x,\qquad A_{fl}(x)_{i,:}=v^{\top}_{i}/s_{i}(x),\qquad\calh_{fl}(x)=A_{fl}(x)^{\top}A_{fl}(x)=\sum_{i}\frac{v_{i}v^{T}_{i}}{s_{i}(x)^{2}},
\]
\[
\Pi_{fl}(x)=A_{fl}(x)\calh_{fl}(x)^{-1}A_{fl}(x)^{T},\qquad\ell_{i}(x)=(\Pi_{fl}(x))_{ii}.
\]
The matrix $\calh_{fl}(x)$ is $n\times n$ and positive definite.
Inside the cube, the two rows $\pm e_{j}$ contribute $e_{j}e^{T}_{j}[(1-x_{j})^{-2}+(1+x_{j})^{-2}]\succeq2e_{j}e^{\top}_{j}$,
because the even convex function $(1-t)^{-2}+(1+t)^{-2}$ has minimum
$2$ on $(-1,1)$, at $t=0$. Hence $\calh_{fl}(x)\succeq2I$. The
leverage scores satisfy $0\le\ell_{i}\le1$ and $\sum_{i}\ell_{i}=\tr\Pi_{fl}=n$,
because $\Pi_{fl}$ is the orthogonal projection onto the $n$-dimensional
column space of $A_{fl}$. The barrier and its restriction to the
current subspace are 
\begin{equation}
\Psi_{fl}(x)=\tfrac{1}{2}\log\det\calh_{fl}(x)-\lambda\sum_{i}\log(s_{i}(x)/\rho_{i}),\qquad\Psi_{S}(y)=\Psi_{fl}(Jy),\qquad\calh_{S}(y)=J^{\top}\calh_{fl}(Jy)J.\label{eq:fullspace-barrier}
\end{equation}
The first term is Vaidya's volumetric barrier; the second is a logarithmic
barrier with the fixed weight $\lambda$. Three features of this definition
matter later. 
\begin{itemize}
\item The determinant always has dimension $n$, whatever the current dimension
$N$ of $S$, and the identity $\sum_{i}\ell_{i}=n$ always refers
to full-space leverage scores. 
\item A row with $J^{\top}v_{i}=0$ stays in $\calh_{fl}$ and in the leverage
scores. On $S$ its slack is the constant $\rho_{i}$, because $v^{\top}_{i}Jy=(J^{\top}v_{i})^{\top}y=0$,
so it contributes nothing to restricted derivatives. 
\item Dividing by the offsets makes every logarithmic term vanish at the
origin, so $\Psi_{fl}(0)=\frac{1}{2}\log\det\calh_{fl}(0)$. 
\end{itemize}
The gradients are 
\[
\nabla\Psi_{fl}(x)=\sum_{i}(\ell_{i}(x)+\lambda)\frac{v_{i}}{s_{i}(x)}=A^{\top}_{fl}(\ell+\lambda\boldsymbol{1}),\qquad\nabla\Psi_{S}(y)=J^{\top}\nabla\Psi_{fl}(Jy).
\]

The Hessians are
\begin{align*}
\nabla^{2}\Psi_{fl}(x) & =A^{\top}_{fl}\bigl[3\Diag(\ell)-2(\Pi_{fl}\circ\Pi_{fl})+\lambda I_{M}\bigr]A_{fl}\\
\nabla^{2}\Psi_{S}(y) & =J^{\top}A^{\top}_{fl}\bigl[3\Diag(\ell)-2(\Pi_{fl}\circ\Pi_{fl})+\lambda I_{M}\bigr]A_{fl}J
\end{align*}

A point $z=Jy$ is a \emph{certified center} if all its slacks are
positive and 
\begin{equation}
\|\nabla\Psi_{S}(y)\|_{\calh_{S}(y)^{-1}}\le\tau,\label{eq:center-tolerance}
\end{equation}
where $\|g\|_{H^{-1}}=(g^{T}H^{-1}g)^{1/2}$. Both conditions are
checked exactly in the arithmetic model. The homogeneous oracle is
called only at certified centers. The first center is the origin:
the cube rows make $\Psi_{fl}$ even and strictly convex, so $0$
is its exact minimizer. After a wide step, Lemma~\ref{lem:wide}
restores certification with a constant number of Newton steps. After
a contraction, the penalty continuation of Section~\ref{sec:update-center}
supplies the next certified center.

\textbf{Potential function.} We use the following potential function
to bound the number of oracle queries and the number of phases of
recentering after contraction. 
\[
\Phi(S,P)=\min_{x\in P\cap S}\Psi_{fl}(x).
\]
The potential is well defined by Lemma~\ref{lem:hybrid-sc}. Initially
the paired cube rows make the barrier even and strictly convex, so
its minimum is at the origin, with value $\frac{n}{2}\log2$. In general
for any $P,S$, the origin is always feasible so $\Phi(S,P)\le\Psi_{fl}(0)$.
With $M$ rows, $\calh_{fl}(0)=\sum^{M}_{i=1}$$\frac{v_{i}v^{\top}_{i}}{\rho^{2}_{i}}\preceq\frac{M}{\delta^{2}}I$
so $\Psi_{fl}(0)\le\frac{n}{2}\log\frac{M}{\delta^{2}}$.

\subsection{One barrier on every section}

Throughout, we refer to $P\cap S$ as a section. The next lemma shows
that the full-space barrier behaves like a standard self-concordant
barrier on every section, with constants that do not depend on the
number of rows.
\begin{lem}[One barrier on every section]
\label{lem:hybrid-sc} Let $c=400/\lambda$ and $W=n+\lambda M$.
For every linear subspace $S$ the restriction $c\Psi_{S}$ is a standard
self-concordant function on the relative interior of $P_{S}$. More
precisely, fix $x\in\operatorname{int}P$ and a direction $h\in S$,
and let 
\[
\eta_{i}=-\frac{v^{T}_{i}h}{s_{i}},\|h\|^{2}_{x}=h^{T}\calh_{fl}(x)h=\sum_{i}\eta^{2}_{i},\qquad Q=\sum_{i}\ell_{i}(x)\eta^{2}_{i}.
\]
Then 
\[
Q+\lambda\|h\|^{2}_{x}\le D^{2}\Psi_{fl}[h,h]\le3Q+\lambda\|h\|^{2}_{x}\le(3+\lambda)\|h\|^{2}_{x},
\]
\[
|D^{3}\Psi_{fl}[h,h,h]|\le\frac{40}{\sqrt{\lambda}}(D^{2}\Psi_{fl}[h,h])^{3/2},\qquad|D\Psi_{fl}[h]|^{2}\le WD^{2}\Psi_{fl}[h,h].
\]
In particular $\lambda\calh_{S}\preceq\nabla^{2}\Psi_{S}\preceq(3+\lambda)\calh_{S}$,
and $\Psi_{fl}$ has a unique minimizer on every section $P_{S}$,
attained in its relative interior. 
\end{lem}

\begin{proof}
Let $A=A_{fl}(x)$, $\calh=\calh_{fl}(x)$, and $s_{i}=s_{i}(x)$,
$V=\frac{1}{2}\log\det\calh_{fl}$ and $L=-\lambda\sum_{i}\log(s_{i}/\rho_{i})$.

\emph{Estimates for the volumetric term.} The bound $Q\le D^{2}V[h,h]\le3Q$
is shown by \cite[Theorem A.4]{Ans96}. The bound $|D^{3}V[h,h,h]|\le30\max_{i}\left|\eta_{i}\right|Q$
is shown by \cite[Theorem 5.1]{Ans00}. 

We also note the following useful formulas. Along a direction $h$
the slacks move as $s_{i}(x+th)=s_{i}(x)\cdot(1+t\eta_{i})$. Expanding
$(1+t\eta_{i})^{-2}$ gives 
\[
\calh_{fl}(x+th)=\calh-2t\,A^{\top}\!\Diag(\eta)A+3t^{2}A^{\top}\!\Diag(\eta^{2})A-4t^{3}A^{\top}\!\Diag(\eta^{3})A+O(t^{4}),
\]
so the first three derivatives of $\calh$ along $h$ are $\dot{\calh}=-2A^{T}\Diag(\eta)A$,
$\ddot{\calh}=6A^{T}\Diag(\eta^{2})A$ and $\dddot{\calh}=-24A^{T}\Diag(\eta^{3})A$.
Let $T_{j}=\calh^{-1/2}A^{T}\Diag(\eta^{j})A\calh^{-1/2}\ j=1,2,3$.
Along any curve of positive definite matrices, by Jacobi's formula
for the derivative of the determinant,
\begin{gather*}
(\log\det\calh)^{\boldsymbol{\cdot}}=\tr(\calh^{-1}\dot{\calh}),\qquad(\log\det\calh)^{\boldsymbol{\cdot}\boldsymbol{\cdot}}=\tr(\calh^{-1}\ddot{\calh})-\tr\bigl((\calh^{-1}\dot{\calh})^{2}\bigr),\\
(\log\det\calh)^{\boldsymbol{\cdot}\boldsymbol{\cdot}\boldsymbol{\cdot}}=\tr(\calh^{-1}\dddot{\calh})-3\tr(\calh^{-1}\dot{\calh}\calh^{-1}\ddot{\calh})+2\tr\bigl((\calh^{-1}\dot{\calh})^{3}\bigr)
\end{gather*}
Substituting and halving, the volumetric term $V=\frac{1}{2}\log\det\calh_{fl}$
satisfies 
\[
DV[h]=-\sum_{i}\ell_{i}\eta_{i},\qquad D^{2}V[h,h]=3\tr T_{2}-2\tr T^{2}_{1}
\]
\[
D^{3}V[h,h,h]=-12\tr T_{3}+18\tr(T_{1}T_{2})-8\tr T^{3}_{1}
\]

\emph{The logarithmic term.} We have 
\begin{align*}
DL[h] & =-\lambda\sum_{i}\eta_{i}\\
D^{2}L[h,h] & =\lambda\sum_{i}\eta^{2}_{i}=\lambda\|h\|^{2}_{x}\\
|D^{3}L[h,h,h]| & =2\lambda\Bigl|\sum_{i}\eta^{3}_{i}\Bigr|\le2\lambda\left\Vert \eta\right\Vert ^{3}_{3}\le2\lambda\left\Vert \eta\right\Vert ^{3}_{2}=2\lambda\|h\|^{3}_{x}.
\end{align*}

\emph{Combining.} Note that the formulas for $DV[h]$ and $DL[h]$
hold for all $h$, and they give the formula for $\nabla\Psi_{fl}$
above. Similarly for $\nabla^{2}\Psi_{fl}$. Adding the bounds for
$V$ and $L$ gives the two-sided bound on $D^{2}\Psi_{fl}[h,h]$,
and $Q\le\sum_{i}\eta^{2}_{i}=\|h\|^{2}_{x}$ (since $\ell_{i}\le1$)
gives the last inequality of the first display. For the third derivative,
use $\|h\|_{x}\le\lambda^{-1/2}(Q+\lambda\|h\|^{2}_{x})^{1/2}$ and
$Q\le Q+\lambda\|h\|^{2}_{x}\le D^{2}\Psi_{fl}[h,h]$: 
\[
|D^{3}\Psi_{fl}[h,h,h]|\le30\|h\|_{x}Q+2\lambda\|h\|^{3}_{x}\le\frac{30+2}{\sqrt{\lambda}}(Q+\lambda\|h\|^{2}_{x})^{3/2}\le\frac{40}{\sqrt{\lambda}}(D^{2}\Psi_{fl}[h,h])^{3/2}.
\]
For the scaled function $c\Psi_{fl}$ this becomes $|D^{3}(c\Psi_{fl})|\le\frac{40}{\sqrt{\lambda c}}(D^{2}(c\Psi_{fl}))^{3/2}=2(D^{2}(c\Psi_{fl}))^{3/2}$,
which is standard self-concordance.

\emph{The gradient bound.} The first derivative is $D\Psi_{fl}[h]=-\sum_{i}(\ell_{i}+\lambda)\eta_{i}$.
The weights $\ell_{i}+\lambda$ are positive and add up to $n+\lambda M=W$,
so Cauchy--Schwarz gives 
\[
|D\Psi_{fl}[h]|^{2}=\left|\sum_{i}(\ell_{i}+\lambda)\eta_{i}\right|^{2}\le W\sum_{i}(\ell_{i}+\lambda)\eta^{2}_{i}=W(Q+\lambda\|h\|^{2}_{x})\le W\,D^{2}\Psi_{fl}[h,h].
\]

\emph{Restriction to $S$.} The fact that $c\Psi_{S}$ is self-concordant
follows from the fact that $c\Psi_{fl}$ is self-concordant and \cite[Section 2.4 rule D]{NT08}.
We include an explanation for completeness. All the inequalities above
hold for every direction $h\in S$. Writing $h=Jk$, the derivatives
of $\Psi_{S}(y)=\Psi_{fl}(Jy)$ in direction $k$ are the derivatives
of $\Psi_{fl}$ in direction $h$, and $k^{T}\calh_{S}k=\|h\|^{2}_{x}$.
So $c\Psi_{S}$ is standard self-concordant, and the second-derivative
bounds $\lambda\|h\|^{2}_{x}\le D^{2}\Psi_{S}[k,k]\le(3+\lambda)\|h\|^{2}_{x}\ \forall k$
give $\lambda\calh_{S}\preceq\nabla^{2}\Psi_{S}\preceq(3+\lambda)\calh_{S}$.

\emph{The section minimum.} The section $P_{S}$ is bounded. Rows
with $J^{T}v_{i}=0$ have the constant slack $\rho_{i}>0$ on $S$,
so at a relative boundary point of $P_{S}$ some row with nonzero
projection has zero slack, and its term $-\lambda\log(s_{i}/\rho_{i})$
tends to $+\infty$ there. The other logarithmic terms are bounded
below on the bounded section, and $\frac{1}{2}\log\det\calh_{fl}\ge\frac{n}{2}\log2$
because $\calh_{fl}\succeq2I$ (due to the cube rows). Hence $\Psi_{S}$
tends to $+\infty$ at the relative boundary and attains its minimum
in the relative interior. Since $\nabla^{2}\Psi_{S}\succeq\lambda\calh_{S}\succeq2\lambda I$,
it is strictly convex, and the minimizer is unique. 
\end{proof}

Moving between the full space and the current subspace requires one
more elementary fact.
\begin{lem}[Full-space and restricted inverse comparison]
\label{lem:restricted-inverse} Let $\calh\in\mathbb{R}^{n\times n}$
be positive definite. For every $b\in S$, 
\[
b^{\top}J(J^{\top}\calh J)^{-1}J^{\top}b\le b^{\top}\calh^{-1}b.
\]
Moreover, at any $x\in\operatorname{int}P$ put $A_{S}=A_{fl}(x)J$,
so that $A^{\top}_{S}A_{S}=\calh_{S}$. Then for every vector $e\in\mathbb{R}^{M}$,
\[
\|A^{\top}_{S}e\|_{\calh^{-1}_{S}}\le\|e\|_{2}.
\]
\end{lem}

\begin{proof}
For a positive definite matrix $G$ and a vector $g$, $g^{\top}G^{-1}g=\max_{k}\{g^{\top}G^{-1}g-(k-G^{-1}g)^{\top}G(k-G^{-1}g)\}=\max_{k}\{2g^{\top}k-k^{\top}Gk\}$,
with the maximum attained at $k=G^{-1}g$. Take $G=J^{\top}\calh J$
and $g=J^{\top}b$. Substituting $h=Jk$ shows that 
\[
b^{\top}J(J^{\top}\calh J)^{-1}J^{\top}b=\max_{k\in\R^{N}}\{2b^{\top}Jk-k^{\top}J^{\top}\calh Jk\}=\max_{h\in S}\{2b^{\top}h-h^{\top}\calh h\}
\]
On the other hand, $b^{\top}\calh^{-1}b=\max_{h\in\R^{n}}\{2b^{\top}h-h^{\top}\calh h\}$
is the same maximum over all $h\in\mathbb{R}^{n}$, so it is at least
as large. For the second claim, $A_{S}\calh^{-1}_{S}A^{\top}_{S}$
is the orthogonal projection onto the column space of $A_{S}$, so
$\|A^{\top}_{S}e\|^{2}_{H^{-1}_{S}}=e^{\top}A_{S}\calh^{-1}_{S}A^{\top}_{S}e\le\|e\|^{2}_{2}$. 
\end{proof}

\subsection{Wide steps}

\label{subsec:Wide-steps}

At a certified center $z=Jy\in\R^{n}$ with compressed weights $w\in\R^{N}$,
the homogeneous oracle returns a vector $a\in\R^{N}$. Assume it is
nonzero; a zero greedy vector ends the run, and a returned graph vector
$e_{u}-e_{v}$ is never zero. Let $\widetilde{b}=\widetilde{a}/\sigma$
be its current unit isometric direction, with $\sigma=\|\widetilde{a}\|_{2}$,
and let $b=J\widetilde{b}\in S$ be the corresponding unit normal
vector of the cutting plane in $\R^{n}$. This $b$ is exactly the
inserted row of the new cutting plane. By construction of the oracle,
$\left\langle a,w\right\rangle \ge0$ at the query weights $w$. In
the full-space coordinates 
\[
\left\langle b,z\right\rangle =\widetilde{b}^{\top}J^{\top}Jy=\frac{\widetilde{a}\cdot y}{\sigma}=\frac{a\cdot w}{\sigma}\ge0.
\]
So the homogeneous cutting plane $\{x:\left\langle b,x\right\rangle \le0\}$
does not contain the center in its interior. The algorithm does not
insert this cutting plane itself. It inserts the weaker parallel cutting
plane $\left\langle b,x\right\rangle \le\rho$, whose offset is chosen
so that the center keeps the prescribed slack $\upsilon$: 
\[
\upsilon=\sqrt{b^{\top}\calh_{fl}(z)^{-1}b/\gamma},\qquad\rho=\upsilon+b^{\top}z\ \ge\upsilon.
\]
The slack $\upsilon$ is computed with the full-space inverse and
makes the leverage score of the new row a fixed constant. The call
is a \emph{wide step} if $\upsilon\ge\delta$, and the row $(b,\rho)$
is then inserted. It is a \emph{thin event} if $\upsilon<\delta$,
and then nothing is inserted.
\begin{lem}
[Controlled insertion has fixed progress]\label{lem:wide} At a certified
center $z=Jy\in P_{S}$, let the oracle return a nonzero vector, with
lifted unit normal $b\in S$ and $b^{\top}z\ge0$, and define $\upsilon$
and $\rho$ as above. If $\upsilon\ge\delta$, append $(b,\rho)$
to the original-space rows. The new body $P_{+}$ still contains $\delta B^{n}_{2}$,
and the new barrier $\Psi_{fl,+}$ satisfies 
\[
\min_{P_{+}\cap S}\Psi_{fl,+}\ge\min_{P\cap S}\Psi_{fl}+\gamma/8.
\]
A constant number of exact full Newton corrections restores \eqref{eq:center-tolerance}. 
\end{lem}

\begin{proof}
\emph{The new row is admissible.} Since $b^{\top}z\ge0$, the offset
satisfies $\rho\ge\upsilon\ge\delta$. As $\|b\|_{2}=1$, every $x$
with $\|x\|_{2}\le\delta$ has $\rho-b^{\top}x\ge\delta-\delta=0$,
so $\delta B^{n}_{2}\subseteq P_{+}$. The new slack at $z$ is $\rho-b^{\top}z=\upsilon>0$,
so $z$ stays strictly feasible.

\emph{The barrier increases at $z$.} At $z$ the new row adds $\bar{b}\bar{b}^{\top}$
to $\calh_{fl}(z)$, where $\bar{b}=b/\upsilon$, and $\bar{b}^{\top}\calh_{fl}(z)^{-1}\bar{b}=\gamma$
by the choice of $\upsilon$. The matrix determinant lemma gives $\det\calh_{fl,+}(z)=(1+\gamma)\det\calh_{fl}(z)$.
The old logarithmic terms do not change at $z$, and the new one equals
$-\lambda\log(\upsilon/\rho)=\lambda\log(\rho/\upsilon)\ge0$. Hence
\[
\Psi_{fl,+}(z)-\Psi_{fl}(z)=\tfrac{1}{2}\log(1+\gamma)+\lambda\log(\rho/\upsilon)\ge\tfrac{1}{2}\log(1+\gamma)\ge\gamma/4,
\]
using $\log(1+\gamma)\ge\gamma/2$ for $\gamma\le1$.

\emph{The leverage scores barely move.} Write $\bar{v}_{i}=v_{i}/s_{i}(z)$
for the old scaled rows and $\calh=\calh_{fl}(z)$. Let $\ell^{+}_{i}$
be the new leverage score of row $i$ and $d_{i}=\ell_{i}-\ell^{+}_{i}$
be the difference between the old leverage score and the new one.
Let $\ell^{+}_{{\rm new}}$ be the leverage score of the new row.
By Sherman-Morrison formula, $\calh_{fl,+}(z)^{-1}=\calh^{-1}-\calh^{-1}\bar{b}\bar{b}^{\top}\calh^{-1}/(1+\gamma)$,
so 
\begin{align*}
d_{i} & =\text{\ensuremath{\bar{v}^{\top}_{i}}}\left(\calh^{-1}-\calh_{fl,+}(z)^{-1}\right)\bar{v}_{i}=\frac{(\bar{v}^{\top}_{i}\calh^{-1}\bar{b})^{2}}{1+\gamma}\ge0,\\
\ell^{+}_{{\rm new}} & =\bar{b}^{\top}\calh_{fl,+}(z)^{-1}\bar{b}=\gamma-\frac{\gamma^{2}}{1+\gamma}=\frac{\gamma}{1+\gamma}
\end{align*}
The leverage scores add up to $n$ both before and after the insertion,
so $\sum_{i}d_{i}=\gamma/(1+\gamma)$. As every $d_{i}\ge0$, this
gives $\|d\|_{2}\le\|d\|_{1}\le\gamma$.

\emph{The old center is nearly central for the new barrier.} Recall
the formula for the gradient $\nabla\Psi_{fl}(z)=\sum_{i}(\ell_{i}+\lambda)\frac{v_{i}}{s_{i}(z)}=A^{\top}_{fl}(\ell+\lambda\boldsymbol{1})$
and by the chain rule, $\nabla\Psi_{S}(y)=J^{\top}A^{\top}_{fl}(\ell+\lambda\boldsymbol{1})$.
The new restricted gradient at $y$ is 
\[
\nabla\Psi_{S,+}(y)=\nabla\Psi_{S}(y)-J^{\top}A_{fl}(z)^{\top}d+\Bigl(\frac{\gamma}{1+\gamma}+\lambda\Bigr)J^{\top}\bar{b}.
\]
Measure all three terms in the old restricted norm $\|\cdot\|_{\calh_{S}(y)^{-1}}$.
The first is at most $\tau$ because $z$ is certified. The second
is at most $\|d\|_{2}\le\gamma$ by the second part of Lemma~\ref{lem:restricted-inverse}.
For the third, the first part of that lemma gives $\|J^{\top}\bar{b}\|_{\calh_{S}(y)^{-1}}\le(\bar{b}^{\top}\calh^{-1}\bar{b})^{1/2}=\sqrt{\gamma}$.
Using $\tau=\lambda\sqrt{\gamma}/16$ and $\gamma\le\lambda\sqrt{\gamma}/10$,
\[
\|\nabla\Psi_{S,+}(y)\|_{\calh_{S}(y)^{-1}}\le\tau+\gamma+(\gamma+\lambda)\sqrt{\gamma}\le2\lambda\sqrt{\gamma}.
\]
Adding a row makes $\calh_{S,+}(y)\succeq\calh_{S}(y)$, so $\nabla^{2}\Psi_{S,+}(y)\succeq\lambda\calh_{S,+}(y)\succeq\lambda\calh_{S}(y)$
by Lemma~\ref{lem:hybrid-sc}. The Newton decrement $\mathfrak{d}$
of the standard self-concordant function $c\Psi_{S,+}$ at $y$ therefore
satisfies 
\[
\mathfrak{d}^{2}=c\,\nabla\Psi^{T}_{S,+}(\nabla^{2}\Psi_{S,+})^{-1}\nabla\Psi_{S,+}\le\frac{c}{\lambda}\|\nabla\Psi_{S,+}(y)\|^{2}_{\calh_{S}(y)^{-1}}\le\frac{400}{\lambda^{2}}\cdot4\lambda^{2}\gamma=1600\gamma,
\]
so $\mathfrak{d}\le40\sqrt{\gamma}<1/8$.

\emph{Net progress.} For a standard self-concordant function with
Newton decrement $\mathfrak{d}<1$ at a point, the gap to the minimum
is at most $\rho(\mathfrak{d})=-\mathfrak{d}-\log(1-\mathfrak{d})$
\cite[Section 2.1]{NT08}, and $\rho(u)\le u^{2}$ for $u\le1/2$.
Dividing by $c$, 
\[
\Psi_{fl,+}(z)-\min_{P_{+}\cap S}\Psi_{fl,+}\le\frac{1600\gamma}{c}=4\lambda\gamma.
\]
Combining with the increase at $z$ and with $\Psi_{ff}(z)\ge\min_{P\cap S}\Psi_{fl}$,
\[
\min_{P_{+}\cap S}\Psi_{fl,+}\ge\Psi_{fl}(z)+\frac{\gamma}{4}-4\lambda\gamma\ge\min_{P\cap S}\Psi_{fl}+\frac{\gamma}{8},
\]
because $\lambda\le1/32$.

\emph{Recentering.} From decrement below $1/8$, full Newton steps
for $c\Psi_{S,+}$ stay feasible and converge quadratically, $\mathfrak{d}_{k+1}\le(\mathfrak{d}_{k}/(1-\mathfrak{d}_{k}))^{2}$,
again by \cite[Section 2.1]{NT08}. Since $\Psi_{S,+}''\preceq(3+\lambda)\calh_{S,+}$,
the restricted residual satisfies $\|\nabla\Psi_{S,+}\|_{\calh^{-1}_{S,+}}\le\sqrt{(3+\lambda)/c}\,\mathfrak{d}$,
so a decrement below the fixed constant $\tau\sqrt{c/(3+\lambda)}$
restores \eqref{eq:center-tolerance}. Reaching it takes a number
of steps that depends only on the fixed constants. 
\end{proof}

\subsection{An unconditional bound on insertions}
\begin{lem}[Insertion and row bounds]
\label{lem:insertion-bound} For every finite execution prefix with
$k$ constraints inserted into the constraint set, 
\begin{equation}
k\le\bar{k}_{n},\qquad M=2n+k\le M_{*}:=2n+\bar{k}_{n}.\label{eq:insertion-bounds}
\end{equation}
A certificate drawn from the inserted constraints has support size
$r\le k\le\bar{k}_{n}$. Moreover, 
\[
\bar{k}_{n}=O(n\log(n+2)),\qquad Q_{n}=\bar{k}_{n}+n<Kp^{2}_{n},\qquad\chi_{n}=Q_{n}\delta<1.
\]
These conclusions do not require a bound on thin events or on the
total number of homogeneous oracle calls.
\end{lem}

\begin{proof}
Each insertion raises $\Phi$ by at least $\gamma/8$ by Lemma~\ref{lem:wide}.
An SCC contraction restricts the same barrier to a smaller subspace,
so it cannot decrease $\Phi$. Witness slides, acyclic implication
updates, and thin events change neither the body nor its subspace
and hence leave $\Phi$ unchanged.

The origin belongs to every section, and all logarithmic terms vanish
there. Since every row has $\|v_{i}\|_{2}\le1$ and $\rho_{i}\ge\delta$,
\[
\Phi(S,P)\le\Psi_{fl}(0)=\tfrac{1}{2}\log\det\sum_{i}\frac{v_{i}v^{T}_{i}}{\rho^{2}_{i}}\le\frac{1}{2}\log\left(\frac{\tr\calh_{fl}(0)}{n}\right)^{n}\le\tfrac{n}{2}\log\left(\frac{M}{n\delta^{2}}\right).
\]
Substitute the exact row count $M=2n+k$. For every finite prefix,
\[
k\,\frac{\gamma}{8}\le\frac{n}{2}\log\frac{2n+k}{2n\delta^{2}},\qquad k\le\frac{4n}{\gamma}\left[\log(\delta^{-2})+\log\!\left(1+\frac{k}{2n}\right)\right].
\]
Put $A=4n/\gamma$, $t=k/A$, and $c=A/(2n)=2/\gamma$. For $t\ge0$,
\[
\log(1+ct)\le\log(1+c)+t/2,
\]
because $1+ct\le(1+c)\max\{1,t\}$ and $\log t\le t/2$ for $t\ge1$.
Thus $t\le\log(\delta^{-2})+\log(1+c)+t/2$, which gives 
\[
k\le2A\log\!\left(\frac{1+c}{\delta^{2}}\right)=\bar{k}_{n}.
\]
This bound applies to every prefix independently of certificate validity.
Every certificate used below draws only from inserted rows, so its
support size is at most $k$.

For the fixed constants, the explicit expression is 
\[
\bar{k}_{n}=2^{31}n[40\log K+40\log p_{n}+\log(1+2^{29})]=O(n\log(n+2)).
\]
Using $\log K<70$, $\log p_{n}\le n$, and $\log(1+2^{29})<21$,
\[
\bar{k}_{n}\le2^{31}n(2821+40n)<2^{43}p^{2}_{n},\qquad Q_{n}=\bar{k}_{n}+n<2^{44}p^{2}_{n}<Kp^{2}_{n}.
\]
Hence $\chi_{n}=Q_{n}\delta\le K^{-19}p^{-18}_{n}<1$. 
\end{proof}

\subsection{Thin events}

\label{subsec:thin-events}

A thin event means that the controlled slack would be smaller than
$\delta$. The next lemma shows that the section is then very thin
in the direction $b$, which is the geometric input of the certificate
construction of Lemma~\ref{lem:center-to-certificate}.
\begin{lem}[A thin event identifies a thin direction]
 \label{lem:floor} At a certified center $z=Jy$, with $W=n+\lambda M$,
let the oracle return a nonzero vector, with lifted unit normal $b\in S$
and $b^{\top}z\ge0$. The current convex body is contained in the
following ellipsoid:
\[
P_{S}\subseteq z+\{h\in S:h^{T}\calh_{fl}(z)h\le(4W/\lambda)^{2}\}.
\]
If $\upsilon=\sqrt{b^{\top}\calh_{fl}(z)^{-1}b/\gamma}<\delta$ then
$\operatorname{width}_{b}(P_{S})\le8W\sqrt{\gamma}\,\delta/\lambda<1$
and the current set of inserted rows contains one whose normal vector
has nonzero projection onto $S$.
\end{lem}

\begin{proof}
\emph{Containment.} Let $x\in P_{S}$ and let $h=x-z\in S$, $\eta_{i}=-v^{\top}_{i}h/s_{i}(z)$
and $\|h\|^{2}_{z}=h^{\top}\calh_{fl}(z)h=\sum_{i}\eta^{2}_{i}$.
Since $s_{i}(x)=s_{i}(z)(1+\eta_{i})\ge0$, every $\eta_{i}\ge-1$.
Because $h\in S$, the restricted gradient controls the first-order
change of $\Psi_{fl}$ along $h$: 
\begin{align*}
\Bigl|\sum_{i}(\ell_{i}+\lambda)\eta_{i}\Bigr| & =|\nabla\Psi_{fl}(z)^{\top}h|=|\nabla\Psi_{fl}(z)^{\top}(JJ^{\top}h)|=|\nabla\Psi_{S}(y)^{\top}(J^{\top}h)|\\
 & \le\|\nabla\Psi_{S}(y)\|_{\calh_{S}(y)^{-1}}\|J^{\top}h\|_{\calh_{S}(y)}\le\tau\|h\|_{z},
\end{align*}
where $\|J^{\top}h\|_{\calh_{S}(y)}=\|h\|_{z}$ because $JJ^{\top}h=h$.
The terms with $\eta_{i}<0$ have $|\eta_{i}|\le1$, and the weights
$\ell_{i}+\lambda$ add up to $W$. Splitting the sum into positive
and negative parts therefore gives 
\[
\sum_{\eta_{i}>0}(\ell_{i}+\lambda)\eta_{i}\le\tau\|h\|_{z}+\sum_{\eta_{i}<0}(\ell_{i}+\lambda)|\eta_{i}|\le\tau\|h\|_{z}+W.
\]
There are at most $M$ terms with $\eta_{i}<0$, each of them has
$\left|\eta_{i}\right|\le$1, so they contribute at most $\sqrt{M}$
to $\|\eta\|_{2}$, and the weights $\ell_{i}+\lambda\ge\lambda$
, so 
\[
\|h\|_{z}=\left\Vert \eta\right\Vert _{2}\le\sqrt{M}+\sum_{\eta_{i}>0}\eta_{i}\le\sqrt{M}+\frac{\tau\|h\|_{z}+W}{\lambda}\le2W/\lambda+\|h\|_{z}/2,
\]
using $\sqrt{M}\le M\le W/\lambda$ and $\tau/\lambda=\sqrt{\gamma}/16\le1/2$.
Hence $\|h\|_{z}\le4W/\lambda$, i.e. $x-z\in\{h\in S:h^{T}\calh_{fl}(z)h\le(4W/\lambda)^{2}\}$.

\emph{Width.} For a unit $b\in S$, the largest value of $b^{\top}h$
over the restricted ellipsoid $\{h\in S:h^{\top}\calh_{fl}(z)h\le(4W/\lambda)^{2}\}$
can be computed as follows. Since $b,h\in S$, let $h=Ja$ and $b=Jq$
i.e. $q=J^{\top}b$. We can rewrite the constraint $h^{\top}\calh_{fl}(z)h=a^{\top}J^{\top}\calh_{fl}(z)Ja=a^{\top}\calh_{S}(y)a\le(4W/\lambda)^{2}$.
Thus, by Cauchy-Schwarz,
\begin{align*}
b^{\top}h & =q^{\top}a\\
 & =(\calh_{S}(y)^{-1/2}q)^{\top}(\calh_{S}(y)^{1/2}a)\\
 & \le\left\Vert \calh_{S}(y)^{-1/2}q\right\Vert _{2}\cdot\left\Vert \calh_{S}(y)^{1/2}a\right\Vert _{2}\\
 & \le(4W/\lambda)\sqrt{(J^{\top}b)^{\top}H_{S}(y)^{-1}J^{\top}b}
\end{align*}
Because the ellipsoid is symmetric, the minimum is the negation of
the maximum and we have
\[
\operatorname{width}_{b}(P_{S})\le\frac{8W}{\lambda}\sqrt{(J^{\top}b)^{\top}\calh_{S}(y)^{-1}J^{\top}b}.
\]
If $\upsilon<\delta$, then $b^{\top}\calh_{fl}(z)^{-1}b=\gamma\upsilon^{2}<\gamma\delta^{2}$,
and the first part of Lemma~\ref{lem:restricted-inverse} gives 
\[
(J^{\top}b)^{\top}\calh_{S}(y)^{-1}(J^{\top}b)\le b^{\top}\calh_{fl}(z)^{-1}b<\gamma\delta^{2}.
\]
Together these give $\operatorname{width}_{b}(P_{S})\le8W\sqrt{\gamma}\,\delta/\lambda$.
By Lemma~\ref{lem:insertion-bound}, we have $M\le M_{*}=2n+\bar{k}_{n}=2n+O(n\log(n+2))$,
so $W=n+\lambda M\le2Q_{n}$. With $8\sqrt{\gamma}/\lambda=1/2$,
the width is at most $Q_{n}\delta=\chi_{n}<1$.

Constraints whose normal vectors have zero projection onto $S$ have
constant slack for all of $S$. If the normal vectors of all inserted
constraints have zero projection onto $S$, $P_{S}$ would be the
section $[-1,1]^{n}\cap S$ of the cube, which contains the unit ball
of $S$ and so has width at least $2$ in the unit direction $b\in S$.
This contradicts the width bound $\chi_{n}<1$. 
\end{proof}

\subsection{The global count of oracle calls}
\begin{thm}[Bound on oracle calls]
\label{thm:fresh} The cutting plane method makes at most $Q_{n}=\bar{k}_{n}+n=O(n\log(n+2))$
homogeneous oracle calls over all contractions.
\end{thm}

\begin{proof}
Lemma~\ref{lem:insertion-bound} already bounds the number of constraints
inserted into the constraint set, independently of the number of thin
events. Each oracle call leads to either a new constraint being inserted
into the constraint set, a thin event, or termination due to a zero
greedy vector. In a thin event, Lemma \ref{lem:floor} shows that
the current body is thin in the direction provided by the oracle and
Lemma~\ref{lem:center-to-certificate} supplies a certificate using
the current set of constraints, and Lemma~\ref{lem:accuracy} proves
the conditions needed to extract arcs from the certificate.

By Lemma~\ref{lem:episode}, after a thin event, the algorithm keeps
that certificate and extracts arcs, making no further homogeneous
call until certifying that the function is nonnegative or the first
SCC contraction (steps~4 and~5 of Section~\ref{sec:algorithm}).
There are at most $n-1$ nonterminal thin events and at most one terminal
call, either a thin event or termination due to a zero greedy vector.
Every other call is a wide step and inserts one constraint row. If
$q$ is the total number of homogeneous calls and $k$ the number
of insertions, it follows that 
\[
q\le k+n\le\bar{k}_{n}+n=Q_{n}=O(n\log(n+2)).
\]
The same argument bounds every finite execution prefix. The count
includes graph-only calls, even though these use no $\EO$. 
\end{proof}

\paragraph{The potential budget.}

The same bookkeeping bounds the total increase of the potential, which
is what pays for the warm starts of Section~\ref{sec:update-center}.
Use the proved insertion and row bounds to define the fixed quantities
\begin{equation}
M_{*}=2n+\bar{k}_{n},\qquad W_{*}=n+\lambda M_{*},\qquad B_{*}=\frac{n}{2}\log(M_{*}/\delta^{2})-\frac{n}{2}\log2.\label{eq:gap-budget}
\end{equation}
The potential starts at $\frac{n}{2}\log2$, never decreases, and
never exceeds $\frac{n}{2}\log(M/\delta^{2})\le\frac{n}{2}\log(M_{*}/\delta^{2})$.
If $\Delta\Phi_{j}$ denotes the increase of $\Phi$ caused by the
$j$th SCC contraction, then, since all other increases are nonnegative,
\[
\Delta\Phi_{j}\ge0,\qquad\sum_{j}\Delta\Phi_{j}\le B_{*}=O(n\log(n+2)).
\]
The continuation of Section~\ref{sec:update-center} uses these proved
fixed bounds, which depend on $n$ alone. Both $M_{*}$ and the actual
row count are $O(n\log(n+2))$. Every thin point may be stored for
historical bookkeeping, but its shifted row is never inserted to the
constraints: the previously inserted constraints already supplies
the certificate.

\section{Turning a thin direction into a certificate}

\label{sec:accuracy} 

In this section, we describe the method to get from a thin direction
to a certificate. Recall that a certificate is a linear combination
of inserted rows with large total coefficients but the length of the
sum is very small. We start from a certified center $z=Jy$ with small
gradient in the $\calh^{-1}_{S}$ restricted norm and a thin direction
$q$, showing that $\left\Vert \calh_{S}(x_{0})\right\Vert _{2}$
is large. The small gradient $\nabla\Psi_{S}(y)$ is already a linear
combination of projected rows that nearly cancel. We project the coefficient
from $\nabla\Psi_{S}(y)$ to the space of linear combination that
exactly cancel. It can be shown that the total change in the coefficients
is small, bounded by the original gradient length. All coefficients
in the linear combination $\nabla\Psi_{S}(y)$ are much larger, so
the modified coefficient are within a factor 2 of the original. We
need to remove the cube rows from the linear combination and show
that the contribution of the cube rows has small length and the total
remaining coefficient is large. The fact that $\left\Vert \calh_{S}(x_{0})\right\Vert _{2}$
is large forces the sum of coefficients to be large.
\begin{lem}[Certificate from center]
\label{lem:center-to-certificate} Let $x_{0}\in\R^{N}$ with $Jx_{0}\in S$
satisfy $\|\nabla\Psi_{S}(x_{0})\|_{\calh_{S}(x_{0})^{-1}}\le\tau$,
and suppose a thin event supplies a unit vector $b=Jq\in S$ with
$q^{\top}\calh_{S}(x_{0})^{-1}q<\gamma\delta^{2}$. Let $\mathcal{R}$
be the set of inserted rows. One can compute positive simplex coefficients
$p_{i}$'s on the nonzero inserted rows $v_{i}$ such that $\sum_{i\in\mathcal{R}}p_{i}=1$
and $\left\Vert \sum_{i\in\mathcal{R}}p_{i}\frac{J^{\top}v_{i}}{\left\Vert J^{\top}v_{i}\right\Vert }\right\Vert _{2}\le\chi_{n}/1024$.
The algorithm takes $\widetilde{O}(n^{\omega})$ arithmetic operations
and no value-oracle calls. 
\end{lem}

\begin{proof}
Let $Z=A_{fl}(Jx_{0})J$ so $\calh_{S}(x_{0})=Z^{\top}Z$. Recall
that $\nabla\Psi_{S}(x_{0})=Z^{\top}(\ell+\lambda\boldsymbol{1})$.
We solve for $d=\calh^{-1}_{S}\nabla\Psi_{S}(x_{0})$. Let $w=\ell+\lambda\boldsymbol{1}-Zd$
and set $\varphi_{i}=\frac{w_{i}}{s_{i}(Jx_{0})}$.The assumption
on $x_{0}$ gives
\[
\left\Vert Zd\right\Vert _{2}=\left\Vert \nabla\Psi_{S}(x_{0})\right\Vert _{\calh^{-1}_{S}}\le\tau
\]

Consequently $\max_{i}\frac{v^{\top}_{i}Jd}{s_{i}(Jx_{0})}\le\left\Vert Zd\right\Vert _{\infty}\le\tau$.
For all $i$, we have $\ell_{i}(Jx_{0})\ge0$ so $w_{i}\ge\lambda-\tau\ge\lambda/2$.
Furthermore, 
\begin{align*}
\sum_{i}\varphi_{i}J^{\top}v_{i} & =J^{\top}\sum_{i}\left((\varphi_{i}s_{i}(Jx_{0}))\frac{v_{i}}{s_{i}(Jx_{0})}\right)\\
 & =Z^{\top}w\\
 & =Z^{\top}\left(\ell+\lambda\boldsymbol{1}\right)-\calh_{S}d=0
\end{align*}

Thus, we have obtained a linear combination of projected rows that
cancel exactly. We need to remove the cube rows from this linear combination
and show that the sum of the remaining coefficients is large. We have
\begin{align*}
\sum_{i}\varphi_{i}\rho_{i} & =\sum_{i}\frac{w_{i}}{s_{i}(Jx_{0})}(s_{i}(Jx_{0})+v^{\top}_{i}Jx_{0})\\
 & =\sum_{i}w_{i}+w^{\top}Zx_{0}\\
 & \le\sum_{i}(\ell_{i}+\lambda)+\left\Vert Zd\right\Vert _{1}\\
 & \le n+\lambda M+\sqrt{M}\tau\le4\lambda Q_{n}
\end{align*}

For the last bound we use $M\le2Q_{n},n\le\lambda Q_{n}$ and $\tau\le\lambda/2$.
Thin event gives $q^{\top}\calh_{S}(x_{0})^{-1}q<\gamma\delta^{2}$
for unit vector $q$ so 
\begin{equation}
\tr\calh_{S}(x_{0})\ge\left\Vert \calh_{S}(x_{0})\right\Vert _{2}\ge1/(\gamma\delta^{2})\label{eq:H_S-trace}
\end{equation}

Let $\mathcal{C}$ be the set of original cube rows and $\mathcal{R}$
the set of inserted rows (greedy and graph vectors). Cube rows have
offsets $\rho_{i}=1$ and $\left\Vert J^{\top}v_{i}\right\Vert \le\left\Vert v_{i}\right\Vert \le1$
so
\[
\sum_{i\in\mathcal{C}}\varphi_{i}\left\Vert J^{\top}v_{i}\right\Vert \le\sum_{i\in\mathcal{C}}\varphi_{i}\rho_{i}\le4\lambda Q_{n}
\]
For inserted rows, let $\nu_{\mathcal{R}}=\sum_{i\in\mathcal{R}}\varphi_{i}\|J^{\top}v_{i}\|$.
We have
\begin{align*}
\sum_{i\in\mathcal{R}}\varphi_{i}\|J^{\top}v_{i}\| & =\sum_{i}\varphi_{i}\|J^{\top}v_{i}\|-\sum_{i\in\mathcal{C}}\varphi_{i}\|J^{\top}v_{i}\|\\
 & \ge\frac{\lambda}{2}\sum_{i}\frac{\|J^{\top}v_{i}\|}{s_{i}(Jx_{0})}-4\lambda Q_{n}\\
 & \ge\frac{\lambda}{2}\sqrt{\sum_{i}\frac{\|J^{\top}v_{i}\|^{2}}{s_{i}(Jx_{0})^{2}}}-4\lambda Q_{n}\\
 & =\frac{\lambda}{2}\sqrt{\tr\calh_{S}(x_{0})}-4\lambda Q_{n}\\
 & \overset{by\ \eqref{eq:H_S-trace}}{\ge}\frac{\lambda}{2\sqrt{\gamma}\delta}-4\lambda Q_{n}
\end{align*}

The constants make this bound for $\nu_{\mathcal{R}}$ positive. We
have $16\sqrt{\gamma}\,Q_{n}\delta<1$, so $\nu_{\mathcal{R}}\ge\frac{\lambda}{4\sqrt{\gamma}\delta}$.

Thus, by setting $p_{i}=\varphi_{i}\|J^{\top}v_{i}\|/\nu_{\mathcal{R}}$for
all nonzero $J^{\top}v_{i}$, we have $\sum_{i\in\mathcal{R}}p_{i}=1$
and 
\[
\left\Vert \sum_{\mathcal{R}}p_{i}\frac{J^{\top}v_{i}}{\left\Vert J^{\top}v_{i}\right\Vert }\right\Vert _{2}=\frac{\|\sum_{\mathcal{R}}\varphi_{i}J^{\top}v_{i}\|_{2}}{\nu_{\mathcal{R}}}\le\frac{4\lambda Q_{n}}{\lambda/(4\sqrt{\gamma}\,\delta)}\le16\sqrt{\gamma}\,Q_{n}\delta=\chi_{n}/1024.
\]
 
\end{proof}

For a stored nonzero current compressed row $a_{i}\in\R^{N}$, let
$\widetilde{a}_{i,C}=a_{i,C}/\sqrt{|C|}$ and $\sigma_{i}=\|\widetilde{a}_{i}\|_{2}$
in the isometric coordinates of Section~\ref{sec:geometry}. Let
$b_{i}=\widetilde{a}_{i}/\sigma_{i}$ so that $\left\Vert b_{i}\right\Vert =1$.
Lemma \ref{lem:center-to-certificate} gives simplex coefficients
$p_{i}\ge0$, $\sum_{i}p_{i}=1$ such that 
\begin{equation}
\left\Vert \sum_{i}p_{i}b_{i}\right\Vert _{2}\le2\chi_{n},\qquad\chi_{n}=Q_{n}\delta.\label{eq:unit-residual}
\end{equation}
Convert them to coefficients for compressed rows by 
\[
Z=\sum_{i}\frac{p_{i}}{\sigma_{i}},\qquad\alpha_{i}=\frac{p_{i}/\sigma_{i}}{Z}.
\]
Every current $\sigma_{i}$ is recomputed after a contraction. Otherwise,
the certificate retains each term's graph/greedy label and original
timestamp.
\begin{lem}[Accuracy and the full-certificate guard]
\label{lem:accuracy} For a certificate drawn from the inserted rows,
whose size is at most $\bar{k}_{n}$ by Lemma~\ref{lem:insertion-bound},
the converted certificate has 
\begin{equation}
\frac{\|y\|_{1}}{L}\le\sqrt{n}\left\Vert \sum_{i}p_{i}b_{i}\right\Vert _{2}\le e_{n}:=2\sqrt{n}\,Q_{n}\delta.\label{eq:physical-residual}
\end{equation}
The total greedy coefficients $\theta$ is positive. Either $\Delta\ge\beta$,
which certifies optimality of the witness set, or the condition \eqref{eq:safety}
holds, enabling arc extraction.
\end{lem}

\begin{proof}
In isometric coordinates, $\widetilde{y}=\sum_{i}\alpha_{i}\widetilde{a}_{i}=(\sum_{i}p_{i}b_{i})/Z$.
When we convert to the compressed coordinates for the combinatorial
part, $y_{C}=\sqrt{\left|C\right|}\widetilde{y}_{C}$ and $a_{i,C}=\sqrt{\left|C\right|}\widetilde{a}_{i,C}$.
Since $\sum_{C}|C|=n$, weighted Cauchy--Schwarz gives 
\[
\|y\|_{1}=\sum_{C}\sqrt{|C|}|\widetilde{y}_{C}|\le\sqrt{n}\|\widetilde{y}\|_{2}.
\]
We have $\|a_{i}\|_{1}\ge\|\widetilde{a}_{i}\|_{2}=\sigma_{i}$, so
$L=\sum_{i}\alpha_{i}\left\Vert a_{i}\right\Vert _{1}\ge\sum_{i}\alpha_{i}\sigma_{i}=1/Z>0$.
Therefore, 
\[
\frac{\|y\|_{1}}{L}\le\sqrt{n}\left\Vert \sum_{i}p_{i}b_{i}\right\Vert _{2}\le2\sqrt{n}\,Q_{n}\delta
\]
If the total greedy coefficients $\theta=0$, Lemma~\ref{lem:no-cancel}
gives $L\le4N\beta$, contradicting $\beta=\left\Vert y\right\Vert _{1}/2\le e_{n}L/2$
and $4N\beta\le2LNe_{n}\le4LKp^{7/2}_{n}\delta<L$.

For $\Delta<\beta$, the proved support bound $r\le\bar{k}_{n}\le Q_{n}<Kp^{2}_{n}$
and $N\le n$ give 
\[
\frac{4N^{2}G+4N\beta}{L}\le(2n^{3}r+2n)e_{n}\le(2n^{3}Q_{n}+2n)\,2\sqrt{n}\,Q_{n}\delta\le8K^{2}p^{15/2}_{n}\delta=8K^{-18}p^{-25/2}_{n}<\frac{1}{2}.
\]

First inequality used $G\le rn\beta\le rn\left\Vert y\right\Vert _{1}/2\le rnLe_{n}$/2.
Since the condition \eqref{eq:safety} holds, we can identify new
arcs as described in Section \ref{subsec:arcs-from-certificate}.
Every size bound used here follows from Lemma~\ref{lem:insertion-bound}. 
\end{proof}

\section{Updating the center after contractions with warm-start}

\label{sec:update-center}

In this section, we describe how to efficiently construct the new
center of the barrier after contracting an SCC formed after adding
arcs. We warm-start the starting point using the previous center.
Let $S$ be the subspace before the contraction and let $J\in\mathbb{R}^{n\times d}$
be the matrix whose columns are the basis vectors $\xi_{C}$ for $S$
as before. Let $S'$ be the new subspace after the contraction, and
let $d-r$ be its dimension. Let $\Gamma\in\mathbb{R}^{r\times d}$
be a matrix with orthonormal rows that span the $r$ coordinates removed
by the contraction. Then $\Gamma\Gamma^{\top}=I_{r}$, $\Gamma^{\top}\Gamma$
is the orthogonal projection onto the removed directions, and 
\[
S'=\left\{ Jx\colon x\in\ker(\Gamma)\right\} =J\ker\Gamma.
\]
In the following, we work in the domain before the contraction, i.e.,
the domain 
\[
\left\{ x\in\mathbb{R}^{d}\colon s_{i}(Jx)>0\text{ for every retained row }i\right\} 
\]
and we capture the contraction via the additional constraints $\Gamma x=0$.

Let $J'$ have orthonormal columns spanning $\ker\Gamma$, and thus
$J_{\mathrm{new}}:=JJ'$ has orthonormal columns spanning $S'$. Starting
with the approximate center $z_{\mathrm{in}}:=Jx_{\mathrm{in}}$ for
the barrier on $P\cap S$, our goal is to obtain a new approximate
center $z_{\mathrm{new}}$ for the barrier on $P\cap S'$. We require
that the center is in $S'$, it has positive slacks, and it satisfies
the $\tau$-approximate centering condition as before, i.e., it satisfies
the following conditions: 
\begin{enumerate}
\item The contraction equalities: $z_{\mathrm{new}}=J_{\mathrm{new}}J'^{\top}\overline{x}$
for a vector $\overline{x}$ satisfying $\Gamma\overline{x}=0$, and
thus $z_{\mathrm{new}}\in S'$, 
\item Positive slacks: $s_{i}(z_{\mathrm{new}})>0$ for every retained row
$i$, 
\item $\tau$-approximate centering: $\|g_{\mathrm{new}}\|_{\calh^{-1}_{\mathrm{new}}}\le\tau$
where we let the columns of $J_{\mathrm{new}}$ form an orthonormal
basis for $S'$, $g_{\mathrm{new}}:=J^{\top}_{\mathrm{new}}\nabla\Psi_{fl}(z_{\mathrm{new}})$,
and $\calh_{\mathrm{new}}:=J^{\top}_{\mathrm{new}}\calh_{fl}(z_{\mathrm{new}})J_{\mathrm{new}}$ 
\end{enumerate}
We refer to a center satisfying these conditions as a certified center.

\subsection{A path-following algorithm for computing the new center}

\label{subsec:recenter-algorithm}

\begin{algorithm}
\caption{Recentering algorithm}
\label{alg:recenter}

\textbf{\textsc{Recenter$(x_{\mathrm{in}})$:}}

\begin{algorithmic}[1]

\STATE\label{line:recenter-basis} Construct $J'$ with orthonormal
columns spanning $\ker\Gamma$, so that $JJ'$ has orthonormal columns
spanning $S'$.

\STATE $x\gets x_{\mathrm{in}}$

\STATE\label{line:recenter-initialize} $c\gets400/\lambda$, $\overline{\tau}\gets\tau/8$,
$\widetilde{\tau}\gets\tau/(32(3+\lambda))$, $\kappa\gets\log(\tau/(8n))$,
$\vartheta_{0}\gets1/(64\sqrt{c})$

\WHILE{$\mathrm{true}$}\label{line:recenter-loop}

\STATE\label{line:recenter-center} $x\gets\textsc{RecenterPhase}(\kappa,x)$

\STATE\label{line:recenter-stage-project} $(\mathrm{ok},\overline{x})\gets\textsc{ProjectAndCheck}(x)$

\IF{$\mathrm{ok}=\mathrm{true}$} \STATE\label{line:recenter-accept}
\textbf{break}

\ENDIF \STATE\label{line:recenter-energy} $\mu\gets e^{\kappa/2}\|\Gamma x\|_{2}$,
$\vartheta\gets\vartheta_{0}/\max\{1,\mu\}$

\STATE\label{line:recenter-increase} $\kappa\gets\kappa+\log(1+\vartheta)$

\ENDWHILE

\STATE\label{line:recenter-install} $J_{\mathrm{new}}\gets JJ'$,
$x_{\mathrm{new}}\gets J'^{\top}\overline{x}$, $z_{\mathrm{new}}\gets J_{\mathrm{new}}x_{\mathrm{new}}$

\RETURN\label{line:recenter-return} $(J_{\mathrm{new}},x_{\mathrm{new}},z_{\mathrm{new}})$.

\end{algorithmic}
\end{algorithm}

\begin{algorithm}
\caption{Algorithm for a single phase}
\label{alg:recenter-phase}

\textbf{$\textsc{RecenterPhase}$$(\kappa,x)$:}

\begin{algorithmic}[1]

\WHILE{true}

\STATE\label{line:phase-evaluate} Evaluate $\nabla\Psi_{\kappa}(x)$,
$\nabla^{2}\Psi_{\kappa}(x)$, and $\mathcal{K}_{\kappa}(x):=\calh_{S}(x)+\frac{e^{\kappa}}{\lambda}\Gamma^{\top}\Gamma$

\IF{$\|\nabla\Psi_{\kappa}(x)\|^{2}_{\mathcal{K}_{\kappa}(x)^{-1}}\leq\overline{\tau}^{2}$}\label{line:phase-stop}

\RETURN\label{line:phase-return} $x$

\ENDIF

\STATE\label{line:phase-newton-step}Update $x\gets x-\left(\nabla^{2}\Psi_{\kappa}(x)\right)^{-1}\nabla\Psi_{\kappa}(x)$.
\COMMENT{Newton method update.}

\ENDWHILE

\end{algorithmic}
\end{algorithm}

\begin{algorithm}
\caption{Projection test}
\label{alg:project-and-check}

\textbf{\textsc{ProjectAndCheck}}\textbf{$(x)$:}

\begin{algorithmic}[1]

\STATE\label{line:project-point} $h\gets-\Gamma^{\top}\Gamma x$,
$\overline{x}\gets x+h$.

\STATE\label{line:project-length} Compute $q\gets\sum^{M}_{i=1}\left(v^{\top}_{i}Jh/s_{i}(Jx)\right)^{2}=\left\Vert h\right\Vert ^{2}_{\calh_{S}(x)}$.

\IF{$q>\widetilde{\tau}^{2}$}\label{line:project-guard}

\RETURN\label{line:project-reject} $(\mathrm{false},\overline{x})$.

\ENDIF

\RETURN\label{line:project-return} $(\mathrm{true},\overline{x})$.

\end{algorithmic}
\end{algorithm}

The algorithm is shown in Algorithm \ref{alg:recenter}. The algorithm
constructs the new center using the following path-following approach.
For a parameter $\kappa\in\mathbb{R}$, consider the following function:
\[
\Psi_{\kappa}(x):=\Psi_{fl}(Jx)+\frac{e^{\kappa}}{2}\|\Gamma x\|^{2}_{2}.
\]
The function $\Psi_{\kappa}$ adds a penalty term $\frac{e^{\kappa}}{2}\|\Gamma x\|^{2}_{2}$
to encourage its minimizer to get closer to satisfying the contraction
equalities $\Gamma x=0$. We refer to $e^{\kappa}$ as the penalty
weight.

The algorithm starts with an initial penalty weight is $e^{\kappa_{0}}=\tau/(8n)$,
and it increases it in phases by a factor of $1+\vartheta$.  In
each phase, $\kappa$ is fixed and the algorithm approximately minimizes
$\Psi_{\kappa}$ using Newton's method, starting with the point output
by the previous phase. Let $x$ be the final iterate constructed by
Newton's method during the current phase. To get a point that exactly
satisfies the contraction equalities $\Gamma x=0$, we let $\overline{x}:=x-\Gamma^{\top}\Gamma x$
which satisfies $\Gamma\overline{x}=0$ since $\Gamma\Gamma^{\top}=I_{r}$.
The algorithm accepts $\overline{x}$ when the projection displacement
$h=-\Gamma^{\top}\Gamma x$ satisfies 
\[
\left\Vert h\right\Vert ^{2}_{\mathcal{H}_{S}(x)}\le\widetilde{\tau}^{2},\qquad\widetilde{\tau}=\frac{\tau}{32(3+\lambda)}.
\]
As we show in Lemma \ref{lem:penalty-projection}, this suffices to
show that $\overline{x}$ satisfies all of the required conditions
for an approximate center of the barrier in $P\cap S'$. 

\subsection{Analysis of $\textsc{RecenterPhase}$}

Let $f_{\kappa}:=c\Psi_{\kappa}$ where $c:=400/\lambda$. The function
$f_{\kappa}$ is self-concordant. The Newton update steps with respect
to $f_{\kappa}$ are the same as those with respect to $\Psi_{\kappa}$.
We let $\mathfrak{d}_{\kappa}(x)$ be the Newton decrement of $f_{\kappa}$
at $x$, i.e. 
\[
\mathfrak{d}_{\kappa}(x):=\sqrt{\nabla f_{\kappa}(x)^{\top}\left(\nabla^{2}f_{\kappa}(x)\right)^{-1}\nabla f_{\kappa}(x)}=\sqrt{c\nabla\Psi_{\kappa}(x)^{\top}\left(\nabla^{2}\Psi_{\kappa}(x)\right)^{-1}\nabla\Psi_{\kappa}(x)}
\]
The penalty derivatives are 
\begin{align*}
\nabla\Psi_{\kappa}(x) & =J^{\top}\nabla\Psi_{fl}(Jx)+e^{\kappa}\Gamma^{\top}\Gamma x,\\
\nabla^{2}\Psi_{\kappa}(x) & =J^{\top}\nabla^{2}\Psi_{fl}(Jx)J+e^{\kappa}\Gamma^{\top}\Gamma.
\end{align*}
Let $\mathcal{K}_{\kappa}(x):=\calh_{S}(x)+\frac{e^{\kappa}}{\lambda}\Gamma^{\top}\Gamma$.
By Lemma~\ref{lem:hybrid-sc}, we have 
\begin{equation}
\lambda\mathcal{K}_{\kappa}(x)\preceq\nabla^{2}\Psi_{\kappa}(x)\preceq(3+\lambda)\mathcal{K}_{\kappa}(x).\label{eq:penalty-hessian}
\end{equation}
In particular, $\nabla^{2}\Psi_{\kappa}(x)$ is positive definite
and thus it is invertible. Since the domain is bounded and the base
barrier tends to infinity at its relative boundary, $\Psi_{\kappa}$
has a unique minimizer $z_{\kappa}$ in the interior for every finite
real $\kappa$.

We first show that the starting point of the algorithm has sufficiently
small Newton decrement, and thus the algorithm starts in the region
of quadratic convergence. 
\begin{lem}[Initial Newton decrement]
\label{lem:recenter-initialization}We have $\mathfrak{d}_{\kappa_{0}}(x_{\mathrm{in}})\leq1/8$. 
\end{lem}

\begin{proof}
Since $\nabla^{2}\Psi_{\kappa_{0}}(x_{\mathrm{in}})\succeq\lambda\calh_{S}$,
we have 
\begin{align*}
\mathfrak{d}_{\kappa_{0}}(x_{\mathrm{in}}) & =\sqrt{c}\|\nabla\Psi_{\kappa_{0}}(x_{\mathrm{in}})\|_{\nabla^{2}\Psi_{\kappa_{0}}(x_{\mathrm{in}})^{-1}}\\
 & \leq\sqrt{\frac{c}{\lambda}}\|\nabla\Psi_{\kappa_{0}}(x_{\mathrm{in}})\|_{\calh_{S}(x_{\mathrm{in}})^{-1}}\\
 & \leq\sqrt{\frac{c}{\lambda}}\left(\|J^{\top}\nabla\Psi_{fl}(Jx_{\mathrm{in}})\|_{\calh_{S}(x_{\mathrm{in}})^{-1}}+e^{\kappa_{0}}\|\Gamma^{\top}\Gamma x_{\mathrm{in}}\|_{\calh_{S}(x_{\mathrm{in}})^{-1}}\right).
\end{align*}
Since $x_{\mathrm{in}}$ is a $\tau$-approximate center for the barrier
on $P\cap S$, 
\[
\|J^{\top}\nabla\Psi_{fl}(Jx_{\mathrm{in}})\|_{\calh_{S}(x_{\mathrm{in}})^{-1}}\leq\tau.
\]
Since $\Gamma^{\top}\Gamma$ is an orthogonal projection and $Jx_{\mathrm{in}}\in\left[-1,1\right]^{n}$,
we have 
\[
\|\Gamma^{\top}\Gamma x_{\mathrm{in}}\|_{2}\leq\|x_{\mathrm{in}}\|_{2}=\|Jx_{\mathrm{in}}\|_{2}\leq\sqrt{n}=:R.
\]
Since $\calh_{S}(x_{\mathrm{in}})\succeq I/(4R^{2})$, we have 
\[
e^{\kappa_{0}}\|\Gamma^{\top}\Gamma x_{\mathrm{in}}\|_{\calh_{S}(x_{\mathrm{in}})^{-1}}\le2e^{\kappa_{0}}R^{2}=\tau/4.
\]
Combining all inequalities, we obtain 
\[
\mathfrak{d}_{\kappa_{0}}(x_{\mathrm{in}})\le\sqrt{\frac{c}{\lambda}}\left(\tau+\frac{\tau}{4}\right)=\frac{25\tau}{\lambda}<\frac{1}{8}.
\]
\end{proof}

The following lemma shows that, if each phase starts with a sufficiently
small Newton decrement, then the phase terminates in a constant number
of Newton steps. The lemma also records additional properties of the
point returned by the phase that will be used later in the analysis. 
\begin{lem}[Newton steps during a phase]
\label{lem:penalty-newton-corrections} Consider a phase with parameter
$\kappa$, and let $z_{\kappa}$ be the unique minimizer of $\Psi_{\kappa}$
and thus $f_{\kappa}$. Let $x_{0}$ be the initial point at the start
of the phase. If $\mathfrak{d}_{\kappa}(x_{0})\le1/8$, $\textsc{RecenterPhase}(\kappa,x_{0})$
terminates in at most $4$ iterations of Newton's method and it returns
a vector $x$ with the following properties: 
\begin{enumerate}
\item Decrement guarantee: Let $\kappa_{+}=\kappa+\log(1+\vartheta)$ be
the parameter of the next phase. We have $\mathfrak{d}_{\kappa_{+}}(x)\leq1/8$. 
\item Slack guarantee: For every retained row, $\tfrac{1}{2}s_{i}(Jz_{\kappa})\le s_{i}(Jx)\le\tfrac{3}{2}s_{i}(Jz_{\kappa}).$ 
\item Penalty guarantee: Let $\Delta\Phi:=\Phi(S',P)-\Phi(S,P)$ and $\varepsilon:=\overline{\tau}^{2}/\lambda$.
We have $\frac{e^{\kappa}}{2}\|\Gamma x\|^{2}_{2}\le\Delta\Phi+\varepsilon$. 
\end{enumerate}
\end{lem}

\begin{proof}
\emph{Quadratic convergence.} Let $x_{k}$ be the Newton method iterates
in the current phase, including the initial solution $x_{0}$. To
simplify notation, let us define $\mathfrak{d}_{k}:=\mathfrak{d}_{\kappa}(x_{k})$.
We can show by induction that, for all $k\geq0$, we have $\mathfrak{d}_{k}\leq1/8$
and 
\begin{equation}
2\mathfrak{d}_{k}\leq(2\mathfrak{d}_{0})^{2^{k}}\leq2^{-2^{k+1}}.\label{eq:newton-convergence}
\end{equation}
We have $\mathfrak{d}_{0}\leq1/8$ and thus the base case $k=0$ follows.
For the inductive step, suppose that $\mathfrak{d}_{k}\leq1/8$ and
$2\mathfrak{d}_{k}\leq(2\mathfrak{d}_{0})^{2^{k}}$. By \cite[Section 9.6.4, Equation 9.55]{BoydVandenberghe2004},
we have 
\[
\mathfrak{d}_{k+1}\leq\left(\frac{\mathfrak{d}_{k}}{1-\mathfrak{d}_{k}}\right)^{2}\leq2\mathfrak{d}^{2}_{k}.
\]
Since $\mathfrak{d}_{k}\leq1/8$, we have $2\mathfrak{d}^{2}_{k}\leq\mathfrak{d}_{k}/4$
and thus $\mathfrak{d}_{k+1}\leq1/8$. Since $2\mathfrak{d}_{k}\leq(2\mathfrak{d}_{0})^{2^{k}}$,
we have $2\mathfrak{d}_{k+1}\leq(2\mathfrak{d}_{0})^{2^{k+1}}$, as
needed. This completes the induction.

\emph{Number of iterations}. Next, we show that the algorithm terminates
in at most $4$ iterations. The algorithm terminates when $\|\nabla\Psi_{\kappa}(x_{k})\|_{\mathcal{K}_{\kappa}(x_{k})^{-1}}\leq\overline{\tau}$.
The Hessian comparisons in (\ref{eq:penalty-hessian}) imply 
\[
\sqrt{\frac{c}{3+\lambda}}\,\|\nabla\Psi_{\kappa}(x)\|_{\mathcal{K}_{\kappa}(x)^{-1}}\le\mathfrak{d}_{\kappa}(x)\le\sqrt{\frac{c}{\lambda}}\,\|\nabla\Psi_{\kappa}(x)\|_{\mathcal{K}_{\kappa}(x)^{-1}}.
\]
Thus it suffices to reach $\mathfrak{d}_{k}\leq\overline{\tau}\sqrt{c/(3+\lambda)}$.
By \eqref{eq:newton-convergence} and the fact that $\overline{\tau}=2^{-31}$
and $\sqrt{c}=640$, we have 
\[
\mathfrak{d}_{4}\leq2^{-33}<\overline{\tau}\sqrt{\frac{c}{3+\lambda}},
\]
and thus the algorithm terminates in at most $4$ iterations.

\emph{Decrement guarantee. }We have 
\begin{align*}
\nabla\Psi_{\kappa_{+}}(x) & =\nabla\Psi_{\kappa}(x)+\vartheta e^{\kappa}\Gamma^{\top}\Gamma x\\
\nabla^{2}\Psi_{\kappa_{+}}(x) & \succeq\nabla^{2}\Psi_{\kappa}(x)\\
\nabla^{2}\Psi_{\kappa_{+}}(x) & \succeq e^{\kappa}\Gamma^{\top}\Gamma,
\end{align*}
because $e^{\kappa_{+}}=(1+\vartheta)e^{\kappa}$. The last inequality
and $\Gamma\Gamma^{\top}=I_{r}$ give $\|e^{\kappa}\Gamma^{\top}\Gamma x\|_{(\nabla^{2}\Psi_{\kappa_{+}}(x))^{-1}}\le e^{\kappa/2}\|\Gamma x\|_{2}=\mu$.
Together with the return condition and $\vartheta\mu\le\vartheta_{0}=1/(64\sqrt{c})$,
these imply that 
\[
\mathfrak{d}_{\kappa_{+}}(x)\leq\mathfrak{d}_{\kappa}(x)+\sqrt{c}\vartheta\mu\leq\frac{1}{64}+\sqrt{\frac{c}{\lambda}}\overline{\tau}<\frac{1}{32}<\frac{1}{8}.
\]
\emph{Slack guarantee.} Using the definition of the slacks and $\calh_{S}(z_{\kappa})=\sum_{i}\frac{J^{\top}v_{i}v^{\top}_{i}J}{s_{i}(Jz_{\kappa})^{2}}$,
we obtain 
\[
\frac{\left|s_{i}(Jx)-s_{i}(Jz_{\kappa})\right|}{s_{i}(Jz_{\kappa})}=\frac{\left|v^{\top}_{i}J(x-z_{\kappa})\right|}{s_{i}(Jz_{\kappa})}\leq\left\Vert x-z_{\kappa}\right\Vert _{\mathcal{H}_{S}(z_{\kappa})}
\]
Since $\nabla^{2}f_{\kappa}(z_{\kappa})\succeq c\lambda\calh_{S}(z_{\kappa})=400\calh_{S}(z_{\kappa})$,
we have 
\[
\left\Vert x-z_{\kappa}\right\Vert _{\calh_{S}(z_{\kappa})}\leq\frac{1}{20}\|x-z_{\kappa}\|_{\nabla^{2}f_{\kappa}(z_{\kappa})}.
\]
Since $\mathfrak{d}_{\kappa}(x)\leq1/64$, by \cite[Equation 2.6c]{NT08}
we have 
\[
r:=\|x-z_{\kappa}\|_{\nabla^{2}f_{\kappa}(x)}\leq\frac{\mathfrak{d}_{\kappa}(x)}{1-\mathfrak{d}_{\kappa}(x)}\leq\frac{1}{63}.
\]
Additionally, 
\[
(1-r)^{2}\nabla^{2}f_{\kappa}(x)\preceq\nabla^{2}f_{\kappa}(z_{\kappa})\preceq\frac{1}{(1-r)^{2}}\nabla^{2}f_{\kappa}(x),
\]
and thus 
\[
\|x-z_{\kappa}\|_{\nabla^{2}f_{\kappa}(z_{\kappa})}\leq\frac{r}{1-r}\leq\frac{1}{62}.
\]
Combining all the inequalities, we obtain 
\[
\frac{\left|s_{i}(Jx)-s_{i}(Jz_{\kappa})\right|}{s_{i}(Jz_{\kappa})}\leq\frac{1}{20}\cdot\frac{1}{62}=\frac{1}{1240}<\frac{1}{2}.
\]
\emph{Penalty guarantee.} We have 
\[
\frac{e^{\kappa}}{2}\|\Gamma x\|^{2}_{2}=\Psi_{\kappa}(x)-\Psi_{fl}(Jx)\leq\Psi_{\kappa}(x)-\Phi(S,P).
\]
Next, we show that $\Psi_{\kappa}(x)\leq\Psi_{\kappa}(z_{\kappa})+\varepsilon$.
Let $\rho(s)=-\ln(1-s)-s$. Since $\mathfrak{d}_{\kappa}(x)<1$, by
\cite[Equation 2.6b]{NT08}, we have 
\[
f_{\kappa}(x)-f_{\kappa}(z_{\kappa})\leq\rho(\mathfrak{d}_{\kappa}(x)).
\]
Since $\rho(s)\leq s^{2}$ for $s\leq1/2$, we obtain 
\[
f_{\kappa}(x)-f_{\kappa}(z_{\kappa})\leq\mathfrak{d}_{\kappa}(x)^{2}
\]
and thus 
\begin{equation}
\Psi_{\kappa}(x)-\Psi_{\kappa}(z_{\kappa})\leq\frac{\mathfrak{d}_{\kappa}(x)^{2}}{c}\leq\frac{\overline{\tau}^{2}}{\lambda}=\varepsilon.\label{eq:penalty-gap}
\end{equation}
Plugging into the previous inequality, we obtain 
\[
\frac{e^{\kappa}}{2}\|\Gamma x\|^{2}_{2}\leq\Psi_{\kappa}(z_{\kappa})-\Phi(S,P)+\varepsilon.
\]
Since the exact center for the barrier on $P\cap S'$ has zero penalty
and is a candidate for $z_{\kappa}$, we have $\Psi_{\kappa}(z_{\kappa})\le\Phi(S',P)$,
and thus 
\[
\frac{e^{\kappa}}{2}\|\Gamma x\|^{2}_{2}\le\Phi(S',P)-\Phi(S,P)+\varepsilon=\Delta\Phi+\varepsilon.
\]
\end{proof}

\subsection{Analysis of \textsc{ProjectAndCheck}}

By the above lemmas, the Newton decrement is at most $1/8$ at the
start of each phase and thus each phase terminates in at most $4$
iterations. Next, we show that, when the algorithm terminates, the
returned point satisfies the approximate $\tau$-center conditions.
Additionally, we show that the algorithm terminates once the parameter
becomes sufficiently large.

We first show the following lemma that will be used to show that the
slacks remain positive after projecting $x$ to $\overline{x}$ on
line \ref{line:project-point} of $\textsc{ProjectAndCheck}$. As
before, we let $W=n+\lambda M$.
\begin{lem}[Slack lower bounds]
\label{lem:penalty-slack-bounds}Consider a phase with parameter
$\kappa$. Let $z_{\kappa}$ be the unique minimizer of $\Psi_{\kappa}$,
and let $x$ be the vector returned by $\textsc{RecenterPhase}$ in
that phase. For every retained row, we have $s_{i}(Jz_{\kappa})\ge\frac{\lambda\rho_{i}}{W}$
and $s_{i}(Jx)\ge\frac{\lambda\rho_{i}}{2W}\ge\frac{\lambda\delta}{2W}$.
Additionally, $\calh_{S}(x)\preceq\frac{4MW^{2}}{\lambda^{2}\delta^{2}}I_{d}$. 
\end{lem}

\begin{proof}
Let $\ell_{i}:=\ell_{i}(Jz_{\kappa})$ be the $i$-th leverage score
evaluate at $Jz_{\kappa}$ in the full original space. We have $\sum_{i}\ell_{i}=n$.
Since $z_{\kappa}$ is the minimizer of $\Psi_{\kappa}$, we have
\[
0=\nabla\Psi_{\kappa}(z_{\kappa})=\sum_{i}(\ell_{i}+\lambda)\frac{J^{\top}v_{i}}{s_{i}(Jz_{\kappa})}+e^{\kappa}\Gamma^{\top}\Gamma z_{\kappa}.
\]
By taking the inner product with $z_{\kappa}$ and using that $\sum_{i}\ell_{i}=n$
and there are $M$ retained rows, we obtain 
\begin{align*}
0 & =\sum_{i}(\ell_{i}+\lambda)\frac{v^{\top}_{i}Jz_{\kappa}}{s_{i}(Jz_{\kappa})}+e^{\kappa}\left\Vert \Gamma z_{\kappa}\right\Vert ^{2}_{2}\\
 & =\sum_{i}(\ell_{i}+\lambda)\frac{\rho_{i}-s_{i}(Jz_{\kappa})}{s_{i}(Jz_{\kappa})}+e^{\kappa}\left\Vert \Gamma z_{\kappa}\right\Vert ^{2}_{2}\\
 & =\sum_{i}(\ell_{i}+\lambda)\frac{\rho_{i}}{s_{i}(Jz_{\kappa})}-\sum_{i}(\ell_{i}+\lambda)+e^{\kappa}\left\Vert \Gamma z_{\kappa}\right\Vert ^{2}_{2}\\
 & =\sum_{i}(\ell_{i}+\lambda)\frac{\rho_{i}}{s_{i}(Jz_{\kappa})}-(n+\lambda M)+e^{\kappa}\left\Vert \Gamma z_{\kappa}\right\Vert ^{2}_{2}\\
 & =\sum_{i}(\ell_{i}+\lambda)\frac{\rho_{i}}{s_{i}(Jz_{\kappa})}-W+e^{\kappa}\left\Vert \Gamma z_{\kappa}\right\Vert ^{2}_{2}.
\end{align*}
Thus 
\[
\sum_{i}(\ell_{i}+\lambda)\frac{\rho_{i}}{s_{i}(Jz_{\kappa})}\leq W.
\]
Since each term in the sum is positive, it follows that, for every
$i$, 
\[
(\ell_{i}+\lambda)\frac{\rho_{i}}{s_{i}(Jz_{\kappa})}\leq W
\]
and thus 
\[
s_{i}(Jz_{\kappa})\geq\frac{(\ell_{i}+\lambda)\rho_{i}}{W}\geq\frac{\lambda\rho_{i}}{W}.
\]
The slack guarantee in Lemma~\ref{lem:penalty-newton-corrections}
gives 
\[
s_{i}(Jx)\geq\frac{1}{2}s_{i}(Jz_{\kappa})\geq\frac{1}{2}\frac{\lambda\rho_{i}}{W}\geq\frac{1}{2}\frac{\lambda\delta}{W}.
\]
As noted above, Lemmas \ref{lem:recenter-initialization} and \ref{lem:penalty-newton-corrections}
guarantee that the Newton decrement is at most $1/8$ at the beginning
of each phase.

Since $J$ is an isometry and $\|v_{i}\|_{2}\le1$, we have 
\[
\calh_{S}(x)=\sum_{i}\frac{(J^{\top}v_{i})(J^{\top}v_{i})^{\top}}{s_{i}(Jx)^{2}}\preceq\frac{4MW^{2}}{\lambda^{2}\delta^{2}}I_{d}.
\]
\end{proof}

Let $\varepsilon=\overline{\tau}^{2}/\lambda$ as in Lemma \ref{lem:penalty-newton-corrections},
and let 
\begin{gather*}
\widetilde{\tau}:=\frac{\tau}{32(3+\lambda)},\\
\kappa_{*}:=\max\!\left\{ \kappa_{0},\log\!\left(\frac{8M_{*}W^{2}_{*}(B_{*}+\varepsilon)}{\lambda^{2}\delta^{2}\widetilde{\tau}^{2}}\right)\right\} ,
\end{gather*}
where $M_{*}$, $W_{*}$, $B_{*}$ are the values defined in \eqref{eq:gap-budget}.
Note that we have $\kappa_{*}-\kappa_{0}=O(\log(n+2))$. Indeed, $e^{\kappa_{0}}=\tau/(8n)$
and the expression inside the logarithm defining $\kappa_{*}$ is
at most polynomial in $n$.

Recall that Algorithm \ref{alg:recenter} constructs a matrix $J'$
with orthonormal columns that span $\ker\Gamma$. Thus the matrix
$J_{\mathrm{new}}:=JJ'$ has orthonormal columns that span $S'$.
For any point $y$ in the old-coordinate domain, let 
\[
G(y):=J'^{\top}J^{\top}\nabla\Psi_{fl}(Jy),\qquad\calh'(y)=J'^{\top}\calh_{S}(y)J'.
\]
At $y=\overline{x}$ these are the gradient and normal matrix in the
new section. They are used only in the analysis of the projection
test; \textsc{ProjectAndCheck} does not evaluate them.
\begin{lem}[Projection certification and termination]
\label{lem:penalty-projection}Consider a phase with parameter $\kappa$.
Let $x$ be vector returned by $\textsc{RecenterPhase}$ and $\overline{x}=x-\Gamma^{\top}\Gamma x$.
If $\left\Vert h\right\Vert ^{2}_{\calh_{S}(x)}\le\widetilde{\tau}^{2}$,
$\textsc{ProjectAndCheck}$ returns $(\mathrm{true},\overline{x})$
at line \ref{line:project-return} and $\overline{x}$ satisfies $\Gamma\overline{x}=0$,
$s_{i}(J\overline{x})>0$ for every retained row $i$, and $\|G(\overline{x})\|_{\calh'(\overline{x})^{-1}}<\tau.$
Furthermore, if $\kappa\geq\kappa_{*}$ then $\left\Vert h\right\Vert ^{2}_{\calh_{S}(x)}\le\widetilde{\tau}^{2}$.
\end{lem}

\begin{proof}
Let us now show that the guarantee $\left\Vert h\right\Vert ^{2}_{\calh_{S}(x)}\le\widetilde{\tau}^{2}$
implies that $\overline{x}$ satisfied the desired conditions. Since
$\Gamma\Gamma^{\top}=I_{r}$, we have $\Gamma\overline{x}=0$ and
$\|h\|_{2}=\|\Gamma x\|_{2}$. The entire segment $x_{t}=x+th$, $0\le t\le1$,
is strictly feasible (i.e., the slacks are strictly positive). Indeed,
we have 
\[
\frac{|v^{\top}_{i}Jh|}{s_{i}(Jx)}\le\|h\|_{\calh_{S}(x)}\le\widetilde{\tau}<1,
\]
and thus we have 
\begin{equation}
\begin{gathered}(1-t\widetilde{\tau})s_{i}(Jx)\le s_{i}(Jx_{t})\le(1+t\widetilde{\tau})s_{i}(Jx),\\
(1+t\widetilde{\tau})^{-2}\calh_{S}(x)\preceq\calh_{S}(x_{t})\preceq(1-t\widetilde{\tau})^{-2}\calh_{S}(x).
\end{gathered}
\label{eq:penalty-projection-metrics}
\end{equation}
Since $\Gamma J'=0$, we have $G(x)=J'^{\top}\nabla\Psi_{\kappa}(x)$
and $J'^{\top}\mathcal{K}_{\kappa}(x)J'=\calh'(x)$. The variational
formula for an inverse quadratic form therefore yields 
\begin{align*}
\|G(x)\|^{2}_{\calh'(x)^{-1}} & =\max_{u\in\ker\Gamma}\{2\nabla\Psi_{\kappa}(x)^{\top}u-u^{\top}\mathcal{K}_{\kappa}(x)u\}\\
 & \le\max_{u\in\mathbb{R}^{d}}\{2\nabla\Psi_{\kappa}(x)^{\top}u-u^{\top}\mathcal{K}_{\kappa}(x)u\}\\
 & =\|\nabla\Psi_{\kappa}(x)\|^{2}_{\mathcal{K}_{\kappa}(x)^{-1}}\le\overline{\tau}^{2},
\end{align*}
where the last inequality follows from the terminating condition on
line~\ref{line:phase-stop} of \textsc{RecenterPhase}.

To bound $\|G(\overline{x})\|_{\calh'(\overline{x})^{-1}}$, we bound
the change in $G$ along the projection segment. Let $B_{S}(y)=J^{\top}\nabla^{2}\Psi_{fl}(Jy)J$.
The fundamental theorem of calculus gives 
\[
G(\overline{x})-G(x)=\int^{1}_{0}J'^{\top}B_{S}(x_{t})h\,dt
\]
and thus 
\[
\|G(\overline{x})\|_{\calh'(x)^{-1}}\leq\|G(x)\|_{\calh'(x)^{-1}}+\int^{1}_{0}\|J'^{T}B_{S}(x_{t})h\|_{\calh'(x)^{-1}}\,dt.
\]
By Lemma~\ref{lem:hybrid-sc} and (\ref{eq:penalty-projection-metrics}),
\[
0\preceq B_{S}(x_{t})\preceq\frac{3+\lambda}{(1-\widetilde{\tau})^{2}}\calh_{S}(x).
\]
Thus 
\[
0\preceq\calh_{S}(x)^{-1/2}B_{S}(x_{t})\calh_{S}(x)^{-1/2}\preceq\frac{3+\lambda}{(1-\widetilde{\tau})^{2}}I
\]
and therefore 
\begin{align*}
\|B_{S}(x_{t})h\|_{\calh_{S}(x)^{-1}} & =\|\calh_{S}(x)^{-1/2}B_{S}(x_{t})h\|_{2}\\
 & \leq\|\calh_{S}(x)^{-1/2}B_{S}(x_{t})\calh_{S}(x)^{-1/2}\|_{2}\|\calh_{S}(x)^{1/2}h\|_{2}\\
 & \leq\frac{3+\lambda}{(1-\widetilde{\tau})^{2}}\|h\|_{\calh_{S}(x)}\leq\frac{3+\lambda}{(1-\widetilde{\tau})^{2}}\widetilde{\tau}.
\end{align*}
For any vector $q$, since $\calh'(x)=J'^{\top}\calh_{S}(x)J'$, we
have 
\begin{align*}
\|J'^{\top}q\|_{\calh'(x)^{-1}} & =\sup_{u\neq0}\frac{\left|q^{\top}J'u\right|}{\|u\|_{\calh'(x)}}=\sup_{u\neq0}\frac{\left|q^{\top}J'u\right|}{\|J'u\|_{\calh_{S}(x)}}\leq\sup_{w\neq0}\frac{\left|q^{\top}w\right|}{\|w\|_{\calh_{S}(x)}}=\|q\|_{\calh_{S}(x)^{-1}}.
\end{align*}
Thus 
\[
\|J'^{T}B_{S}(x_{t})h\|_{\calh'(x)^{-1}}\le\|B_{S}(x_{t})h\|_{\calh_{S}(x)^{-1}}\leq\frac{3+\lambda}{(1-\widetilde{\tau})^{2}}\widetilde{\tau},
\]
and therefore 
\begin{align*}
\|G(\overline{x})\|_{\calh'(x)^{-1}} & \leq\overline{\tau}+\frac{3+\lambda}{(1-\widetilde{\tau})^{2}}\widetilde{\tau}.
\end{align*}
The Hessian comparison in \eqref{eq:penalty-projection-metrics} gives
$\calh'(\overline{x})\succeq(1+\widetilde{\tau})^{-2}\calh'(x)$,
and thus 
\[
\|G(\overline{x})\|_{\calh'(\overline{x})^{-1}}\leq(1+\widetilde{\tau})\|G(\overline{x})\|_{\calh'(x)^{-1}}\leq(1+\widetilde{\tau})\left(\overline{\tau}+\frac{3+\lambda}{(1-\widetilde{\tau})^{2}}\widetilde{\tau}\right)<\frac{65}{288}\tau<\tau,
\]
where we used that $\overline{\tau}=\frac{\tau}{8}$ and $\widetilde{\tau}=\frac{\tau}{32(3+\lambda)}<\frac{1}{4}$.

Finally, we verify that $\kappa\geq\kappa_{*}$ implies that $\left\Vert h\right\Vert _{\calh_{S}(x)}\le\widetilde{\tau}$.
Using Lemma~\ref{lem:penalty-slack-bounds}, the fact that $\|h\|_{2}=\|\Gamma x\|_{2}$,
and the penalty guarantee in Lemma~\ref{lem:penalty-newton-corrections},
we obtain
\begin{align*}
\|h\|^{2}_{\calh_{S}(x)} & \leq\frac{4MW^{2}}{\lambda^{2}\delta^{2}}\left\Vert h\right\Vert ^{2}_{2}=\frac{4MW^{2}}{\lambda^{2}\delta^{2}}\|\Gamma x\|^{2}_{2}\le\frac{8MW^{2}(\Delta\Phi+\varepsilon)}{\lambda^{2}\delta^{2}e^{\kappa}}.
\end{align*}
Suppose that $\kappa\ge\kappa_{*}$. The upper bound $M\leq M_{*}$
given by Lemma \ref{lem:insertion-bound}, the upper bound $\Delta\Phi\leq B_{*}$
shown in \eqref{eq:gap-budget}, and the the definition of $\kappa_{*}$
imply that the last expression above is at most $\widetilde{\tau}^{2}$.
Thus we have $\|h\|^{2}_{\calh_{S}(x)}\leq\widetilde{\tau}^{2}$.
\end{proof}

\subsection{Charging all phases to the potential}

In this section, we upper bound the number of phases, i.e., the number
of times line~\ref{line:recenter-center} in \textsc{Recenter} is
executed. Lemma~\ref{lem:penalty-projection} bounds the range of
the penalty parameter $\kappa$, but the relative increments at line~\ref{line:recenter-energy}
can be small when $\mu_{\kappa}$ is large. We therefore count parameter
increases in two ways. Increases with $\mu_{\kappa}\ge1$ consume
a fixed amount of potential; the remaining increases advance $\kappa$
by a fixed amount. For a fixed body and contraction, define 
\[
F(\kappa)=\Psi_{\kappa}(z_{\kappa}),\qquad E(\kappa)=e^{\kappa}\|\Gamma z_{\kappa}\|^{2}_{2}.
\]
The exact centers $z_{\kappa}$ are as in Lemma~\ref{lem:penalty-newton-corrections}.
The functions $F$ and $E$ are used only in the analysis. 
\begin{lem}[Energy along the exact penalty path]
\label{lem:penalty-path-energy} For finite $\kappa$, the functions
$F$ and $E$ satisfy 
\[
\Phi(S,P)\le F(\kappa)\le\Phi(S',P),\qquad\frac{dF}{d\kappa}=\frac{E(\kappa)}{2},\qquad E'(\kappa)\ge-E(\kappa).
\]
In particular $F$ is nondecreasing, and for every $\vartheta\ge0$
its increase satisfies 
\begin{equation}
F(\kappa+\log(1+\vartheta))-F(\kappa)\ge\frac{E(\kappa)}{2}\cdot\frac{\vartheta}{1+\vartheta}.\label{eq:penalty-energy-charge}
\end{equation}
At any computed outgoing point $x$, 
\[
|\mu_{\kappa}-\sqrt{E(\kappa)}|\le\sqrt{2\varepsilon}.
\]
\end{lem}

\begin{proof}
\emph{The exact penalty minima lie between the two section minima.}
The penalty is nonnegative, so $F(\kappa)\ge\Phi(S,P)$. The exact
center of the new section is a candidate with zero penalty, so $F(\kappa)\le\Phi(S',P)$.
Hence 
\begin{equation}
\Phi(S,P)\le F(\kappa)\le\Phi(S',P)\quad\text{for every finite }\kappa.\label{eq:penalty-minimum-interval}
\end{equation}
All these exact centers and values appear only in the analysis.

\emph{Differentiating the exact path.} First order optimality condition
for $z_{\kappa}$ gives 
\[
J^{\top}\nabla\Psi_{fl}(Jz_{\kappa})+e^{\kappa}\Gamma^{\top}\Gamma z_{\kappa}=0.
\]
Let $J_{\kappa}=\nabla^{2}\Psi_{\kappa}(z_{\kappa})$. Differentiating
the above equation with respect to $\kappa$ gives: 
\begin{align*}
\underbrace{\left(J^{\top}\nabla^{2}\Psi_{fl}(Jz_{\kappa})J+e^{\kappa}\Gamma^{\top}\Gamma\right)}_{J_{\kappa}}z_{\kappa}'(\kappa) & =-e^{\kappa}\Gamma^{\top}\Gamma z_{\kappa}\\
z_{\kappa}'(\kappa) & =-e^{\kappa}J^{-1}_{\kappa}\Gamma^{\top}\Gamma z_{\kappa}
\end{align*}
We have 
\begin{align*}
\frac{dF}{d\kappa} & =\underbrace{\left(J^{\top}\nabla\Psi_{fl}(Jz_{\kappa})+e^{\kappa}\Gamma^{\top}\Gamma z_{\kappa}\right)^{\top}}_{0}z_{\kappa}'(\kappa)+\frac{e^{\kappa}}{2}\|\Gamma z_{\kappa}\|^{2}_{2}\\
 & =E(\kappa)/2\ge0
\end{align*}
Thus $F$ is nondecreasing. Differentiating $E(\kappa)$ gives 
\begin{align*}
\frac{dE}{d\kappa} & =E(\kappa)+2e^{\kappa}(\Gamma z_{\kappa})^{\top}\Gamma z_{\kappa}'(\kappa)\\
 & =E(\kappa)-2e^{2\kappa}(\Gamma z_{\kappa})^{\top}\Gamma J^{-1}_{\kappa}\Gamma^{\top}\Gamma z_{\kappa}
\end{align*}
Because $J_{\kappa}\succeq e^{\kappa}\Gamma^{\top}\Gamma$ and $J_{\kappa}\succ0$,
we have $I\succeq e^{\kappa}J^{-1/2}_{\kappa}\Gamma^{\top}\Gamma J^{-1/2}_{\kappa}$.
Thus, $I\succeq e^{\kappa}\Gamma J^{-1}_{\kappa}\Gamma^{\top}$. Substitute
this inequality into the expression for $\frac{dE}{d\kappa}$, we
obtain 
\[
\frac{dE}{d\kappa}\ge-E(\kappa)
\]
\emph{Energy yields an increase in the exact minimum.} The inequality
$\frac{dE}{d\kappa}\ge-E(\kappa)$ means $\frac{d}{d\kappa}(e^{\kappa}E(\kappa))=e^{\kappa}\left(\frac{dE}{d\kappa}+E(\kappa)\right)\ge0$
so $e^{\kappa}E(\kappa)$ is nondecreasing. Therefore, for $s\ge0$,
$E(\kappa+s)\ge e^{-s}E(\kappa)$. Integrating the identity for $\frac{dF}{d\kappa}$
over one parameter increase gives 
\begin{align*}
F(\kappa+\log(1+\vartheta))-F(\kappa) & =\frac{1}{2}\int^{\log(1+\vartheta)}_{0}E(\kappa+s)\,ds\\
 & \ge\frac{E(\kappa)}{2}\int^{\log(1+\vartheta)}_{0}e^{-s}\,ds.
\end{align*}
Evaluating the last integral proves (\ref{eq:penalty-energy-charge})\emph{.}

\emph{The computed energy approximates the exact energy.} Line~\ref{line:recenter-energy}
uses $\mu_{\kappa}=e^{\kappa/2}\|\Gamma x\|_{2}$ at the point returned
by \textsc{RecenterPhase}. Its return condition supplies the gap estimate
of Lemma~\ref{lem:penalty-newton-corrections}. The base barrier
$\Psi_{S}$ is convex. 
\[
\Psi_{S}(x)-\Psi_{S}(z_{\kappa})\ge\nabla\Psi_{S}(z_{\kappa})^{\top}(x-z_{\kappa})
\]
We also have 
\[
\left\Vert \Gamma x\right\Vert ^{2}-\left\Vert \Gamma z_{\kappa}\right\Vert ^{2}=2\left(\Gamma z_{\kappa}\right)^{\top}\Gamma(x-z_{\kappa})+\left\Vert \Gamma(x-z_{\kappa})\right\Vert ^{2}
\]
Combining the above two equations yield 
\begin{align*}
\Psi_{\kappa}(x)-\Psi_{\kappa}(z_{\kappa}) & \ge\underbrace{\left(\nabla\Psi_{S}(z_{\kappa})+e^{\kappa}\Gamma^{\top}\Gamma z_{\kappa}\right)^{\top}}_{0}(x-z_{\kappa})+\frac{e^{\kappa}}{2}\left\Vert \Gamma(x-z_{\kappa})\right\Vert ^{2}\\
 & =\frac{e^{\kappa}}{2}\left\Vert \Gamma(x-z_{\kappa})\right\Vert ^{2}
\end{align*}
Combining this inequality with the outgoing gap (\ref{eq:penalty-gap}),
and then using the reverse triangle inequality, proves 
\begin{equation}
\begin{gathered}\frac{e^{\kappa}}{2}\|\Gamma(x-z_{\kappa})\|^{2}_{2}\le\varepsilon,\\
|\mu_{\kappa}-\sqrt{E(\kappa)}|\le e^{\kappa/2}\|\Gamma(x-z_{\kappa})\|_{2}\le\sqrt{2\varepsilon}.
\end{gathered}
\label{eq:penalty-computed-energy}
\end{equation}
\end{proof}

\begin{lem}[Penalty energy pays for the warm paths]
\label{lem:penalty-energy-stages} \textsc{Recenter} executes $O(\log(n+2)+\Delta\Phi)$
phases for one contraction and $O(n\log(n+2))$ phases over all contractions.
The total number of Newton steps has the same respective bounds. Here
a phase is one call to \textsc{RecenterPhase} at line~\ref{line:recenter-center}. 
\end{lem}

\begin{proof}
The procedure initializes $\kappa$ at line~\ref{line:recenter-initialize},
and each execution of the outer loop makes one call to \textsc{RecenterPhase}
and one call to \textsc{ProjectAndCheck}. Every failed projection
is followed by exactly one parameter update at line~\ref{line:recenter-increase},
while a successful projection exits at line~\ref{line:recenter-accept}.
Thus the number of stages is one plus the number of executed updates.
We bound these updates using Lemma~\ref{lem:penalty-path-energy}.

\emph{Counting increases with $\mu_{\kappa}\ge1$.} Since $\varepsilon<1/8$,
we have $\sqrt{2\varepsilon}<1/2$. For $\mu_{\kappa}\ge1$, (\ref{eq:penalty-computed-energy})
implies $\sqrt{E(\kappa)}\ge\mu_{\kappa}-1/2\ge\mu_{\kappa}/2$, so
$E(\kappa)\ge\mu^{2}_{\kappa}/4$. Line~\ref{line:recenter-energy}
chooses $\vartheta=\vartheta_{0}/\mu_{\kappa}\le\vartheta_{0}<1$.
Equation~(\ref{eq:penalty-energy-charge}) consequently gives 
\[
F(\kappa_{+})-F(\kappa)\ge\frac{E(\kappa)\vartheta_{0}/\mu_{\kappa}}{2(1+\vartheta_{0}/\mu_{\kappa})}\ge\frac{\vartheta_{0}\mu_{\kappa}}{16}\ge\frac{\vartheta_{0}}{16}.
\]
The parameter intervals for successive increases are disjoint and
$F$ is nondecreasing. The sum of their increases is at most $\Phi(S',P)-\Phi(S,P)=\Delta\Phi$
by (\ref{eq:penalty-minimum-interval}). Thus there are at most $16\Delta\Phi/\vartheta_{0}$
parameter increases of this type, even when other increases occur
between them.

\emph{Counting increases with $\mu_{\kappa}<1$.} At every remaining
execution of line~\ref{line:recenter-energy}, $\log(1+\vartheta)=\log(1+\vartheta_{0})\ge\vartheta_{0}/2$,
so line~\ref{line:recenter-increase} advances the parameter by $\kappa_{+}-\kappa\ge\vartheta_{0}/2$.

By Lemma~\ref{lem:penalty-projection}, the call at line~\ref{line:recenter-stage-project}
succeeds at the first outgoing stage with $\kappa\ge\kappa_{*}$.
The main loop then exits before another parameter update. Any increase
crossing that threshold starts below it and has relative size at most
$\log(1+\vartheta_{0})\le\vartheta_{0}$. Thus every parameter value
reached by the pseudocode is at most $\kappa_{*}+\vartheta_{0}$,
and the parameter advance from $\kappa_{0}$ in every finite execution
prefix is at most $\vartheta_{0}+\kappa_{*}-\kappa_{0}$. Since all
advances are positive, the number of increases with $\mu_{\kappa}<1$
is at most 
\[
\frac{2}{\vartheta_{0}}\left(\vartheta_{0}+\kappa_{*}-\kappa_{0}\right)=O(\log(n+2)).
\]
This argument also counts the last increase if it crosses the threshold.

Adding the one initialization stage proves the explicit bound 
\[
1+\frac{16\Delta\Phi}{\vartheta_{0}}+\frac{2}{\vartheta_{0}}\left(\vartheta_{0}+\kappa_{*}-\kappa_{0}\right)=O(\Delta\Phi+\log(n+2)).
\]
The argument bounds every finite execution prefix. Each helper call
terminates, so these uniform bounds exclude an infinite outer loop:
\textsc{Recenter} eventually reaches its return at line~\ref{line:recenter-return}.
Acceptance at an earlier outgoing point only reduces this count.

\emph{Summing over all contractions.} There are at most $n-1$ contraction
events, because each removes at least one dimension. The retained
body may differ between events, but each event's potential increase
is nonnegative and the common budget (\ref{eq:gap-budget}) gives
$\sum_{j}\Delta\Phi_{j}\le B_{*}=O(n\log(n+2))$. Summing the per-contraction
bound therefore gives 
\[
O\!\left((n-1)\log(n+2)+\sum_{j}\Delta\Phi_{j}\right)=O(n\log(n+2))
\]
executed phases. Since each phase needs at most four Newton steps,
by Lemma~\ref{lem:penalty-newton-corrections}, the same bounds hold
for the number of executions of line~\ref{line:phase-newton-step}
in all calls to $\textsc{RecenterPhase}$. 
\end{proof}

\subsection{Analysis of the arithmetic operations}

In this section, we analyze the number of arithmetic operations performed
by the algorithm. 
\begin{lem}[Algorithm implementation]
\label{lem:penalty-dense-implementation} Each call made by \textsc{Recenter}
to \textsc{RecenterPhase} costs $\widetilde{O}((M+n)^{\omega})$ arithmetic
operations. The same bound holds for a Newton step or a wide/thin
test. A call to \textsc{ProjectAndCheck} costs $O(Mn)$. All these
calculations use stored data and make no value-oracle calls. 
\end{lem}

\begin{proof}
At an interior point $Jx$, we compute $A_{fl}$, $\calh_{fl}$, $\Pi_{fl}$,
and the leverage scores as in Section~\ref{sec:geometry}, then restrict
the barrier derivatives to the current basis. . The penalty adds
$e^{\kappa}\Gamma^{\top}\Gamma x$ to the gradient, $e^{\kappa}\Gamma^{\top}\Gamma$
to the Hessian, and $(e^{\kappa}/\lambda)\Gamma^{\top}\Gamma$ to
$\mathcal{K}_{\kappa}(x)$. Thus computing $\nabla\Psi_{\kappa}(x)$
and $\nabla^{2}\Psi_{\kappa}(x)$ require only a constant number of
matrix products and linear system solves of order at most $M+n$.
Padding rectangular products to this order costs $\widetilde{O}((M+n)^{\omega})$;
row scaling and entrywise squaring cost $O((M+n)^{2})$.

We note that each $\tau$-approximate centering test (line \ref{line:phase-stop})
and each Newton step (line \ref{line:phase-newton-step}) in Algorithm
\ref{alg:recenter-phase} can be implemented using a linear system
solve. For the former, we solve the system $\mathcal{K}_{\kappa}(x)w=\nabla\Psi_{\kappa}(x)$
and check whether $\left\langle w,\nabla\Psi_{\kappa}(x)\right\rangle \leq\overline{\tau}^{2}$.
For the latter, we solve the system $\nabla^{2}\Psi_{\kappa}(x)w=\nabla\Psi_{\kappa}(x)$
and update $x\gets x-w$. By Lemma~\ref{lem:penalty-newton-corrections},
a phase uses at most four Newton steps and five $\tau$-approximate
centering tests. 

A single run of $\textsc{ProjectAndCheck}$ computes $h$ using the
stored contraction projector $\Gamma^{\top}\Gamma$ in $O(n^{2})$
operations. It then computes $q$ from the cached projected rows and
current slacks in $O(Mn)$ operations. Since $M\ge2n$, these costs
total $O(Mn)$.  
\end{proof}

\begin{lem}[Basis and row transfer after a contraction]
\label{lem:penalty-basis-transfer} Preparing the new block basis
and contraction projector costs $O(n^{2})$ per contraction; transferring
the center and retained projected rows costs $O(Mn)$. The total over
all contractions is $\widetilde{O}(n^{3})$. 
\end{lem}

\begin{proof}
For old blocks $C$ and new blocks $D$, define 
\[
J'_{C,D}=\begin{cases}
\sqrt{|C|/|D|}, & C\subseteq D,\\
0, & \text{otherwise}.
\end{cases}
\]
Its columns have disjoint supports and unit norm, and $JJ'=(\xi_{D})_{D}$.
Each old block belongs to one new block, so $J'$ has only $d$ nonzero
entries. Thus forming $J_{\mathrm{new}}=JJ'$ costs $O(nd)$, while
applying $J'^{\top}$ to the center and the $M$ cached projected
rows costs $O((M+1)d)$. The original rows and offsets stay fixed.
The contraction calculations use 
\[
\Gamma^{\top}\Gamma=I_{d}-J'J'^{\top},\qquad\|\Gamma x\|_{2}=\|(I_{d}-J'J'^{\top})x\|_{2},
\]
so constructing the projector in $O(d^{2})$ operations suffices.
Since $d\le n$, there are at most $n-1$ contractions, and $M\le M_{*}=O(n\log(n+2))$,
the total cost is $O(n^{3}+M_{*}n^{2})=\widetilde{O}(n^{3})$. 
\end{proof}

By Lemma~\ref{lem:penalty-energy-stages}, there are $T=O(n\log(n+2))$
phases over all contractions. Each phase makes one call to each helper,
and there are no initial projection tests. Using $M\le M_{*}=O(n\log(n+2))$,
the two preceding lemmas give the total recentering cost 
\[
\widetilde{O}\!\left(T(M_{*}+n)^{\omega}+TM_{*}n+n^{3}\right)=\widetilde{O}(n^{\omega+1})
\]
for every admissible $\omega\in[2,3]$. 

\section{The complete algorithm and termination}

\label{sec:algorithm} Initialize an empty implication graph. Initialize
the minimum value $m$, the witness set $W$, the sliding parameter
$t$ (Section \ref{sec:calibration}). Initialize caches for calibration-set
values, indexed by original elements (Section \ref{sec:calibration}
and Lemma \ref{lem:cal-cache}). Initialize $P=\left[-1,1\right]^{n}$;
its exact initial center is zero. Calibrate the initial empty DAG
(enforce~\eqref{eq:calibration}) and update $m$, $W$, and $t$
as necessary before the first query to the separation oracle (Section
\ref{subsec:separation-oracle}). Thereafter execute the following
procedures with the current quotient DAG and original-space constraint
rows. 
\begin{enumerate}
\item After an SCC contraction, reaggregate every compressed stored vector
returned by the separation oracle and prepare its corresponding isometric
vector. Keep the constraints rows and offsets unchanged, and retain
the incoming old basis and center for the recentering algorithm (Section
\ref{subsec:recenter-algorithm}). A zero greedy vector certifies
that the witness is optimal, and the algorithm terminates. Update
the center using the algorithm in Section~\ref{sec:update-center} and \ref{subsec:recenter-algorithm}.
After constructing the new center, set the basis to $J_{\mathrm{new}}$
and represent the center in its new coordinates (Section \ref{sec:update-center}).
 Recalibrate using~\eqref{eq:calibration}. 
\item If the current DAG has only one block, terminate since the witness
is optimal. Otherwise, call the separation oracle (Section \ref{subsec:separation-oracle})
at the compressed weights of the center and receive a vector. If the
returned vector is zero, terminate since the witness is optimal. Otherwise,
keep the vector as a potential new row with its historical greedy
order, partition, DAG, and timestamp.
\item Wide step: Consider the vector in step 2. Lift its current unit isometric
direction to a unit normal $b$ in the current original-space subspace.
Compute $\upsilon=\sqrt{b^{\top}\calh_{fl}(z)^{-1}b/\gamma}$, where
$z$ is the center. If $\upsilon\ge\delta$, insert the vector above
as a new row with offset $\rho=\upsilon+b^{\top}z$, restore the center
with exact Newton steps, and return to step 2. 
\item Thin event: If $\upsilon<\delta$, form the almost canceling convex
combination of inserted rows of Lemma~\ref{lem:center-to-certificate}
and convert them to a combinatorial certificate using Lemma~\ref{lem:accuracy}.
The vector from line 2 is not inserted into the constraint set. 
\item If the certificate's $\Delta\ge\beta$, terminate since the witness
is optimal. Otherwise choose a target from Lemma~\ref{lem:target},
extract one arc by Lemma~\ref{lem:extract}, and insert it. If this
creates an SCC, return immediately to step 1 before any DAG calibration.
Otherwise recalibrate, keep the certificate unchanged, and repeat
this step.
\end{enumerate}
\begin{prop}[Correctness and finiteness]
\label{prop:correctness} This algorithm terminates and its returned
witness minimizes $h$.
\end{prop}

\begin{proof}
Initialization satisfies \eqref{eq:invariant}. Sliding parameter
updates increase $t$ without exceeding $t^{*}$, and preserve the
invariant \eqref{eq:invariant}. Each new arc excludes only current
closed sets with strictly positive $H_{t}$, by Lemma~\ref{lem:extract},
and stays valid for future parameters. SCC contraction retains all
invariant sets as unions of the new blocks. Historical cutting planes
remain valid by Lemma~\ref{lem:historical-cuts-valid}.

We can certify that $H_{t}$ is nonnegative on closed sets for a zero
greedy vertex, a certificate with $\Delta\ge\beta$, or the final
one-block DAG. The number of wide steps is bounded. Each nonterminal
thin event runs in finite time and leads to a contraction, by Lemma~\ref{lem:episode}.
At termination every closed set has original value at least $m$;
the original endpoints also have value at least $m$. By \eqref{eq:invariant},
an original minimizer is among those sets. The stored witness attains
$m$, proving its optimality. 
\end{proof}

\begin{proof}[Proof of the query part of Theorem~\ref{thm:main}]
By Theorem~\ref{thm:fresh} there are $q=O(n\log(n+2))$ homogeneous
calls. Each greedy vector is fully evaluated once, using at most $n-1$
proper prefix values. Sliding that chain reuses its cached original
values. All later projections, geometric computations, Newton's steps,
and coefficient inversion use no EO. There are $O(n^{2})$ arc extractions,
each of which costs $O(\log(n+2))$ EO by Lemma~\ref{lem:extract}.
All calibrations cost $O(n^{2})$ EO by Lemma~\ref{lem:cal-cache}.
Endpoints cost two. Thus the complete $\EO$ count is $O(nq)+O(n^{2}\log(n+2))+O(n^{2})+2=O(n^{2}\log(n+2)).$
\end{proof}

\section{Complete deterministic time bound}\label{sec:cubic-bookkeeping}
\begin{lem}
Let $q$ be the total number of stored oracle vectors (greedy and
graph). There is a data structure with $O(n^{2})$ amortized time
per oracle query and $O(r+n\log(n+2))$ amortized time per arc extraction.
\end{lem}

\begin{proof}
To get an arc, first we scan the blocks to find $z$ with $\theta c_{z}>G$
and compute $\beta_{A_{z}}$, which takes $O(n)$ time. The main issue
with the naive approach is the time it takes to find a term in the
certificate with large gap. We describe a data structure to quickly
identify the target with large gap so that we can extract a new arc.
When an original row $a_{i}$ is inserted with blocks $C$ at that
time, construct the vector $\mathsf{x}_{i}\in\R^{n}$ with $\mathsf{x}_{i}(u)=a_{i}(C)/|C|$.
We maintain 
\[
S_{i,z}=\sum_{u\in\Anc(z)}\mathsf{x}_{i}(u),
\]
where $\Anc(z)$ is the original ancestor set: all original items
$u$ that can reach $z$ in the graph of the original ground set,
including $z$ and its SCC. Initialize all target sums when a row
is installed in $O(n^{2})$ work. Every newly acquired original reachability
pair $(u,z)$ adds $\mathsf{x}_{i}(u)$ to $S_{i,z}$ for each installed
row. There are at most $n^{2}$ such pair events, so all initializations
and updates cost $O(qn^{2})$ arithmetic and $O(qn)$ space. Neither
SCC changes nor a new certificate reset this cache. Original items
in one SCC have the same ancestor set. For $A_{z}=U\setminus\Anc(z)$,
the sum is $a_{i}(A_{z})=a_{i}(U)-S_{i,z}=-S_{i,z}$, since every
$a_{i}(U)=0$. Thus a greedy gap equals $\alpha_{i}[c_{z}-(t-t_{i})+S_{i,z}]$
and a graph gap equals $\alpha_{i}S_{i,z}$. Selecting the largest
gap for the chosen target $z$ takes $O(r)$ time. For a selected
greedy row, restrict its old order to $A_{z}$, build coordinate and
baseline prefix sums, and assign each current positive baseline to
one old block that is part of its current block, in $O(n)$ work.
Every queried prefix is the intersection of two old closed sets so
it is closed. Each binary-search value therefore needs only $O(n)$
to form its explicit original set, plus one $\EO$.
\end{proof}

\textbf{Time for the geometric portion.}

The potential argument bounds the number of homogeneous oracle queries
by $O(n\log(n+2))$. Evaluating their greedy chains and all arc extraction
and calibration cost $O(n^{2}\log(n+2))$ original $\EO$.

Lemma~\ref{lem:penalty-energy-stages} bounds all contraction paths
by $O(n\log(n+2))$ stages, each with a constant number of exact dense
Newton corrections. Insertions of new constraint rows require another
$O(n\log(n+2))$ corrections. There are at most $n$ thin events,
each requiring $O(n^{\omega})$ time. Hence all matrix work costs
\[
\widetilde{O}\!\left([n\log(n+2)](n\log(n+2))^{\omega}\right)=\widetilde{O}(n^{\omega+1}).
\]

\textbf{Time for the combinatorial portion.} Let $q$ denote the total
number of stored vectors returned by the separation oracle. Theorem~\ref{thm:fresh}
gives $q=O(n\log(n+2))$ overall. Greedy generation, graph-consistency
scans, and original set lifting cost $O(n^{2})$ per fresh oracle
call, or $O(n^{3}\log(n+2))$ in total. At each of at most $n-1$
contractions, re-aggregate every stored row in $O(nq)$ arithmetic
operations and normalize its current projection. This costs $O(n^{2}q)=O(n^{3}\log(n+2))$
overall. Current parameter credits $t-t_{i}$ are updated from timestamps
attached to all stored points. The total time for maintaining reachability
among original items, SCCs and quotient graphs is $O(n^{3})$. The
total time for calibration throughout the algorithm is $O(n^{3})$
by Lemma \ref{lem:cal-cache}.

A complete arc extraction costs $O(r+n\log(n+2))$ arithmetic operations.
At most $n^{2}$ arc extractions, together with the preceding cache
updates, cost 
\begin{equation}
O(qn^{2}+n^{3}\log(n+2))=O(n^{3}\log(n+2)).\label{eq:bookkeeping}
\end{equation}
The complete deterministic time is therefore $O(n^{2}\log(n+2)\,\EO)+\widetilde{O}(n^{\omega+1}).$

\section{Improving arithmetic runtime using randomized leverage-score maintenance}

\label{sec:randomized-maintenance}

The implementation above recomputes a projection matrix and a barrier
Hessian after every contraction. We now give a randomized implementation
that avoids these computations. It maintains full-space leverage scores
and uses approximate linear solves for local corrections. All representations
of inserted rows and offsets, and the certificate requirements remain
the same as before. The barrier used in the potential argument is
still the single function $\Psi_{fl}$ of~\eqref{eq:fullspace-barrier}.
\begin{thm}[Randomized arithmetic bound]
\label{thm:randomized-time} There is a randomized implementation
of the algorithm that, with high probability, returns an exact minimizer
of $h$, makes $O(n^{2}\log(n+2))$ original value-oracle calls, and
uses $\widetilde{O}(n^{3})$ arithmetic operations. Its total time
is therefore 
\[
O(n^{2}\log(n+2)\,\EO)+\widetilde{O}(n^{3}).
\]
The probability refers to the internal maintenance data structures;
on their simultaneous success event, the output is an exact minimizer.
It replaces the dense geometric implementation, including the refinement
procedure, by the procedures below. 
\end{thm}

\subsection{Maintenance primitives and the required accuracy}

Let $V$ be the matrix with rows $v^{\top}_{i}$, let $D(x)=\Diag(s_{i}(x)^{-2})$,
and write 
\[
A=A_{fl}(x)=D(x)^{1/2}V,\qquad\calh=A^{\top}A=\calh_{fl}(x),\qquad Z=AJ,\qquad\calh_{S}=Z^{\top}Z.
\]
All leverage scores in this section are those of $A$ in the original
$n$-dimensional space and they sum up to $n$.

We state the two maintenance data structures from previous work. In
the two statements below, $\mathsf{R}$ is an inserted row matrix
with $d_{\mathrm{ds}}$ columns and at most $m_{\mathrm{ds}}$ row
slots. Write $\mathsf{r}^{\top}_{i}$ for row $i$ and, for its current
weights $\mathbf{w}$, let 
\[
\mathsf{G}=\mathsf{R}^{\top}\Diag(\mathbf{w})\mathsf{R},\qquad\ell_{\mathsf{R},i}(\mathbf{w})=w_{i}\mathsf{r}^{\top}_{i}\mathsf{G}^{-1}\mathsf{r}_{i}.
\]
These generic dimensions are distinct from the original ground-set
dimension $n$ and the current section dimension $N$.
\begin{lem}[Leverage-score maintenance interface, \cite{JLSW20}]
\label{lem:random-score-interface} Suppose the active weights are
positive, $\mathsf{R}$ has full column rank, and the active row count
is $O(d_{\mathrm{ds}})$.Let $c_{\mathrm{ls}}>0$ be a fixed constant,
$\delta_{\mathrm{fail}}$ be the failure probability, and $q_{\mathrm{max}}$
be an upper bound on the number of operations. Theorem~5.1 of~\cite{JLSW20}
gives a randomized data structure with the following operations: 
\begin{enumerate}
\item Initialization costs $O(d^{\omega+o(1)}_{\mathrm{ds}})$. 
\item A weight update on a fixed row set is permitted when 
\begin{equation}
\|\log\mathbf{w}^{\mathrm{new}}-\log\mathbf{w}^{\mathrm{old}}\|_{2}\le0.01.\label{eq:random-score-update-interface}
\end{equation}
A row $\mathsf{r}^{\top}_{\mathrm{new}}$ with weight $w_{\mathrm{new}}$
may instead be inserted when 
\begin{equation}
w_{\mathrm{new}}\mathsf{r}_{\mathrm{new}}\mathsf{r}^{\top}_{\mathrm{new}}\preceq0.01\mathsf{G}_{\mathrm{old}}.\label{eq:random-score-insert-interface}
\end{equation}
Each permitted update costs amortized $O(d^{2}_{\mathrm{ds}}\log(q_{\mathrm{max}}/\delta_{\mathrm{fail}}))$. 
\item A query costs $O(d_{\mathrm{ds}}\log(q_{\mathrm{max}}/\delta_{\mathrm{fail}}))$
and returns a score vector $\widehat{\ell}_{\mathsf{R}}$ with additive
error 
\begin{equation}
\|\widehat{\ell}_{\mathsf{R}}-\ell_{\mathsf{R}}(\mathbf{w})\|_{2}\le C_{c_{\mathrm{ls}}}\log^{-c_{\mathrm{ls}}}(d_{\mathrm{ds}}+2),\label{eq:random-score-output-interface}
\end{equation}
where $C_{c_{\mathrm{ls}}}$ depends only on the fixed accuracy exponent.
\end{enumerate}
\end{lem}

\begin{rem}
Theorem~5.1 of~\cite{JLSW20} only bounds the accumulated variance
of each outer phase by $O\left(n^{-0.1}\right)$. Doob's inequality
shows that their data structure succeeds on all steps of a single
outer phase with constant probability. One can boost the success probability
to $1-1/n^{2}$ by having $O(\log n)$ copies with fresh randomness
for every outer phase. In each iteration, we return the coordinate-wise
median of the leverage score vectors returned by the $O(\log n)$
copies. The copies are coupled since the adaptive update depends on
the joint output of the copies in prior iterations so the probability
amplification requires a bit of care. We sketch the argument next.
Suppose there are $T$ iterations in one outer phase. For copy $i$,
its estimate $\widehat{\ell}_{i,t}$ at time $t$ of the true leverage
scores $\ell_{t}$ has the decomposition $\widehat{\ell}_{i,t}=\ell_{t}+M_{i,t}+b_{i,t}$
where$M_{i,t}$ be the accumulated mean-zero error of the randomized
updates and $b_{i,t}$ is the accumulated biases and errors of the
deterministic inner and middle updates. Let $B$ be an upper bound
on $\sup_{i,t}\left\Vert b_{i,t}\right\Vert _{2}$. Suppose that in
iteration $t$, the variance of the error is bounded by the predictable
conditional bound $v_{t}$ and $\sum_{t}v_{t}\le V$, where $V$ is
deterministic. For copy $i$, let $M_{i,t}$ be the accumulated mean-zero
error of copy $i$ up to iteration $t$. Let $\tau_{i}=\inf\{t:\left\Vert M_{i,t}\right\Vert ^{2}\ge100V\}$
be the stopping time when the error $\left\Vert M_{i,t}\right\Vert $
first exceeds $10\sqrt{V}$. Define the nonnegative supermartingale
$Y_{i,t}=\left\Vert M_{i,t\wedge\tau_{i}}\right\Vert ^{2}+V-\sum^{t}_{s=1}v_{s}$.
Let $\mathcal{F}_{t}$ contain the history of all copies up to time
$t$. Note that conditioned on their joint past, the increments to
$M_{i,t}$ are independent across copies. We have $\E[Y_{i,t+1}|\mathcal{F}_{t}]\le Y_{i,t}$
and $\E[\prod_{i}Y_{i,t+1}|\mathcal{F}_{t}]\le\prod_{i}Y_{i,t}$.
For any set $S$ of copies, the probability that they all stop before
$T$ is bounded by $\Pr[\prod_{i\in S}Y_{i,T}\ge(100V)^{|S|}]\le\frac{\prod_{i\in S}Y_{i,0}}{(100V)^{|S|}}=\frac{1}{100^{|S|}}$.
If we have $k=\Omega(\log(q_{\max}/\delta_{\mathrm{fail}}))$ copies
then by union bounding over all subsets of $k/10$ copies, we can
show that the probability at least $k/10$ of the copies failing is
bounded by $\delta_{\mathrm{fail}}/q_{\max}$. If $9k/10$ copies
have accumulated mean-zero error at most $10\sqrt{V}$, each coordinate-wise
median is closer to the truth than $4k/10$ good copies. Thus, the
$\ell_{2}$ error of the coordinate-wise median is at most $\sqrt{9/4}(10\sqrt{V}+B)$.
Boosting the success probability is necessary for us because each
outer phase only lasts $T=O(n^{\omega-2})$ updates and we need to
union bound over $\widetilde{O}(q_{\max}n^{2-\omega})$ outer phases.
This introduces an extra $\log(q_{\mathrm{max}}/\delta_{\mathrm{fail}})$
factor to the time compared with the bound in \cite{JLSW20}.
\end{rem}

\begin{lem}[Inverse-maintenance interface, \cite{LS15}]
\label{lem:random-inverse-interface} Consider $T_{\mathrm{ds}}$
updates of positive definite normal matrices $\mathsf{G}_{t}=\mathsf{R}^{\top}\Diag(\mathbf{w}^{(t)})\mathsf{R}$.
Unrevealed rows have weight zero and are supplied when their weight
first becomes positive. Assume a fixed $\beta_{\mathrm{ds}}=\operatorname{poly}(m_{\mathrm{ds}})$
satisfies 
\begin{equation}
\beta^{-1}_{\mathrm{ds}}\mathsf{G}_{0}\preceq\mathsf{G}_{t}\preceq\beta_{\mathrm{ds}}\mathsf{G}_{0}\quad\text{for every update }t.\label{eq:random-inverse-range-interface}
\end{equation}
Each update must be one of the following: 
\begin{enumerate}
\item On an unchanged positive support, taking logarithms only on that support
and writing $\mathbf{u}^{(t)}=\log\mathbf{w}^{(t)}-\log\mathbf{w}^{(t-1)}$,
\begin{equation}
\left(\sum_{i}\ell_{\mathsf{R},i}(\mathbf{w}^{(t)})(u^{(t)}_{i})^{2}\right)^{1/2}\le0.1,\qquad\|\mathbf{u}^{(t)}\|_{\infty}\le0.1.\label{eq:random-inverse-update-interface}
\end{equation}
\item At most one weight changes arbitrarily, including a zero-to-positive
change revealing a new row. 
\end{enumerate}
The single-coordinate version of Definition~18 and Theorem~19 of~\cite{LS15}
maintains solvers with high probability in total time 
\begin{equation}
\widetilde{O}\!\left(d^{\omega}_{\mathrm{ds}}+T_{\mathrm{ds}}(\operatorname{nnz}(\mathsf{R})+d^{2}_{\mathrm{ds}})\right).\label{eq:random-inverse-maintenance-cost}
\end{equation}
For a right-hand side $b_{\mathrm{ds}}$ and $0<\eta\le1/2$, a solve
returns $\widehat{z}$ satisfying 
\begin{equation}
\|\widehat{z}-\mathsf{G}^{-1}_{t}b_{\mathrm{ds}}\|_{\mathsf{G}_{t}}\le\eta\|\mathsf{G}^{-1}_{t}b_{\mathrm{ds}}\|_{\mathsf{G}_{t}}\label{eq:random-inverse-solve-interface}
\end{equation}
in $\widetilde{O}((\operatorname{nnz}(\mathsf{R})+d^{2}_{\mathrm{ds}})\log(1/\eta))$
time. Lemma~16 and Theorem~17 of \cite{LS15} provide the protected
solver used for adaptive queries. 
\end{lem}

The stronger Euclidean update condition $\|\mathbf{u}^{(t)}\|_{2}\le0.01$
implies both conditions in~\eqref{eq:random-inverse-update-interface},
because every leverage score lies in $[0,1]$ and $\|\mathbf{u}^{(t)}\|_{\infty}\le\|\mathbf{u}^{(t)}\|_{2}$.
Thus the same small slack-weight updates can drive both interfaces.

The number of rows of $A$ is $M\le M_{*}=\widetilde{O}(n)$, rather
than $O(n)$. To apply the stated leverage-score result, fix an integer
$d_{0}\ge M_{*}$ with $d_{0}=\widetilde{O}(n)$ and pad $A$ to 
\[
\begin{pmatrix}A & 0\\
0 & I_{d_{0}-n}
\end{pmatrix}.
\]
This is the scaled matrix: the data structure stores the row matrix
$\Diag(V,I_{d_{0}-n})$ and the diagonal weights $\Diag(D,I_{d_{0}-n})$.
It has $d_{0}$ columns and $O(d_{0})$ rows, and the original rows
have exactly their original leverage scores. The additional diagonal
weights are fixed at one. Padding introduces only logarithmic factors
into the update cost. Its initialization costs $d^{\omega+o(1)}_{0}$,
which is within $\widetilde{O}(n^{3})$ using a fixed known exponent
strictly below $3$.

Let $\overline{\tau}=\tau/8$ and fix the sufficiently small constant
\[
\epsilon_{0}=\lambda\overline{\tau}/2^{20}.
\]
The algorithm maintains estimates $\widehat{\ell}$, clipped coordinatewise
to $[0,1]$, of the true leverage scores $\ell$ and they satisfy
\begin{equation}
\|\widehat{\ell}-\ell\|_{2}\le\epsilon_{0}.\label{eq:random-score-accuracy}
\end{equation}
Clipping cannot increase the error. To maintain $\widehat{\ell}$,
we apply Lemma \ref{lem:random-score-interface} with $C_{c_{\mathrm{ls}}}\log^{-c_{\mathrm{ls}}}(d_{0}+2)\le\epsilon_{0}$.

By Lemma~\ref{lem:restricted-inverse}, the surrogate gradient $\widehat{g}=Z^{\top}\left(\widehat{\ell}+\lambda\boldsymbol{1}\right)$
satisfies 
\begin{equation}
\|\widehat{g}-\nabla\Psi_{S}\|_{\calh^{-1}_{S}}=\|Z^{\top}(\widehat{\ell}-\ell)\|_{\calh^{-1}_{S}}\le\|\widehat{\ell}-\ell\|_{2}\le\epsilon_{0}.\label{eq:random-gradient-error}
\end{equation}
For a penalty problem using the objective $\Psi_{\kappa}(x)=\Psi_{S}(x)+\frac{e^{\kappa}}{2}\left\Vert \Gamma x\right\Vert ^{2}$,
we can approximate the gradient using $Z^{\top}\left(\widehat{\ell}+\lambda\boldsymbol{1}\right)+e^{\kappa}\Gamma^{\top}\Gamma x$.
Since $\mathcal{K}_{\kappa}\succeq\calh_{S}$, the same error bound
holds in the $\mathcal{K}^{-1}_{\kappa}$ norm: $\|\widehat{g}-\nabla\Psi_{\kappa}\|_{\mathcal{K}^{-1}_{\kappa}}\le\epsilon_{0}$.
Computing these gradients from the maintained leverage scores takes
$O(Mn)$ arithmetic operations.

\subsection{Restricted linear system solves without rebuilding after contractions}

\textbf{Intuition of the method.} Instead of the exact Newton steps
which requires the expensive $\nabla^{2}\Psi_{fl}$, we can use preconditioned
gradient descent with $\calh_{fl}$, which is a good spectral approximation
of $\nabla^{2}\Psi_{fl}$. We can use inverse maintenance and leverage
score maintenance to maintain an approximate solver for $\calh_{fl}$.
However, we also need to solve linear systems in the subspace $S$,
which involves $\calh_{S}=J^{\top}\calh_{fl}J$. It is not immediate
how to obtain a fast solver for $J^{\top}\calh_{fl}J$. Recall that
in the center update after contraction, we enforce the constraint
$\Gamma x=0$ by adding a huge penalty term $\frac{e^{\kappa}}{2}\left\Vert \Gamma x\right\Vert ^{2}$
to the barrier function. Using a similar idea, we maintain an approximate
solver for $F$ that is the sum of the original $\calh_{fl}$ plus
the Hessian of all the penalty terms $\left\Vert \Gamma x\right\Vert ^{2}$
accumulated through the center update steps with a sufficiently high
coefficient $\varpi/2$. The accumulated sum of penalty terms amount
to $\frac{\varpi}{2}\left\Vert Qx\right\Vert ^{2}$ where the rows
of $Q$ form an orthonormal basis of $S^{\perp}$. With a large coefficient
on the penalty term, it can be shown that $J^{\top}F^{-1}J$ is a
good spectral approximation of $\left(J^{\top}\calh_{fl}J\right)^{-1}$.

First fix the polynomial bound 
\begin{equation}
L_{*}:=\frac{64M_{*}W^{2}_{*}}{\lambda^{2}\delta^{2}},\qquad\varpi:=L^{2}_{*}.\label{eq:random-spectral-cap}
\end{equation}
At exact penalty centers, Lemma~\ref{lem:penalty-slack-bounds} gives
$s_{i}\ge\lambda\delta/W_{*}$. At phase starts and local correction
iterates, the slack comparison with that center loses only a fixed
factor. The same holds at certified centers after insertions and at
the local projections used below. Consequently, throughout these procedures,
\begin{equation}
2I\preceq\calh\preceq L_{*}I.\label{eq:random-H-range}
\end{equation}
The local correction argument below also justifies the stated slack
comparison for the replacement of Newton's method.

Let $Q$ have orthonormal rows spanning $S^{\perp}$, and during an
active contraction put $C=\Gamma J^{\top}$. Maintain solvers for
$\calh$ and for 
\begin{equation}
F=\calh+\varpi Q^{\top}Q+\varrho_{\kappa}C^{\top}C,\qquad\varrho_{\kappa}=e^{\kappa}/\lambda.\label{eq:random-augmented-matrix}
\end{equation}
The last term $\varrho_{\kappa}C^{\top}C$ is only present during
an active contraction. During recentering after a row insertion, we
omit the term $\varrho_{\kappa}C^{\top}C$. These are weighted normal
matrices with rows drawn from $V$, $Q$, and $C$. The rows of $Q$
are mutually orthogonal to those of $C$. When a contraction removes
$r$ directions, its $r$ active rows become permanent rows of $Q$
after the projection is accepted, and their weights change from $\varrho_{\kappa}$
to $\varpi$. Because the subspaces are nested, at most $n-1$ such
directions are ever introduced. No constraint row or maintained full-space
leverage score is changed by this transfer of new rows to $Q$.
\begin{lem}[An original-space preconditioner for a restricted solve]
\label{lem:random-restricted-solver} Let $K=J^{\top}\calh J+\varrho_{\kappa}\Gamma^{\top}\Gamma$,
with the penalty term $\varrho_{\kappa}\Gamma^{\top}\Gamma$ omitted
when inactive. Under~\eqref{eq:random-H-range}, 
\begin{equation}
K^{-1}\preceq J^{\top}F^{-1}J\preceq2K^{-1}.\label{eq:random-preconditioner}
\end{equation}
Suppose that we have an approximate solver for $F$ such that for
input $r\in\R^{N}$ and $w=F^{-1}Jr$, it returns a solution $w'$
such that $\left\Vert w'-w\right\Vert _{F}\le\varepsilon_{\mathrm{sol}}\left\Vert w\right\Vert _{F}$.
For $\varepsilon_{\mathrm{sol}}\le1/4$ and $0<\eta<1$, there is
an algorithm that on input $b$, it returns $d$ satisfying $\left\Vert d-K^{-1}b\right\Vert _{K}\le\eta\left\Vert K^{-1}b\right\Vert _{K}$
in $O(1+\log(1/\eta))$ iterations. Each iteration uses $O(Mn+n^{2})$
arithmetic operations and one call to the approximate $F$ solver. 
\end{lem}

\begin{proof}
Complete $J$ to an orthogonal basis of $\R^{n}$ using rows of $Q$.
Note that $JJ^{\top}+Q^{\top}Q=I$. In this basis, 
\[
F=\begin{pmatrix}K & B\\
B^{\top} & E+\varpi I
\end{pmatrix},\qquad\|B\|_{2}\le\left\Vert \calh\right\Vert _{2}\le L_{*},\quad E\succeq0.
\]
The upper-left block of $F^{-1}$ is the inverse of the Schur complement
$K-B(E+\varpi I)^{-1}B^{\top}$. Since $K\succeq2I$ and $\varpi=L^{2}_{*}$,
\[
0\preceq B(E+\varpi I)^{-1}B^{\top}\preceq I\preceq K/2.
\]
Inverting the resulting inequalities proves~\eqref{eq:random-preconditioner}.

Suppose we want to solve $Kd=b$ by minimizing $\frac{1}{2}\left\Vert K^{1/2}d-K^{-1/2}b\right\Vert ^{2}$.
Let $d^{*}=K^{-1}b$. We use Richardson iteration/preconditioned gradient
descent, where the ideal next iterate is $d-\tfrac{1}{2}J^{\top}F^{-1}J(Kd-b)$.
Instead of $w=F^{-1}J(Kd-b)$, the approximate solver gives $w+e_{F}$
where $\left\Vert e_{F}\right\Vert _{F}\le\epsilon_{\mathrm{sol}}\left\Vert w\right\Vert _{F}$.
Our actual new iterate is $d_{+}=d-\tfrac{1}{2}J^{\top}(w+e_{F})$.
The $K$-norm error of the solution follows the recurrence
\begin{align*}
K^{1/2}(d_{+}-d^{*}) & =K^{1/2}(d-d^{*})-\frac{1}{2}K^{1/2}J^{\top}F^{-1}JK(d-K^{-1}b)-\frac{1}{2}K^{1/2}J^{\top}e_{F}\\
 & =(I-\frac{1}{2}K^{1/2}J^{\top}F^{-1}JK^{1/2})K^{1/2}(d-d^{*})-\frac{1}{2}K^{1/2}J^{\top}e_{F}
\end{align*}
Because $\left\Vert I-\frac{1}{2}K^{1/2}J^{\top}F^{-1}JK^{1/2}\right\Vert \le1/2$
we have
\[
\left\Vert d_{+}-d^{*}\right\Vert _{K}\le\frac{1}{2}\left\Vert d-d^{*}\right\Vert _{K}+\frac{1}{2}\left\Vert J^{\top}e_{F}\right\Vert _{K}
\]
We need to bound $\left\Vert J^{\top}e_{F}\right\Vert _{K}$ based
on $\left\Vert e_{F}\right\Vert _{F}\le\varepsilon_{\mathrm{sol}}\left\Vert w\right\Vert _{F}$.
We have $JKJ^{\top}=JJ^{\top}\calh JJ^{\top}+\varrho_{\kappa}C^{\top}C$.
For any $x$, we have $JJ^{\top}x=x-Q^{\top}Qx$ and $\left\Vert JJ^{\top}x\right\Vert ^{2}_{\calh}\le2\left\Vert x\right\Vert ^{2}_{\calh}+2\left\Vert Q^{\top}Qx\right\Vert ^{2}_{\calh}$
so 
\begin{align*}
JJ^{\top}\calh JJ^{\top} & \preceq2\calh+2Q^{\top}Q\calh Q^{\top}Q\preceq2\calh+2L_{*}Q^{\top}Q\\
JKJ^{\top} & \preceq2\calh+2\varpi Q^{\top}Q+\varrho_{\kappa}C^{\top}C\preceq2F
\end{align*}
where we used $L_{*}\le\varpi$. Thus, $\|J^{\top}e_{F}\|_{K}\le\sqrt{2}\|e_{F}\|_{F}$.
On the other hand, $\|w\|^{2}_{F}=\left\Vert Kd-b\right\Vert ^{2}_{J^{\top}F^{-1}J}\le2\|Kd-b\|^{2}_{K^{-1}}$.
Therefore, $\|J^{\top}e_{F}\|_{K}\le2\varepsilon_{\mathrm{sol}}\|Kd-b\|_{K^{-1}}=2\varepsilon_{\mathrm{sol}}\left\Vert d-d^{*}\right\Vert _{K}$

Thus, after each iteration the distance $\left\Vert d-d^{*}\right\Vert _{K}$
gets multiplied with a factor at most $1/2+\varepsilon_{\mathrm{sol}}$,
which is at most $3/4$ when $\varepsilon_{\mathrm{sol}}\le1/4$.
Let $d_{t}$ be the iterate at iteration $t$. Starting with $d_{0}=0$
gives $\left\Vert d_{t}-d^{*}\right\Vert _{K}\le(3/4)^{t}\left\Vert d^{*}\right\Vert _{K}$,
proving the guarantee of the Lemma.
\end{proof}

The inverse-maintenance spectral-range hypothesis holds for both matrices.
The penalty analysis uses $e^{\kappa}$ that is at most a polynomial
in $n$. For example, a fixed bound on all active weights is 
\[
\varrho_{\max}:=\frac{e^{\vartheta_{0}}}{\lambda}\max\!\left\{ \frac{\tau}{8n},\frac{8M_{*}W^{2}_{*}(B_{*}+\overline{\tau}^{2}/\lambda)}{\lambda^{2}\delta^{2}\widetilde{\tau}^{2}}\right\} =\operatorname{poly}(n).
\]
Here $\vartheta_{0}=1/(64\sqrt{c})$ and $\widetilde{\tau}=\tau/(32(3+\lambda))$
are the penalty-path constants. Thus, putting $U_{\mathrm{ds}}=L_{*}+\varpi+\varrho_{\max}$,
\[
2I\preceq\calh,F\preceq U_{\mathrm{ds}}I,\qquad U_{\mathrm{ds}}=\operatorname{poly}(n).
\]
Initially both matrices equal $2I$. Stable updates are subdivided
until their log-weight changes have infinity norm at most $0.01$,
so even coordinatewise intermediate states lie between $e^{-0.01}\mathsf{G}_{\mathrm{old}}$
and $e^{0.01}\mathsf{G}_{\mathrm{old}}$. A single-coordinate exceptional
update is monotone in the positive semidefinite order between its
endpoints. Therefore all supplied intermediate states lie in $[I,2U_{\mathrm{ds}}I]$.
Taking $\beta_{\mathrm{ds}}=\max\{2,U_{\mathrm{ds}}\}$ verifies \eqref{eq:random-inverse-range-interface}
relative to the initial $2I$. 

\subsection{Local corrections and stable maintenance updates}

We replace an exact Newton phase by preconditioned gradient descent
iterations. Throughout this analysis assume that the maintained scores
satisfy~\eqref{eq:random-score-accuracy} and that the linear solves
meet their requested tolerances. The error bounds below are deterministic
under these assumptions and do not require independence between successive
correction errors. During one phase, keep $S$, $J$, the retained
rows, and $\kappa$ fixed. Let $x_{0}\in\R^{N}$ be its starting point
and freeze 
\[
K_{0}=\mathcal{K}_{\kappa}(x_{0})=\calh_{S}(x_{0})+\varrho_{\kappa}\Gamma^{\top}\Gamma.
\]
We will mainly describe the error propagation during a recenter phase
after contraction. For the recentering after a new constraint is inserted,
the argument is similar, and can be obtained by omitting every penalty
term $\varrho_{\kappa}\Gamma^{\top}\Gamma$.

\emph{The starting neighborhood and comparison of metrics.} The bound
shown in the proof of Lemma \ref{lem:wide} and Lemma~\ref{lem:recenter-initialization}
put the starting Newton decrement of $c\Psi_{\kappa}$ below $1/8$.
For later penalty phases, the same follows from the true outgoing
residual of the previous phase, as verified below. Since $\nabla^{2}(c\Psi_{\kappa})(x_{0})\succeq400K_{0}$,
the self-concordant distance bound~\cite[Equation 2.6c]{NT08} gives,
for the exact minimizer $x^{*}$ of this phase's objective, 
\begin{equation}
\|x^{*}-x_{0}\|_{K_{0}}\le\frac{1}{\sqrt{400}}\frac{\mathfrak{d}_{\kappa}(x_{0})}{1-\mathfrak{d}_{\kappa}(x_{0})}\le\frac{1}{140}.\label{eq:random-local-start}
\end{equation}
Set $R_{\mathrm{loc}}=1/20$ and $\mathcal{B}_{0}=\{x:\|x-x_{0}\|_{K_{0}}\le R_{\mathrm{loc}}\}$.
For every $x\in\mathcal{B}_{0}$ and every retained row $i$, 
\[
\frac{|s_{i}(Jx)-s_{i}(Jx_{0})|}{s_{i}(Jx_{0})}=\frac{\left|v^{\top}_{i}J(x-x_{0})\right|}{s_{i}(Jx_{0})}\le\|x-x_{0}\|_{\calh_{S}(x_{0})}\le\|x-x_{0}\|_{K_{0}}\le R_{\mathrm{loc}}.
\]
In particular, this entire ball is feasible, and comparison of the
individual scaled rows gives $(1+R_{\mathrm{loc}})^{-2}\frac{v_{i}v^{\top}_{i}}{s_{i}(Jx_{0})^{2}}\preceq\frac{v_{i}v^{\top}_{i}}{s_{i}(Jx)^{2}}\preceq(1-R_{\mathrm{loc}})^{-2}\frac{v_{i}v^{\top}_{i}}{s_{i}(Jx_{0})^{2}}$
and
\begin{equation}
\begin{split}(1+R_{\mathrm{loc}})^{-2}\calh_{S}(x_{0}) & \preceq\calh_{S}(x)\preceq(1-R_{\mathrm{loc}})^{-2}\calh_{S}(x_{0}),\\
(1+R_{\mathrm{loc}})^{-2}K_{0} & \preceq\mathcal{K}_{\kappa}(x)\preceq(1-R_{\mathrm{loc}})^{-2}K_{0}.
\end{split}
\label{eq:random-local-metrics}
\end{equation}
The second line holds because the positive semidefinite penalty $\varrho_{\kappa}\Gamma^{\top}\Gamma$
is fixed throughout the phase. Applying~\eqref{eq:penalty-hessian},
or Lemma~\ref{lem:hybrid-sc} when unpenalized, yields 
\begin{equation}
m_{0}K_{0}\preceq\nabla^{2}\Psi_{\kappa}(x)\preceq L_{0}K_{0},\qquad m_{0}=\frac{\lambda}{(1+R_{\mathrm{loc}})^{2}},\quad L_{0}=\frac{3+\lambda}{(1-R_{\mathrm{loc}})^{2}}.\label{eq:random-local-hessian}
\end{equation}
These constants depend on $\lambda$ alone; in particular $m_{0}\ge\lambda/2$
and $L_{0}<4$.

\emph{Exact contraction and the two sources of error.} Write $g_{t}=\nabla\Psi_{\kappa}(x_{t})$,
and form $\widehat{g}_{t}=Z(x_{t})^{\top}(\widehat{\ell}_{t}+\lambda\boldsymbol{1})+e^{\kappa}\Gamma^{\top}\Gamma x_{t}$,
omitting the last term when inactive. By~\eqref{eq:random-gradient-error}
and~\eqref{eq:random-local-metrics}, 
\begin{equation}
\|\widehat{g}_{t}-g_{t}\|_{K^{-1}_{0}}\le\frac{\epsilon_{0}}{1-R_{\mathrm{loc}}}=\frac{20}{19}\epsilon_{0}<2\epsilon_{0}.\label{eq:random-local-gradient-error}
\end{equation}
The approximate solver can compute a direction $d_{t}$ satisfying
\begin{equation}
\|d_{t}-K^{-1}_{0}\widehat{g}_{t}\|_{K_{0}}=\|\widehat{g}_{t}-K_{0}d_{t}\|_{K^{-1}_{0}}\le\epsilon_{0},\qquad x_{t+1}=x_{t}-L^{-1}_{0}d_{t}.\label{eq:random-local-direction}
\end{equation}
To see the contraction explicitly, suppose $x_{t}\in\mathcal{B}_{0}$
and let 
\[
\overline{B}_{t}=\int^{1}_{0}\nabla^{2}\Psi_{\kappa}(x^{*}+u(x_{t}-x^{*}))\,du.
\]
The segment lies in the convex ball $\mathcal{B}_{0}$, $g_{t}=\nabla\Psi_{\kappa}(x_{t})=\overline{B}_{t}(x_{t}-x^{*})$,
and $m_{0}I\preceq K^{-1/2}_{0}\overline{B}_{t}K^{-1/2}_{0}\preceq L_{0}I$.
Thus the exact preconditioned gradient update has error 
\[
K^{1/2}_{0}(x_{t}-L^{-1}_{0}K^{-1}_{0}g_{t}-x^{*})=\left(I-L^{-1}_{0}K^{-1/2}_{0}\overline{B}_{t}K^{-1/2}_{0}\right)K^{1/2}_{0}(x_{t}-x^{*}),
\]
whose norm is at most $(1-m_{0}/L_{0})\|x_{t}-x^{*}\|_{K_{0}}$. The
score error and solver error perturb this update by at most $L^{-1}_{0}(2\epsilon_{0}+\epsilon_{0})$.
Consequently, with $\Delta_{t}=\|x_{t}-x^{*}\|_{K_{0}}$, 
\begin{equation}
\begin{split}\Delta_{t+1} & \le(1-m_{0}/L_{0})\Delta_{t}+L^{-1}_{0}\left\Vert \widehat{g}_{t}-g_{t}\right\Vert _{K^{-1}_{0}}+L^{-1}_{0}\left\Vert d_{t}-K^{-1}_{0}\widehat{g}_{t}\right\Vert _{K_{0}}\\
 & \le(1-m_{0}/L_{0})\Delta_{t}+3\epsilon_{0}/L_{0},\\
\Delta_{t} & \le(1-m_{0}/L_{0})^{t}\Delta_{0}+\frac{3\epsilon_{0}}{m_{0}}\bigl(1-(1-m_{0}/L_{0})^{t}\bigr).
\end{split}
\label{eq:random-local-recurrence}
\end{equation}

\emph{Feasibility of every correction.} The limit of the right hand
side in~\eqref{eq:random-local-recurrence} satisfies 
\[
\frac{3\epsilon_{0}}{m_{0}}=\frac{3(21/20)^{2}}{2^{20}}\,\overline{\tau}<\frac{1}{140}.
\]
Starting from~\eqref{eq:random-local-start}, induction gives $\Delta_{t}\le1/140$
and hence 
\[
\|x_{t}-x_{0}\|_{K_{0}}\le\Delta_{t}+\|x^{*}-x_{0}\|_{K_{0}}\le\frac{1}{70}<R_{\mathrm{loc}}
\]
for every computed iterate. Indeed, assuming the current iterate is
in the ball justifies the contraction estimate, which then puts the
next iterate strictly inside the ball. This proves feasibility without
assuming that a future iterate is feasible. It also justifies all
the metric and gradient-error comparisons used above.

\emph{A cheap stopping certificate.} We can verify if the randomized
approximate $K_{0}$ solver has achieved the required error early
as follows. We run preconditioned gradient descent as in Lemma \ref{lem:random-restricted-solver}
to compute $d_{t}$ until its exact residual $r_{t}=\widehat{g}_{t}-K_{0}d_{t}$
satisfies $\|r_{t}\|_{2}\le\sqrt{2}\epsilon_{0}$. Since $K_{0}\succeq2I$,
we have$\left\Vert d_{t}-K^{-1}_{0}\widehat{g}_{t}\right\Vert _{K_{0}}=\left\Vert r_{t}\right\Vert _{K^{-1}_{0}}\le\left\Vert r_{t}\right\Vert _{2}/\sqrt{2}\le\epsilon_{0}$This
implies the guarantee required in \eqref{eq:random-local-direction}.
Next we show the solver of Lemma~\ref{lem:random-restricted-solver}
can reach $\|r_{t}\|_{2}\le\sqrt{2}\epsilon_{0}$ within a logarithmic
number of iterations. For the zero initial solution, the initial error
is $\|\widehat{g}_{t}\|_{K^{-1}_{0}}\le L_{0}\Delta_{t}+\left\Vert \widehat{g}_{t}-g_{t}\right\Vert _{K^{-1}_{0}}\le L_{0}/140+2\epsilon_{0}<1$.
Note that $\|K_{0}\|_{2}\le U_{\mathrm{ds}}$so the error bound $\|d_{t}-K^{-1}_{0}\widehat{g}_{t}\|_{K_{0}}\le\sqrt{2}\epsilon_{0}/\sqrt{U_{\mathrm{ds}}}$
suffices, since 
\[
\|r_{t}\|_{2}\le\sqrt{U_{\mathrm{ds}}}\,\|d_{t}-K^{-1}_{0}\widehat{g}_{t}\|_{K_{0}}.
\]
The value $\sqrt{2}\epsilon_{0}/\sqrt{U_{\mathrm{ds}}}$ is inverse
polynomial in $n$, so only a logarithmic number of inverse maintenance
calls is needed.

Using the same $d_{t}$, compute 
\[
U^{2}_{t}=2\widehat{g}^{\top}_{t}d_{t}-d^{\top}_{t}K_{0}d_{t}+\tfrac{1}{2}\|r_{t}\|^{2}_{2}.
\]
The residual identity 
\[
\|\widehat{g}_{t}\|^{2}_{K^{-1}_{0}}=2\widehat{g}^{\top}_{t}d_{t}-d^{\top}_{t}K_{0}d_{t}+r^{\top}_{t}K^{-1}_{0}r_{t}
\]
and the fact $K^{-1}_{0}\preceq I/2$ prove that $U^{2}_{t}\ge0$
and 
\begin{equation}
\|\widehat{g}_{t}\|_{K^{-1}_{0}}\le U_{t}\le\|\widehat{g}_{t}\|_{K^{-1}_{0}}+\epsilon_{0}.\label{eq:random-local-certificate}
\end{equation}

The algorithm return $x_{t}$ as soon as $U_{t}\le\overline{\tau}/4$.
By~\eqref{eq:random-local-metrics} and the bound $\left\Vert \nabla\Psi_{\kappa}(x_{t})-\widehat{g}_{t}\right\Vert _{\mathcal{K}_{\kappa}(x_{t})^{-1}}\le\epsilon_{0}$
from \eqref{eq:random-gradient-error}, this certifies the true outgoing
residual 
\begin{equation}
\begin{split}\|\nabla\Psi_{\kappa}(x_{t})\|_{\mathcal{K}_{\kappa}(x_{t})^{-1}} & \le(1+R_{\mathrm{loc}})\|\widehat{g}_{t}\|_{K^{-1}_{0}}+\epsilon_{0}\\
 & \le\frac{21}{20}\frac{\overline{\tau}}{4}+\epsilon_{0}<\overline{\tau}.
\end{split}
\label{eq:random-phase-tolerance}
\end{equation}

\emph{A uniform bound on the number of corrections.} Integrating the
Hessian along the segment from $x^{*}$ gives 
\begin{align*}
\|g_{t}\|_{K^{-1}_{0}} & =\left\Vert \int^{1}_{0}K^{-1/2}_{0}\nabla^{2}\Psi_{\kappa}(x^{*}+u(x_{t}-x^{*}))K^{-1/2}_{0}K^{1/2}_{0}(x_{t}-x^{*})\,du\right\Vert _{2}\\
 & \le\int^{1}_{0}\left\Vert K^{-1/2}_{0}\nabla^{2}\Psi_{\kappa}(x^{*}+u(x_{t}-x^{*}))K^{-1/2}_{0}\right\Vert \cdot\left\Vert K^{1/2}_{0}(x_{t}-x^{*})\right\Vert \,du\\
 & \le L_{0}\Delta_{t}
\end{align*}
Combining this with \eqref{eq:random-local-gradient-error}--\eqref{eq:random-local-certificate}
and~\eqref{eq:random-local-recurrence} yields 
\begin{align*}
U_{t} & \overset{\eqref{eq:random-local-certificate}}{\le}\|\widehat{g}_{t}\|_{K^{-1}_{0}}+\epsilon_{0}\\
 & \overset{\eqref{eq:random-local-gradient-error}}{\le}\|g_{t}\|_{K^{-1}_{0}}+3\epsilon_{0}\\
 & \le L_{0}\Delta_{t}+3\epsilon_{0}\\
 & \overset{\eqref{eq:random-local-recurrence}}{\le}\frac{L_{0}}{140}(1-m_{0}/L_{0})^{t}+\frac{3L_{0}\epsilon_{0}}{m_{0}}+3\epsilon_{0}.
\end{align*}
Since $m_{0}\ge\lambda/2$, $L_{0}<4$, and $\epsilon_{0}=\lambda\overline{\tau}/2^{20}$,
the last two terms are at most $27\overline{\tau}/2^{20}<\overline{\tau}/8$.
Therefore the stopping condition $U_{t}\le\overline{\tau}/4$ must
pass after at most 
\begin{equation}
N_{\mathrm{loc}}=\left\lceil \frac{L_{0}}{m_{0}}\log\left(1+\frac{8L_{0}}{140\overline{\tau}}\right)\right\rceil \label{eq:random-local-iteration-count}
\end{equation}
updates, using $(1-m_{0}/L_{0})^{t}\le e^{-tm_{0}/L_{0}}$. This bound
depends only on the fixed constants $\lambda$ and $\tau$, and is
independent of $n$, $M$, and the penalty parameter $\kappa$.

\emph{Preserving the phase guarantees.} Both $x_{t}$ and $x^{*}$
lie in $\mathcal{B}_{0}$, so their slacks satisfy 
\[
\frac{19}{21}\le\frac{s_{i}(Jx_{t})}{s_{i}(Jx^{*})}\le\frac{21}{19}.
\]
In particular the slack comparison with the exact center needed for
\eqref{eq:random-H-range} holds throughout the phase. At the returned
point $x$, strong convexity in the fixed norm $\left\Vert \cdot\right\Vert _{K_{0}}$,
applied on the segment to $x^{*}$, also gives 
\begin{align*}
\Psi_{\kappa}(x)-\Psi_{\kappa}(x^{*}) & \le\frac{\|\nabla\Psi_{\kappa}(x)\|^{2}_{K^{-1}_{0}}}{2m_{0}}\\
 & \le\frac{(1+R_{\mathrm{loc}})^{2}}{2(1-R_{\mathrm{loc}})^{2}}\frac{\overline{\tau}^{2}}{\lambda}<\frac{\overline{\tau}^{2}}{\lambda}.
\end{align*}
Thus the same objective-gap and penalty-energy budgets as in \eqref{eq:penalty-gap}
hold. In particular, during a contraction, $e^{\kappa}\|\Gamma x\|^{2}_{2}/2\le\Delta\Phi+\overline{\tau}^{2}/\lambda$.
The next-phase Newton decrement estimate in the proof of Lemma~\ref{lem:penalty-newton-corrections}
uses only the true residual and the prescribed penalty increment.
It therefore still gives 
\[
\mathfrak{d}_{\kappa_{+}}(x)\le\sqrt{c/\lambda}\,\overline{\tau}+\frac{1}{64}<\frac{1}{32}<\frac{1}{8}.
\]
This closes the induction on phase starts. The stage-count and computed-energy
arguments consequently apply with a constant number $N_{\mathrm{loc}}$
of iterations per phase.

\emph{Synchronizing the maintenance instances.} Before freezing $K_{0}$,
install any newly inserted row and the phase's current penalty weights
in the inverse-maintenance instances. The inverse-maintenance instances
for $\calh$ and $F$ may hold their phase starting weights until
the correction finishes; only the leverage-score instance follows
the intermediate iterates. They then update to the outgoing slacks.
This supplies the frozen-$K_{0}$ solves without copying or rebuilding
a data structure. At every phase start they are synchronized with
its starting point and current penalty. After a phase ends, replay
the leverage score instance's slack updates during the phase to the
two frozen inverse-maintenance instances so that their weights match
at the end of a phase.

The projection test of Section~\ref{sec:update-center} described
in \textsc{ProjectAndCheck} does not involve any matrix inverse and
thus, it not affected by the approximate solver.  

Next we account for the number of inverse and leverage score maintenance
updates. Let $z_{+}$ and $z$ be two consecutive iterates of the
preconditioned gradient descent. If $z_{+}=z+p$ and $r=\|p\|_{\calh}<1$,
put $\eta_{i}=-v^{\top}_{i}p/s_{i}(z)$. Then $\|\eta\|_{2}=r$, $s_{i}(z_{+})=s_{i}(z)(1+\eta_{i})$,
and 
\begin{equation}
\|\log D(z_{+})-\log D(z)\|_{2}=2\|\log(1+\eta)\|_{2}\le\frac{2r}{1-r}.\label{eq:random-log-drift}
\end{equation}
For a local correction within the region where the Newton decrement
is bounded by $1/8$, the invariant above ($\Delta_{t}\le1/140$)
gives $\|x_{t+1}-x_{t}\|_{K_{0}}\le\Delta_{t+1}+\Delta_{t}\le1/70$.
Hence~\eqref{eq:random-local-metrics} bounds its current $\calh_{S}$-length
by $(20/19)/70=2/133$, so its log-weight drift is at most $4/131$.
The direct segment from $x_{0}$ to the phase endpoint has $K_{0}$-length
at most $1/70$ as well. Each local correction, its phase endpoint
update, and each accepted projection therefore takes a constant number
of stable diagonal updates, by subdividing a feasible segment when
necessary. This subdivision makes no oracle calls.

If a contraction removes $r_{j}$ dimensions, changing its penalty
parameter by $\Delta\kappa$ changes the $\ell_{2}$ norm of its penalty
log-weights by $\sqrt{r_{j}}\Delta\kappa$. The whole path requires
\begin{equation}
O\!\left(T_{j}+\sqrt{r_{j}}(\kappa_{\max,j}-\kappa_{0})\right)=O(T_{j}+\sqrt{r_{j}}\log(n+2))\label{eq:random-penalty-update-count}
\end{equation}
internal stable updates. As $\sum_{j}r_{j}\le n-1$, $\sum_{j}\sqrt{r_{j}}\le n-1$,
and $\sum_{j}T_{j}=O(n\log(n+2))$, the total is $O(n\log(n+2))$.
Introducing these directions and subsequently changing their individual
weights to $\varpi$ uses only $O(n)$ single-coordinate updates.
These counts include contractions that remove many dimensions at once.

\subsection{Approximate wide/thin tests}

Let $z=Jy$ be the current certified center and let $b\in S$ be the
unit vector supplied by a nonzero homogeneous oracle call. Thus $\|b\|_{2}=1$
and $b^{\top}z\ge0$. The original wide/thin rule uses quantity $q=b^{\top}\calh^{-1}b$,
with $\calh=\calh_{fl}(z)$. We compute an upper estimate satisfying
\begin{equation}
q\le q^{+}\le\tfrac{9}{8}q.\label{eq:random-direction-estimate}
\end{equation}
The bounds $2I\preceq\calh\preceq L_{*}I$ imply 
\[
\frac{1}{L_{*}}\le q\le\frac{1}{2}.
\]

Let $d^{*}=\calh^{-1}b$. Use the maintained approximate solver for
$\calh$ to compute $d$ satisfying
\begin{equation}
\|d-d^{*}\|_{\calh}\le\frac{1}{2L_{*}}.\label{eq:random-direction-solve-accuracy}
\end{equation}
This is an inverse-polynomial target so it can be reached with an
overhead of $O(\log(n+2))$ steps as shown in the runtime of Lemma
\ref{lem:random-inverse-interface}. Let $r_{b}=b-\calh d$ be the
residual error.

\emph{Certifying $q^{+}$ from $d$ and its residual.} For the returned
$d$, compute 
\begin{equation}
q^{+}=2b^{\top}d-d^{\top}\calh d+\tfrac{1}{2}\|r_{b}\|^{2}_{2}.\label{eq:random-direction-qplus}
\end{equation}
Expanding $r_{b}=b-\calh d$ gives the exact identity 
\begin{align*}
q & =\left(\calh d+r_{b}\right)^{\top}\calh^{-1}\left(\calh d+r_{b}\right)\\
 & =2r^{\top}_{b}d+d^{\top}\calh d+r^{\top}_{b}\calh^{-1}r_{b}\\
 & =2b^{\top}d-d^{\top}\calh d+r^{\top}_{b}\calh^{-1}r_{b}.
\end{align*}
Since $0\preceq\calh^{-1}\preceq I/2$, 
\begin{equation}
0\le q^{+}-q=\tfrac{1}{2}\|r_{b}\|^{2}_{2}-r^{\top}_{b}\calh^{-1}r_{b}\le\tfrac{1}{2}\|r_{b}\|^{2}_{2}\le\frac{1}{8L_{*}}\le\frac{q}{8}.\label{eq:random-direction-certificate-error}
\end{equation}
This proves~\eqref{eq:random-direction-estimate}.

\emph{The wide decision and maintenance of the inserted row.} Set
$\widehat{\upsilon}=\sqrt{q^{+}/\gamma}$. The exact $\upsilon=\sqrt{q/\gamma}$
is used only in the analysis. Equation \eqref{eq:random-direction-estimate}
gives 
\[
\upsilon\le\widehat{\upsilon}\le\sqrt{9/8}\,\upsilon.
\]
If $\widehat{\upsilon}\ge\delta$, insert the row normal $b$ with
offset $\rho=b^{\top}z+\widehat{\upsilon}$. Its slack at $z$ is
$\widehat{\upsilon}>0$, and $\rho\ge\widehat{\upsilon}\ge\delta$.
Thus $z$ remains strictly feasible and the new body still contains
$\delta B^{n}_{2}$, as in Lemma~\ref{lem:wide}.

Write $\bar{b}=b/\widehat{\upsilon}$ for the new scaled row. Its
pre-insertion leverage is 
\begin{equation}
\gamma_{\mathrm{eff}}=\bar{b}^{\top}\calh^{-1}\bar{b}=\frac{q}{\widehat{\upsilon}^{2}}=\gamma\frac{q}{q^{+}}\in[8\gamma/9,\gamma].\label{eq:random-effective-leverage}
\end{equation}
Cauchy--Schwarz in the $\calh$ norm gives $(\bar{b}^{\top}v)^{2}\le\left\Vert \bar{b}\right\Vert ^{2}_{\calh^{-1}}\left\Vert v\right\Vert ^{2}_{\calh}=\gamma_{\mathrm{eff}}\,v^{\top}\calh v$
for every vector $v$. Consequently, 
\[
\bar{b}\bar{b}^{\top}\preceq\gamma_{\mathrm{eff}}\calh\preceq\gamma\calh\preceq0.01\calh.
\]
This meets the small-row insertion requirement of the leverage score
maintenance interface. The two inverse-maintenance instances each
receive the row vector $b$ with weight $\widehat{\upsilon}^{-2}$
as a single-row update.

After insertion, $\calh_{+}=\calh+\bar{b}\bar{b}^{\top}$. Sherman--Morrison
gives the new row's leverage score as 
\[
\bar{b}^{\top}\calh^{-1}_{+}\bar{b}=\frac{\gamma_{\mathrm{eff}}}{1+\gamma_{\mathrm{eff}}}.
\]

\emph{Recentering and potential progress after a wide decision.} Let
$\Delta\ell_{i}=\ell_{i}-\ell^{+}_{i}$ for the old rows. Sherman--Morrison
and preservation of the leverage score sum give 
\[
\Delta\ell_{i}\ge0,\qquad\|\Delta\ell\|_{2}\le\|\Delta\ell\|_{1}=\frac{\gamma_{\mathrm{eff}}}{1+\gamma_{\mathrm{eff}}}\le\gamma.
\]
Also $\|J^{\top}\bar{b}\|_{\calh_{S}(y)^{-1}}\le\sqrt{\gamma_{\mathrm{eff}}}\le\sqrt{\gamma}$.
The gradient calculation of Lemma~\ref{lem:wide} therefore gives
\[
\|\nabla\Psi_{S,+}(y)\|_{\calh_{S}(y)^{-1}}\le\tau+\gamma+(\gamma+\lambda)\sqrt{\gamma}\le2\lambda\sqrt{\gamma}.
\]
It follows that the starting decrement for $c\Psi_{S,+}$ is at most
$40\sqrt{\gamma}<1/8$. The local corrections of the preceding subsection
therefore restore the true center tolerance in a constant number of
corrections. The same decrement estimate bounds the gap from the barrier
value at $z$ to its new exact minimum by $4\lambda\gamma$.

On the other hand, the determinant lemma and the nonnegative new logarithmic
term give 
\[
\Psi_{fl,+}(z)-\Psi_{fl}(z)=\tfrac{1}{2}\log(1+\gamma_{\mathrm{eff}})+\lambda\log(\rho/\widehat{\upsilon})\ge\frac{\gamma_{\mathrm{eff}}}{4}\ge\frac{2\gamma}{9}.
\]
Combining the gain with the gap bound yields 
\begin{equation}
\min_{P_{+}\cap S}\Psi_{fl,+}-\min_{P\cap S}\Psi_{fl}\ge(2/9-4\lambda)\gamma\ge\gamma/8.\label{eq:random-wide-progress}
\end{equation}
The last inequality uses the fixed $\lambda=2^{-10}$. Thus the original
potential argument still pays for every inserted row, preserving the
insertion and row budgets.

\emph{The thin decision.} If $\widehat{\upsilon}<\delta$, we do not
insert the row. Since the estimate is an upper bound, 
\[
q\le q^{+}=\gamma\widehat{\upsilon}^{2}<\gamma\delta^{2}.
\]
Let $q_{S}=(J^{\top}b)^{\top}\calh_{S}(y)^{-1}(J^{\top}b)$. Lemma~\ref{lem:restricted-inverse}
gives $q_{S}\le q$, so the argument of Lemma~\ref{lem:floor} gives
\begin{equation}
\begin{split}\operatorname{width}_{b}(P\cap S) & \le\frac{8W}{\lambda}\sqrt{q_{S}}<\frac{8W\sqrt{\gamma}}{\lambda}\delta\le\chi_{n},\\
\operatorname{tr}\calh_{S}(y) & \ge\frac{1}{q_{S}}>\frac{1}{\gamma\delta^{2}}.
\end{split}
\label{eq:random-thin-consequences}
\end{equation}
Here $\|J^{\top}b\|_{2}=1$, and the trace bound follows from $(q_{S})((J^{\top}b)^{\top}\calh_{S}(y)(J^{\top}b))\ge1$.
These are the width and trace estimates used by the certificate construction
in the next subsection.

\subsection{A certificate from constant-accuracy leverage scores}

We give the modified version of Lemma \ref{lem:center-to-certificate}
using the approximate solver for $\calh_{S}$ from Lemma \ref{lem:random-restricted-solver}
instead of an exact solver.
\begin{lem}[Correcting the certificate weights]
\label{lem:random-certificate} At a certified thin-event point,
estimates satisfying \eqref{eq:random-score-accuracy} and an approximate
$\calh_{S}$ solver suffice to compute simplex coefficients on the
nonzero projected inserted rows satisfying \eqref{eq:unit-residual}.
The cost is $\widetilde{O}(n^{2})$ arithmetic operations and no value-oracle
calls. 
\end{lem}

\begin{proof}
Let $u=\widehat{\ell}+\lambda\boldsymbol{1}$. During the procedure,
hold $z$, $J$, $Z$, $\calh_{S}$ and this chosen vector $u$ fixed.
Since $J$ has orthonormal columns, 
\[
2I\preceq\calh_{S}=J^{\top}\calh J\preceq L_{*}I,
\]
so the target $d^{*}=\calh^{-1}_{S}Z^{\top}u$ exists and is unique.
That the thin-event point is a certified center and \eqref{eq:random-gradient-error}
give the initial-error bound 
\[
\|d^{*}\|_{\calh_{S}}=\|Z^{\top}u\|_{\calh^{-1}_{S}}\le\|\nabla\Psi_{S}\|_{\calh^{-1}_{S}}+\|\nabla\Psi_{S}-Z^{\top}u\|_{\calh^{-1}_{S}}\le\tau+\epsilon_{0}.
\]
By applying Lemma \ref{lem:random-restricted-solver} without the
penalty term, we obtain $d$ satisfying $\left\Vert d-d^{*}\right\Vert _{\calh_{S}}\le\frac{\delta}{\sqrt{L_{*}}}$after
at most 
\[
N_{\mathrm{cert}}=\left\lceil \frac{\log\max\{1,\sqrt{L_{*}}(\tau+\epsilon_{0})/\delta\}}{\log(4/3)}\right\rceil =O(\log(n+2))
\]
preconditioned gradient descent steps. Since $\calh_{S}\preceq L_{*}I$,
we can bound the residual $\zeta=\calh_{S}(d^{*}-d)=Z^{\top}u-\calh_{S}d$:
\[
\left\Vert \zeta\right\Vert _{2}\le\sqrt{L_{*}}\left\Vert d-\calh^{-1}_{S}Z^{\top}u\right\Vert _{\calh_{S}}\le\delta
\]

Define corrected weights 
\[
w=u-Zd,\qquad\varphi_{i}=w_{i}/s_{i}(z).
\]
Since $Z^{\top}Z=\calh_{S}\succeq2I$, certification of $z$ and \eqref{eq:random-gradient-error}
give 
\[
\|Zd\|_{2}=\|\calh_{S}d\|_{\calh^{-1}_{S}}\le\|Z^{\top}u\|_{\calh^{-1}_{S}}+\left\Vert \zeta\right\Vert _{\calh^{-1}_{S}}\le\tau+\epsilon_{0}+\delta/\sqrt{2}<3\tau.
\]
Clipping of $\widehat{\ell}_{i}$ to $[0,1]$ ensures $u_{i}\ge\lambda$.
Hence 
\begin{equation}
w_{i}\ge\lambda-3\tau\ge\lambda/2,\qquad\sum_{i}w_{i}\le\underbrace{\sum_{i}(\ell_{i}+\lambda)}_{\le W}+\underbrace{\left\Vert \widehat{\ell}-\ell\right\Vert _{1}}_{\le\sqrt{M}\epsilon_{0}}+\underbrace{\left\Vert Zd\right\Vert _{1}}_{\le3\sqrt{M}\tau}\le2W,\qquad Z^{\top}w=\zeta.\label{eq:random-positive-weights}
\end{equation}
Writing $R=\sqrt{n}$ and using $\|y\|_{2}\le R$, the radial identity
is 
\begin{align}
T_{\rho} & :=\sum_{i}\varphi_{i}\rho_{i}=\sum_{i}\frac{w_{i}}{s_{i}(z)}(s_{i}(z)+v^{\top}_{i}Jy)\label{eq:random-radial-identity}\\
 & =\sum_{i}w_{i}+y^{\top}Z^{\top}w\\
 & =\sum_{i}w_{i}+y^{\top}\zeta\le2W+R\delta.
\end{align}
All terms $\varphi_{i}\rho_{i}$ are positive.

Let $\mathcal{C}$ denote the cube rows and $\mathcal{R}$ the inserted
rows. Cube rows have $\rho_{i}=1$ and $\|J^{\top}v_{i}\|_{2}\le1$,
so $\sum_{i\in\mathcal{C}}\varphi_{i}\|J^{\top}v_{i}\|_{2}\le T_{\rho}$.
The thin test and Lemma~\ref{lem:restricted-inverse} give us a unit
vector $J^{\top}b\in\R^{N}$ such that $\left(J^{\top}b\right)^{\top}\calh^{-1}_{S}\left(J^{\top}b\right)<\gamma\delta^{2}$
, which implies $\sqrt{\tr\calh_{S}}\ge\sqrt{\left\Vert \calh_{S}\right\Vert _{2}}>1/(\sqrt{\gamma}\,\delta)$.
Therefore the mass of the inserted rows satisfies 
\begin{align*}
\Lambda_{\mathcal{R}} & :=\sum_{i\in\mathcal{R}}\varphi_{i}\|J^{\top}v_{i}\|_{2}\\
 & =\sum_{i}\varphi_{i}\|J^{\top}v_{i}\|_{2}-\sum_{i\in\mathcal{C}}\varphi_{i}\|J^{\top}v_{i}\|_{2}\\
 & \ge\frac{\lambda}{2}\sum_{i}\frac{\|J^{\top}v_{i}\|_{2}}{s_{i}(z)}-T_{\rho}\\
 & \ge\frac{\lambda}{2}\sqrt{\tr\calh_{S}}-T_{\rho}\ge\frac{\lambda}{2\sqrt{\gamma}\delta}-(2W+R\delta)\ge\frac{\lambda}{4\sqrt{\gamma}\,\delta}>0.
\end{align*}
The last line uses $T_{\rho}\le2W+R\delta$, $W\le2Q_{n}$, and $2W+R\delta\le\frac{\lambda}{4\sqrt{\gamma}\delta}$.
Set $p_{i}=\varphi_{i}\|J^{\top}v_{i}\|_{2}/\Lambda_{\mathcal{R}}$
on the nonzero projected inserted rows. These are positive and sum
to one. Removing the cube rows from $Z^{\top}w=\zeta$ gives 
\begin{align*}
\left\Vert \sum_{i\in\mathcal{R}}p_{i}\frac{J^{\top}v_{i}}{\|J^{\top}v_{i}\|_{2}}\right\Vert _{2} & \le\frac{\delta+T_{\rho}}{\Lambda_{\mathcal{R}}}\\
 & \le\frac{4\sqrt{\gamma}\,\delta}{\lambda}\bigl(2W+(R+1)\delta\bigr)\\
 & =\frac{W\delta}{2}+\frac{(R+1)\delta^{2}}{4}<2Q_{n}\delta=2\chi_{n},
\end{align*}
where $\sqrt{\gamma}/\lambda=1/16$. Thus the conversion from isometric
coordinates to compressed coordinates and all conditions of Lemma~\ref{lem:accuracy}
apply unchanged.

Finally, each refinement uses one protected approximate $F$ solve
and $O(Mn+n^{2})$ work for matrix--vector products. Together with
the explicit $O(\log(n+2))$ bound above and the inverse-maintenance
interface, this costs $\widetilde{O}(n^{2})$ arithmetic operations.
Constructing the weights and checking each residual uses $O(Mn)$
work. The geometry remains fixed throughout, and no refinement step
uses the value oracle. 
\end{proof}

{\small}{\small\par}

\subsection{Total cost for the randomized algorithm}
\begin{proof}[{\small Proof of Theorem~\ref{thm:randomized-time}}]
{\small We maintain the leverage-score structure and the two inverse-maintenance
instances for $\calh$ and $F$, using independent randomness. There
are adaptive queries so we use the protected adaptive solvers of~\cite[Section 5]{LS15}.
The output perturbation method for inverse maintenance solver can
condition on the other structures without revealing the internal randomness
of the instance being protected. There are still $O(n\log(n+2))$
homogeneous oracle calls and penalty phases, and at most $n$ thin
events.}{\small\par}

{\small Equations~\eqref{eq:random-log-drift} and the penalty-weight
accounting give $\widetilde{O}(n)$ maintenance rounds. Revealing
the inserted and contraction rows also costs only $\widetilde{O}(n)$
single-row rounds. There are $\widetilde{O}(n)$ local corrections,
directional tests, and certificate solves, each using at most a logarithmic
number of near-quadratic solves and matrix--vector products. The
complete matrix cost is therefore 
\[
\widetilde{O}\!\left(n^{\omega}+n\cdot n^{2}\right)=\widetilde{O}(n^{3}).
\]
Initialization of the padded leverage score structure also fits this
bound. The combinatorial bookkeeping already costs $O(n^{3}\log(n+2))$
by~\eqref{eq:bookkeeping}. The original $O(n^{2}\log(n+2))$ $\EO$
bound still holds.}{\small\par}
\end{proof}

\bibliographystyle{plain}
\bibliography{sfm-refs}

\end{document}